\documentclass[journal]{IEEEtran}
\UseRawInputEncoding 
\usepackage{cite}
\usepackage{hyperref} 

\ifCLASSINFOpdf
 \usepackage[pdftex]{graphicx}
  \graphicspath{{./figures/}}
  \usepackage{svg}

\else
\fi
\usepackage{amsmath, amssymb, amsfonts}
\usepackage{textcomp}
\usepackage{mathtools}

\usepackage{algorithmic}
\ifCLASSOPTIONcompsoc
\else
\fi
\usepackage{siunitx}
\usepackage{blindtext}
\usepackage[utf8]{inputenc} 
\usepackage[export]{adjustbox}
\usepackage[font=footnotesize]{caption}
\usepackage[labelformat=simple]{subcaption}
\usepackage{booktabs}
\usepackage[flushleft]{threeparttable}
\usepackage{makecell}

\usepackage{color}

\renewcommand{\figurename}[1]{Fig. #1}
\renewcommand{\tablename}{Table }
\DeclareCaptionLabelSeparator{dotspace}{. }
\usepackage{url}

\def\BibTeX{{\rm B\kern-.05em{\sc i\kern-.025em b}\kern-.08em
    T\kern-.1667em\lower.7ex\hbox{E}\kern-.125emX}}

\begin{document}
%
\title{Improved Decoupled Control of Modular Multilevel Converter
under Constraint of Nearest Level \\Modulation via Disturbance Observer Design}
%
%
%

\author{Jaeyeon~Park,~
       Dongjoon~Kim,~Seungjun Lee,~\IEEEmembership{Graduate Student Member,~IEEE,}
       and Shenghui~Cui,~\IEEEmembership{Member,~IEEE}

\vspace{-1em}

\thanks{
J. Park, D. Kim, S. Lee, and S. Cui 
are with the Department of Electrical and Computer Engineering,
Seoul National University, Seoul 08826, Republic of Korea (e-mail: cuish@snu.ac.kr)}

}

\maketitle
\global\csname @topnum\endcsname 0
\global\csname @botnum\endcsname 0
\begin{abstract}
Nearest level modulation (NLM) is an attractive modulation method
for its implementation simplicity in modular multilevel converter (MMC).
However, it introduces significant voltage and current distortion 
when the number of submodules (SMs) per arm is small,
as in medium-voltage applications.
While indirect modulation
offers fully decoupled control 
of ac-side current, dc-side current and SM capacitor energy,
its performance is fundamentally reliant on accurate arm voltage synthesis, 
making it incompatible with the large quantization error inherent in NLM. 
To resolve this conflict,
this paper proposes a new control strategy 
based on a disturbance observer (DOB). 
The key idea is
to estimate and actively compensate 
for the inevitable arm voltage synthesis error induced by NLM,
thereby enabling fully decoupled control of 
indirect-modulated MMC even under NLM operation with a small number of SMs.
A key advantage is its ease of implementation, 
as it requires no modifications to the conventional NLM and decoupled control structure.
The validity and effectiveness of the proposed method 
in improving current quality and decoupled SM energy control
are verified through both simulation and experimental results.
\end{abstract}
\section{Introduction}  \label{sec:Introduction}

\IEEEPARstart{M}{odular} multilevel converter (MMC) 
has been highlighted as a scalable voltage-source converter (VSC) for
high-power and high-voltage applications, such as 
high-voltage dc (HVDC) transmission system and 
medium-voltage dc (MVDC) distribution system,
owing to its modularity and excellent quality of voltage and current waveforms \cite{Cui2019modular, ansari2020mmc, Cui2020fault, Perez2021modular, sun2022beyond}.

In medium-voltage applications,
such as MMC-based static synchronous compensator\cite{wei2014research,Kim2025new} and motor drive \cite{makato2010medium, Jung2015control},
where the number of submodules (SMs) in each arm is relatively small
compared to HVDC applications,
the carrier-based pulse-width modulation (PWM) 
scheme \cite{Rohner2010modulation,Yi2018nearest,Nguyen2019phase,Yin2024improved} 
is widely used
to generate arm voltage following the given modulation reference,
because it can minimize the quantization issue of arm voltage induced by the small number of SMs.
However, this approach
introduces two primary challenges: high switching losses due to the high-switching frequency operation of SMs, 
and increased control hardware complexity to generate synchronized and high-time resolution carrier signals for each SM \cite{Leon2017energy}.

Due to these challenges associated with PWM schemes,
the nearest level modulation (NLM) scheme, also known as nearest level control (NLC), which only determines the number of inserted SMs at each control period \cite{Meshram2015simplified},
is often considered a more feasible option when implementation simplicity and efficiency are prioritized.
Especially MMC with a large number of SMs, such as those in HVDC applications,
can effectively utilize NLM because the quantization issue of arm voltage is less pronounced with a sufficient number of levels.
However, when the number of SMs is small,
NLM scheme leads to a significant error in the arm voltage synthesis
due to the rounding operation of the modulation reference to the nearest voltage level.
This issue brings a significant challenge in practice to address as follows,
since most vendors nowadays in industry are expanding business from HVDC applications to MVDC applications
and not the other way around:
The sophisticated controller system was developed and well established in orientation to HVDC applications,
therefore the firmware is developed to be feasible only with NLM to reduce communication burdens.
While developing a new controller system dedicated to MVDC applications enabling PWM functionality in the firmware
not only calls for significant investment but also prolongs the delivery time.

To address this limitation of NLM in MMC with a small number of SMs,
various NLM schemes have been proposed 
to increase the effective level of arm voltage and reduce the harmonic distortion of output voltage and current
\cite{Pengfei2015level,Lin2016improved,Nguyen2020nearest,Yin2021variable}.
In \cite{Pengfei2015level},
the rounding threshold of the NLM scheme is adjusted from 0.5 to 0.25 
to increase the effective number of ac-side voltage levels.
However, this method inherently modifies the average voltage of SM capacitor in steady state 
and does not suppress the circulating current.
These issues are resolved in \cite{Lin2016improved},
by altering the small offset between the positive and negative value
at double fundamental frequency. 
However, the method's performance is limited by its dependency on the ac-side power factor. 
Specifically, the initial phase of the offset must be adjusted according to the load, 
and consequently, effective circulating current suppression cannot be guaranteed across the entire range of power factors.
In \cite{Nguyen2020nearest} and \cite{Yin2021variable},
the model predictive control approach is applied to the NLM scheme 
to satisfy multiple control objectives, such as reducing the distortion of output current
and suppressing the circulating current.
Although both methods show improved performance,
they are not suitable for practical applications because of the lack of stability analysis for the control loop.

Focusing the way to generate modulation reference,
there are two main approaches:
direct modulation and indirect modulation
(which is also known as compensated modulation) \cite{Suman2015operation}.
The direct modulation generates the modulation reference based on the constant nominal SM capacitor voltage,
while the indirect modulation generates it based on the instantaneously measured SM capacitor voltage \cite{antonopoulos2009dynamics}.
Although the direct modulation inherently balances the SM capacitor voltage in steady state,
it results in considerable second-order circulating currents which must be suppressed by 
extra passive \cite{Tu2010parameter,Jacobson2010VSCHVDC} or active \cite{Tu2011reduced} means.
The dc-link voltage of the MMC is tightly coupled to the average SM capacitor voltage,
and it results in considerable second-order oscillation voltages in the dc link in the case of ac-side single-line-to-ground (SLG)
fault and can lead to malfunction of protection systems \cite{Tu2012suppressing}.
Moreover, direct modulation based control of MMC
results in complicated impedance characteristics on the ac side in low frequency regions
due to inherently incorrect arm voltage synthesis \cite{Heng22018ac,Luca2019method},
and it leads to instability issues due to interactions with ac grids.
Such instability issue has been intensively reported in academia and industry in recent years
\cite{Lyu2017subsynchronous,Lyu2018optimal},
and sophisticated harmonic state space analysis has to be employed to analyze the 
impedance characteristics of the MMC to enhance the stability \cite{Heng2020impedance}.
On the other hand, 
the indirect modulation based decoupled control 
can inherently avoid second-order circulating currents,
and dc-link voltage can be fully decoupled from the average SM capacitor voltage \cite{antonopoulos2009dynamics,Sekiguchi2014grid,Cui2014comprehensive}.
Therefore, inherently,
no second-order oscillation voltage appears 
in the dc link in the case of a grid side SLG fault.
Different from two-level converters or MMC under direct modulation based control,
the dc-link current can be directly controlled \cite{Cui2018comprehensive}.
Moreover, the impedance characteristics of the MMC on ac side becomes 
as simple as two-level converter due to accurate arm voltage synthesis,
which significantly facilitates the impedance analysis to enhance the stability \cite{Luca2019effects}.

However, this decoupled control capability comes with a critical prerequisite. 
Since indirect modulation utilizes current control loops not only to control grid-side and dc-link currents but also to balance the SM capacitor energy, 
its performance fundamentally relies on the accurate synthesis of the arm voltage references. 
Consequently, this approach is typically considered effective only for MMCs that can ensure minimal voltage distortion,
such as those employing PWM \cite{Sekiguchi2014grid} or those with a sufficiently large number of SMs \cite{Leon2017energy}.

A common observation of the aforementioned improved NLM schemes \cite{Pengfei2015level,Lin2016improved,Nguyen2020nearest,Yin2021variable} is that
they are all developed based on the direct modulation approach.
Therefore, they inherit its fundamental limitation described above.
To the best of the authors' knowledge,
there is no existing
proper NLM scheme for indirect-modulated MMC with a small number of SMs 
that can fully decouple ac-side current control, dc-side current control, and SM capacitor energy balancing.

To address this research gap,
this paper proposes a new control strategy for NLM operation of indirect-modulated MMC with a small number of SMs per arm.
The key idea of the proposed method 
is employing a disturbance observer (DOB)
to estimate and compensate for the arm voltage synthesis error induced by the rounding operation of NLM.
A key advantage of the proposed method is that the DOB acts as an add-on component to the existing control structure,
enabling the effective use of the well established controller system with a conventional NLM scheme and decoupled control structure 
without requiring any modifications to them.
This characteristic significantly simplifies the implementation of the proposed method in practical applications.

The rest of the paper is organized as follows.
Section \ref{sec:TheoreticalBackground}
introduces the theoretical background of the proposed method,
including the system model of MMC, overview of the decoupled control strategy and NLM operation for indirect-modulated MMC,
and the principle of DOB.
Section \ref{sec:proposed} 
analyzes the arm voltage synthesis error of NLM and presents the proposed control strategy.
To ensure the stability of the overall system when the proposed method is integrated, 
a stability analysis is also provided.
In Section \ref{sec:simulation} and Section \ref{sec:experiment},
the effectiveness of the proposed method is validated through simulation and experimental results, respectively.
Finally, Section \ref{sec:conclusion} concludes the paper.
\section{Theoretical Background} \label{sec:TheoreticalBackground}
 This section provides an overview of the modeling and decoupled control strategies for 
indirect-modulated MMC, 
which serve as the foundational control schemes in this study, 
following the approach in \cite{Cui2014comprehensive}. 
Since the proposed method, which is discussed in the following sections, 
is based on the DOB,
this section also introduces the principle and stability conditions of DOB,
as described in \cite{Shim2009almost} and \cite{Shim2020disturbance}.

\subsection{Decoupled Modeling and Control of MMC}\label{subsec:DecoupledModelingAndControlOfMMC}
\begin{figure}[t!]
    \centering
    \includegraphics[width=1\linewidth]{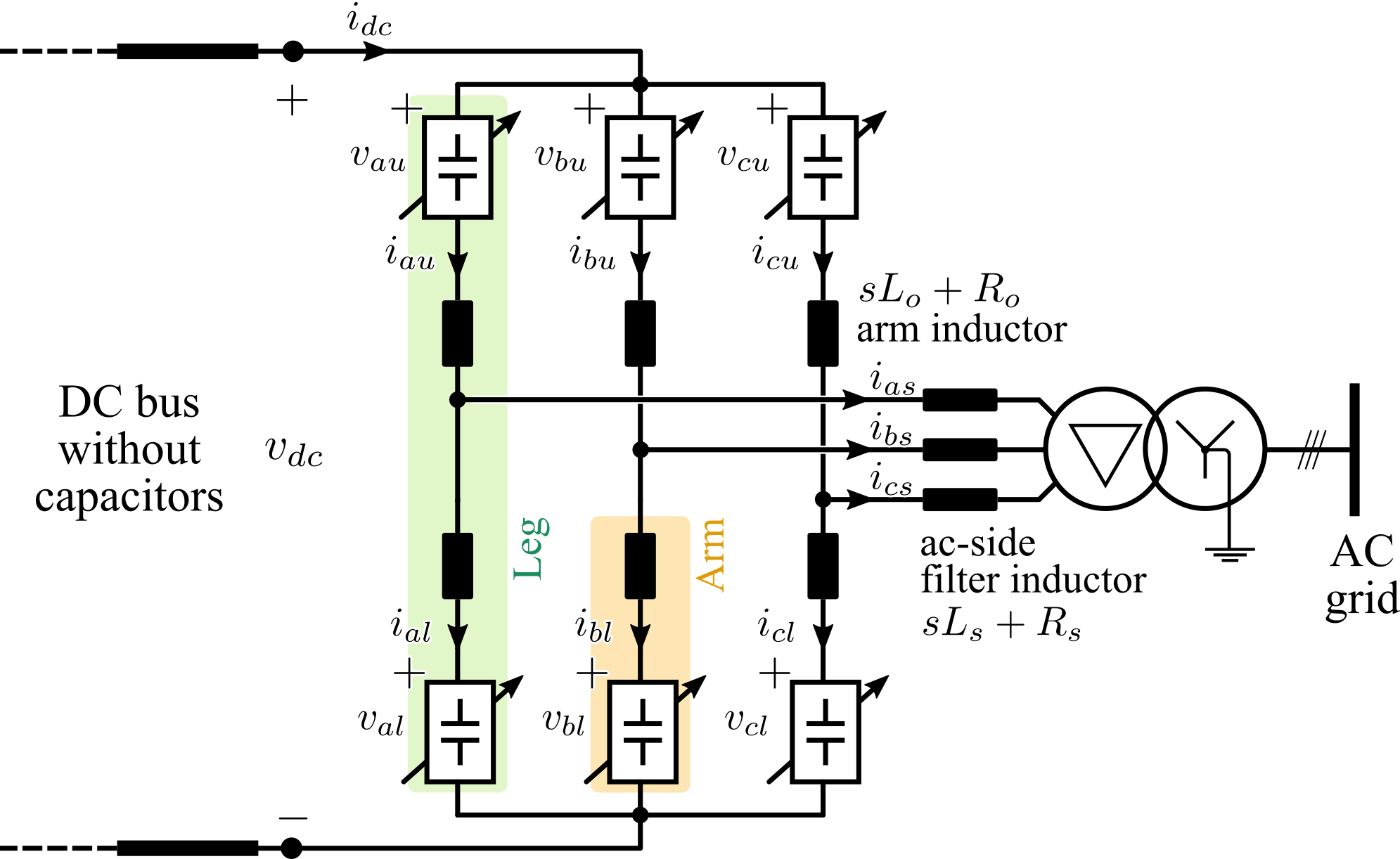}
    \caption{Modular multilevel converter with ac grid and dc bus.}
    \label{fig:mmc}
    \vspace{-0em}
\end{figure}

\begin{figure}[t]
    \vspace{-0.0em}  
    \centering
    \begin{subfigure}[b]{0.18\linewidth}
        \includegraphics[width=1.0\linewidth,center]{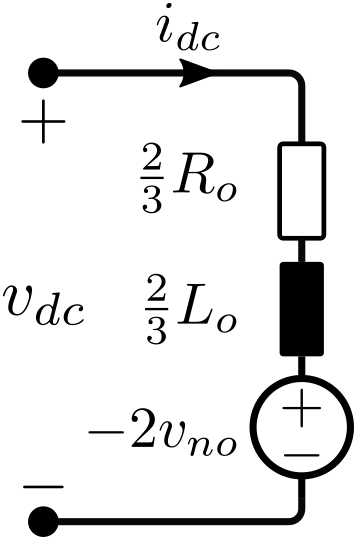}
        \vspace{0.8em}
        \caption{}
        \label{fig:mmc_model_dc}
    \end{subfigure}
    \hfill
    \begin{subfigure}[b]{0.35\linewidth}
        \includegraphics[width=1.0\linewidth,center]{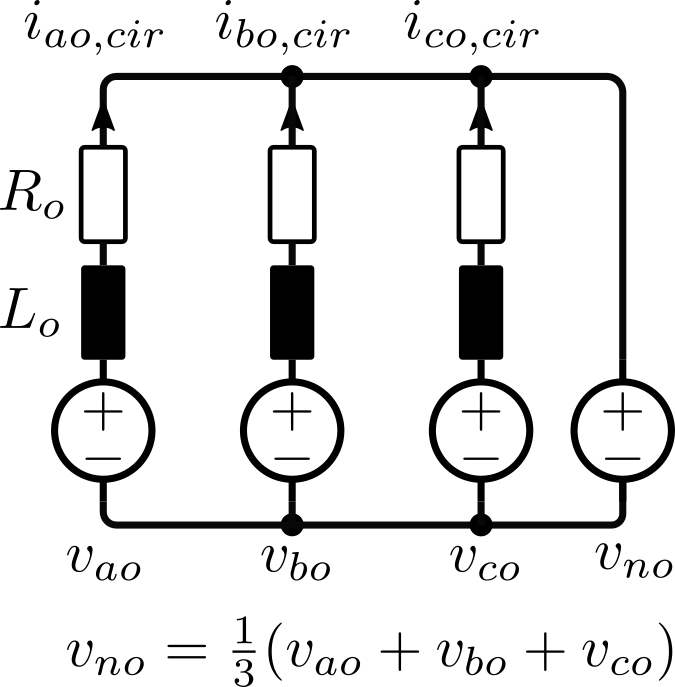}
        \caption{}
        \label{fig:mmc_model_cir}
    \end{subfigure}
    \hfill
    \begin{subfigure}[b]{0.4\linewidth}
        \includegraphics[width=1.0\linewidth,center]{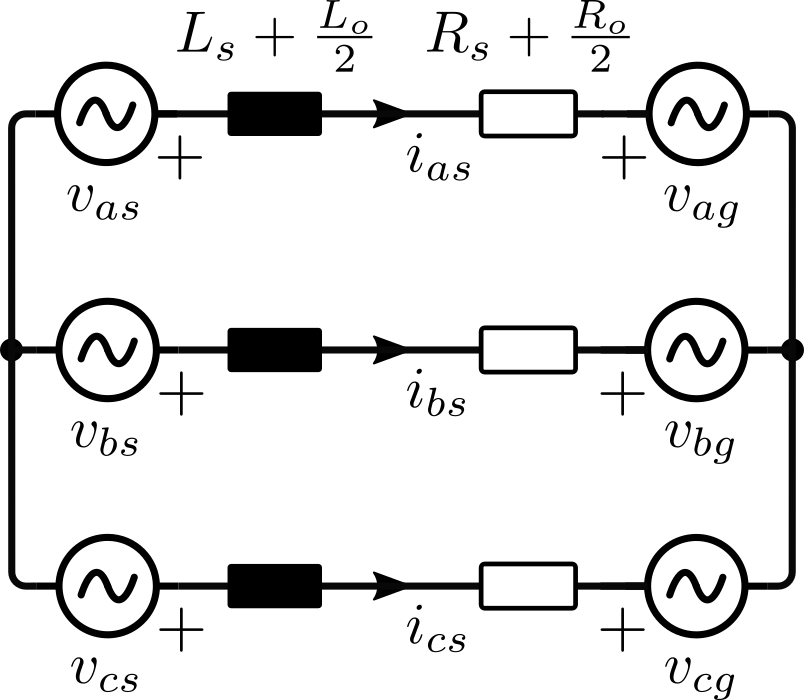}
        \vspace{-0.5em}
        \caption{}
        \label{fig:mmc_model_ac}
    \end{subfigure}
    \caption{Decoupled equivalent models of MMC. (a) dc-side current, (b) circulating current, (c) ac-side current. $v_{xg}$ is the $x$-phase ac grid voltage.}
    \label{fig:mmc_model}
   \vspace{0.0em}  
\end{figure}

\begin{figure}[t]
    \centering

    \includegraphics[width=1\linewidth]{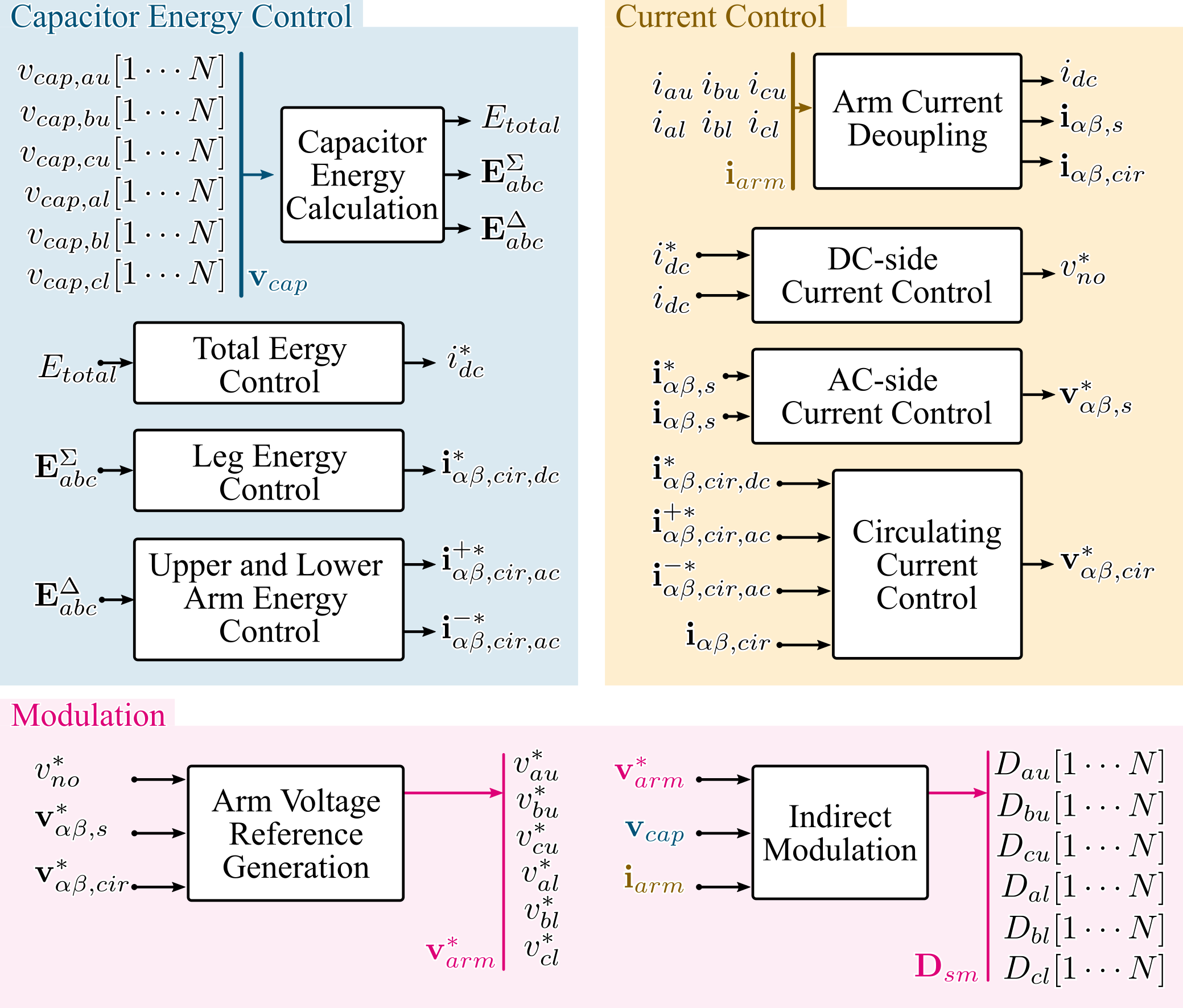}
    \caption{Overview of the decoupled control strategy for inverter mode MMC.
    $\mathbf{E}_{abc}^\Sigma$ and $\mathbf{E}_{abc}^\Delta$ are defined as 
    $\begin{bsmallmatrix}E_{a}^\Sigma&E_{b}^\Sigma&E_{c}^\Sigma\end{bsmallmatrix}^\top$ and 
    $\begin{bsmallmatrix}E_{a}^\Delta&E_{b}^\Delta&E_{c}^\Delta\end{bsmallmatrix}^\top$, 
    where $E_{x}^\Sigma$ is the sum and $E_{x}^\Delta$ is 
    the difference of the upper and lower arm capacitor energies for the $x$-phase leg.
    $D_{xu}$ and $D_{xl}$ are the duty cycles of the SMs in the $x$-phase upper and lower arms, respectively.
    }
    \label{fig:mmc_ctrl}
   \vspace{0.0em}  
\end{figure}

\begin{figure}[t]
    \vspace{0em}  
    \centering
    \begin{subfigure}[b]{0.325\linewidth}
        \includegraphics[width=1\linewidth,center]{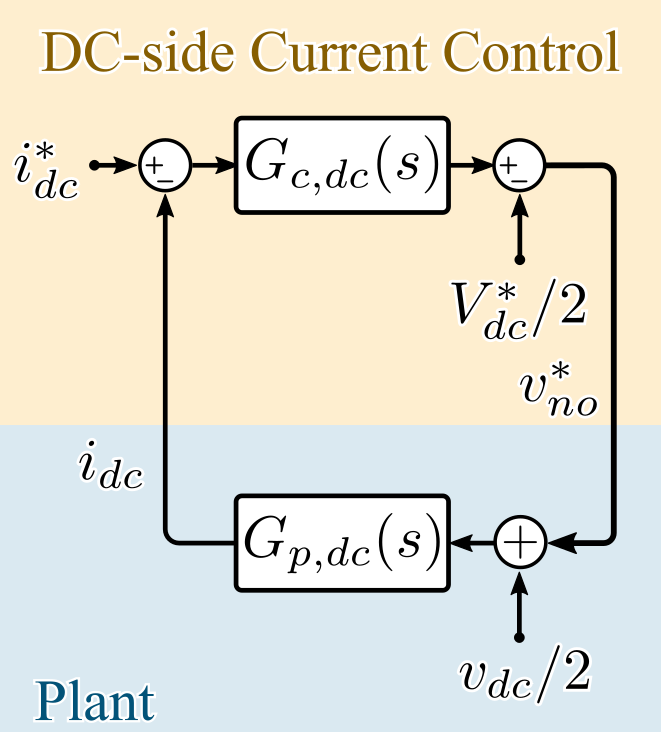}
        \caption{}
        \label{fig:mmc_inner_loop_dc_cc}
    \end{subfigure}
    \hfill
    \begin{subfigure}[b]{0.325\linewidth}
        \includegraphics[width=1\linewidth,center]{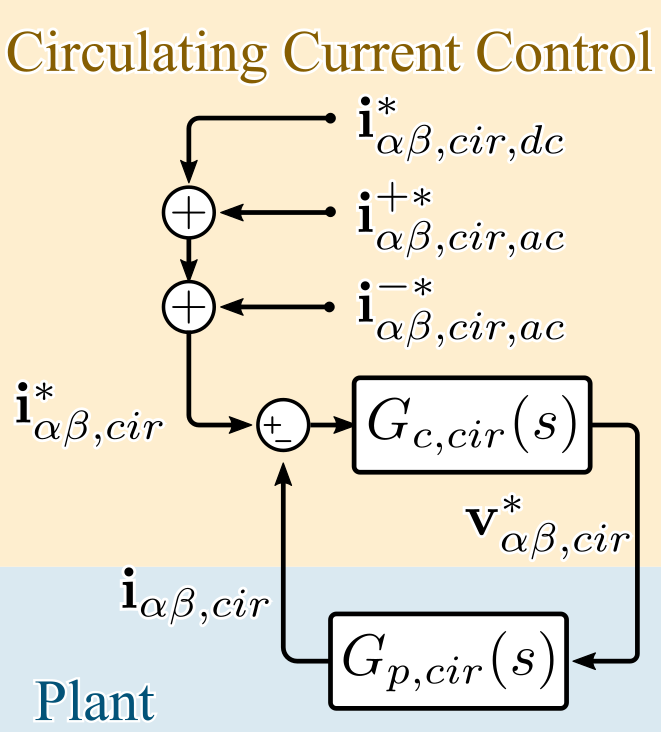}
        \caption{}
        \label{fig:mmc_inner_loop_cir_cc}
    \end{subfigure}
    \hfill
    \begin{subfigure}[b]{0.325\linewidth}
        \includegraphics[width=1\linewidth,center]{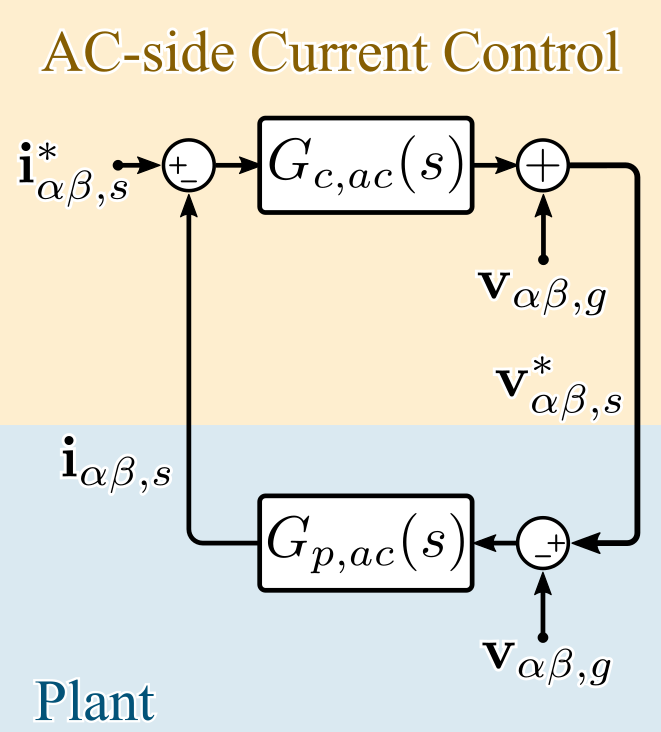}
        \caption{}
        \label{fig:mmc_inner_loop_ac_cc}
    \end{subfigure}
    \caption{Block diagrams of decoupled current controls for indirect-modulated MMC and 
    plant transfer functions for each current component. 
    (a) dc-side current, (b) circulating current, (c) ac-side current control loops. }
    \label{fig:mmc_inner_loop}
   \vspace{0.0em}  
\end{figure}

\figurename{\ref{fig:mmc}} illustrates the schematic of MMC with ac grid and dc bus.
$i_{xu}$, $i_{xl}$, and $i_{xs}$ represent
the $x$-phase current of upper arm, lower arm, and ac side, respectively.
$i_{dc}$ stands for the dc current coming from the dc bus.
$v_{xu}$ and $v_{xl}$ mean $x$-phase arm voltage of upper arm and lower arm.
The arm voltage is the summed output voltage of SMs in an arm.
$L_o$ stands for the inductance of arm inductor, $R_o$ for the resistance of arm inductor,
$L_s$ for the inductance of ac-side filter inductor, and $R_s$ for the resistance of ac-side filter.

From arm voltages,
the ac-side output voltage, 
$\mathbf{v}_s=[v_{as}\;\;v_{bs}\;\;v_{cs}]^\top$
and leg internal voltage,
$\mathbf{v}_o=[v_{ao}\;\;v_{bo}\;\;v_{co}]^\top$
are defined as follows:
\begin{equation}
    \mathbf{v}_{s}=-\frac{1}{2}(\mathbf{v}_{u}-\mathbf{v}_{l}),\quad
    \mathbf{v}_{o}=-\frac{1}{2}(\mathbf{v}_{u}+\mathbf{v}_{l})\label{eq:mmc-voltage}
\end{equation}
where 
$\mathbf{v}_u=[v_{au}\;\;v_{bu}\;\;v_{cu}]^\top$
and
$\mathbf{v}_l=[v_{al}\;\;v_{bl}\;\;v_{cl}]^\top$. 

From arm currents,
the ac-side current,
$\mathbf{i}_s=[i_{as}\;\;i_{bs}\;\;i_{cs}]^\top$,
and
the average of the upper and lower arm currents,
$\mathbf{i}_o=[i_{ao}\;\;i_{bo}\;\;i_{co}]^\top$,
are defined as follows:
\begin{equation}
    \mathbf{i}_{s}=\mathbf{i}_{u}-\mathbf{i}_{l},\quad
    \mathbf{i}_{o}=\frac{1}{2}(\mathbf{i}_{u}+\mathbf{i}_{l})\label{eq:mmc-current}
\end{equation}
where
$\mathbf{i}_u=[i_{au}\;\;i_{bu}\;\;i_{cu}]^\top$
and
$\mathbf{i}_l=[i_{al}\;\;i_{bl}\;\;i_{cl}]^\top$.
From \eqref{eq:mmc-current},
the circulating current, 
$\mathbf{i}_{o,cir}=[i_{ao,cir}\;\;i_{bo,cir}\;\; i_{co,cir}]^\top$,
is defined as follows:
\begin{equation}
    \mathbf{i}_{o,cir} = \mathbf{i}_{o}-\frac{1}{3}
    \begin{bmatrix}
        i_{dc} & i_{dc} & i_{dc}
    \end{bmatrix}^\top. \label{eq:mmc-circulating-current}
\end{equation}

By applying Kirchhoff's laws to the MMC circuit, 
the dynamics of the MMC can be represented by 
decoupled equivalent models for the dc-side current, circulating current, and ac-side current, 
as illustrated in \figurename{\ref{fig:mmc_model}} (see also \cite{Perez2011generalized,Lizana2012capacitor}, and \cite{Cui2014comprehensive}).
In the complex frequency domain,
the dynamics of the dc-side current shown in \figurename{\ref{fig:mmc_model_dc}} can be formulated as follows:
\begin{equation}
    i_{dc}(s)=\underbrace{\frac{1}{s\left(\frac{1}{3}L_o\right)+\left(\frac{1}{3}R_o\right)}}_{\text{\normalsize $G_{p,dc}(s)$}}\left(\frac{v_{dc}(s)}{2}+v_{no}(s)\right), 
    \label{eq:dc-side-circuit-s}
\end{equation}
where
$v_{no}=(v_{ao}+v_{bo}+v_{co})/3$.

The dynamics of the circulating current shown in \figurename{\ref{fig:mmc_model_cir}} 
and ac-side current shown in \figurename{\ref{fig:mmc_model_ac}}
can be expressed in the $\alpha\beta$-frame as follows:
\begin{align}
    \mathbf{i}_{\alpha\beta,cir}(s)&=\underbrace{\frac{1}{sL_o+R_o}}_{\text{\normalsize $G_{p,cir}(s)$}}\mathbf{v}_{\alpha\beta,cir}(s), 
    \label{eq:circulating-current-circuit-alpha-beta-s} \\
    \mathbf{i}_{\alpha\beta,s}(s)&=
    \underbrace{\frac{1}{sL_{ac}+R_{ac}}}_{
        \text{\normalsize $G_{p,ac}(s)$}
    }
    (\mathbf{v}_{\alpha\beta,s}(s)-\mathbf{v}_{\alpha\beta,g}(s)),
    \label{eq:ac-side-circuit-alpha-beta-s}
\end{align}
where $L_{ac}=L_s+L_o/2$ and $R_{ac}=R_s+R_o/2$.
The subscript `$\alpha\beta$' denotes the $\alpha\beta$-frame components obtained by $\alpha\beta$ transformation,
$\mathbf{T}=${\small$\begin{bsmallmatrix}
    2/3 & -1/3 & -1/3 \\
    0 & 1/\sqrt{3} & -1/\sqrt{3} \\
\end{bsmallmatrix}$} as follows:
\begin{gather}
    \mathbf{v}_{\alpha\beta,cir} = \mathbf{T}\mathbf{v}_o, \quad
    \mathbf{i}_{\alpha\beta,cir} = \mathbf{T}\mathbf{i}_{o,cir},\\
    \mathbf{v}_{\alpha\beta,s} = \mathbf{T}\mathbf{v}_s, \quad
    \mathbf{i}_{\alpha\beta,s} = \mathbf{T}\mathbf{i}_{s}, \quad
    \mathbf{v}_{\alpha\beta,g} = \mathbf{T}\mathbf{v}_g, 
    \label{eq:clarke_transformation}
\end{gather}
where $\mathbf{v}_{g}$ is the grid voltage defined as $[v_{ag}\;\;v_{bg}\;\; v_{cg}]^\top$.

Based on the decoupled equivalent models shown in \figurename{\ref{fig:mmc_model}},
the decoupled control strategy for the MMC, which is
presented in \cite{Cui2014comprehensive},
enables independent regulation of the dc-side current, ac-side current, and capacitor energies in the MMC.

An overview of the decoupled control strategy is illustrated in \figurename{\ref{fig:mmc_ctrl}}.
Without loss of generality, the inverter-mode operation is considered in this paper as an example.
For the outer control loops, capacitor energy controllers\,—\,including
total energy, leg energy, and upper/lower arm energy controllers\,—\,are 
implemented to balance the capacitor energy of each arm.
These energy controllers generate references for $i_{dc}$ and $\mathbf{i}_{\alpha\beta,cir}$ denoted by superscript `$*$' in \figurename{\ref{fig:mmc_ctrl}}.
By regulating dc-side current, $i_{dc}$, the total capacitor energy of all SMs is controlled.
The dc component of the circulating current, $\mathbf{i}_{\alpha\beta,cir,dc}$, is controlled to
balance the capacitor energy between the legs.
The positive and negative ac components of the circulating current, 
$\mathbf{i}_{\alpha\beta,cir,ac}^{+}$ and $\mathbf{i}_{\alpha\beta,cir,ac}^{-}$, are controlled to
balance the capacitor energy between the upper and lower arms.
The ac-side current, $\mathbf{i}_{\alpha\beta,s}$, is regulated 
to control the active and reactive power to the ac grid.
Since the capacitor energy control 
is built upon the independent current control of the MMC,
the current controllers for $i_{dc}$, $\mathbf{i}_{\alpha\beta,cir}$, and $\mathbf{i}_{\alpha\beta,s}$
are implemented as inner control loops.

\figurename{\ref{fig:mmc_inner_loop}} illustrates the block diagrams of the decoupled current controls for the indirect-modulated MMC.
To achieve high current control performance,
it is assumed that 
voltage references from the current controllers are accurately applied to the decoupled models of the MMC.
It can be realized by precisely synthesizing the arm voltage references, $v_{xu}^*$ and $v_{xl}^*$,
based on the measured capacitor voltages of the SMs, 
which is known as indirect modulation \cite{Sekiguchi2014grid, Cui2014comprehensive}.
Note that
the synthesis error in the arm voltage 
is assumed to be negligible in the case of decoupled control of the indirect-modulated MMC.

\subsection{Nearest Level Modulation in Indirect-Modulated MMC}\label{subsec:NLM_Challenges_SmallSM}
\begin{figure}[t]
    \centering

    \includegraphics[width=1.0\linewidth]{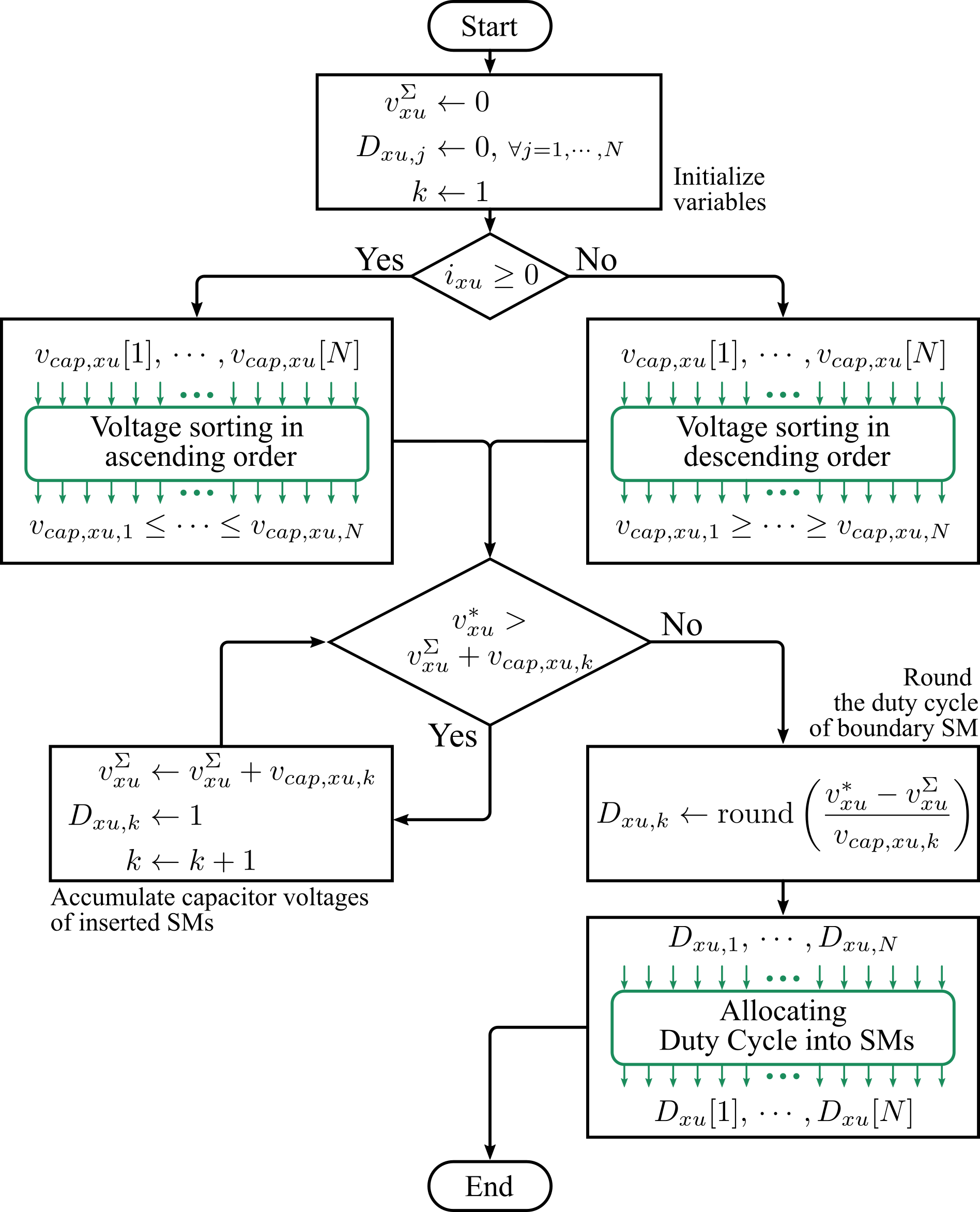}
    \vspace{0.5em}
    \caption{Operation flowchart of NLM for indirect-modulated MMC
    to determine duty cycles of SMs in $x$-phase upper arm.
    Duty cycles of SMs in lower arm can be determined in the same way.
    Half-bridge SM is assumed to be used in this flowchart.}
    \label{fig:mmc_nlm_flowchart}
   \vspace{0.0em}  
\end{figure}

\begin{figure}[t]
    \vspace{-0.0em}  
    \centering
    \begin{subfigure}[b]{1.0\linewidth}
        \includegraphics[width=1.0\linewidth,center]{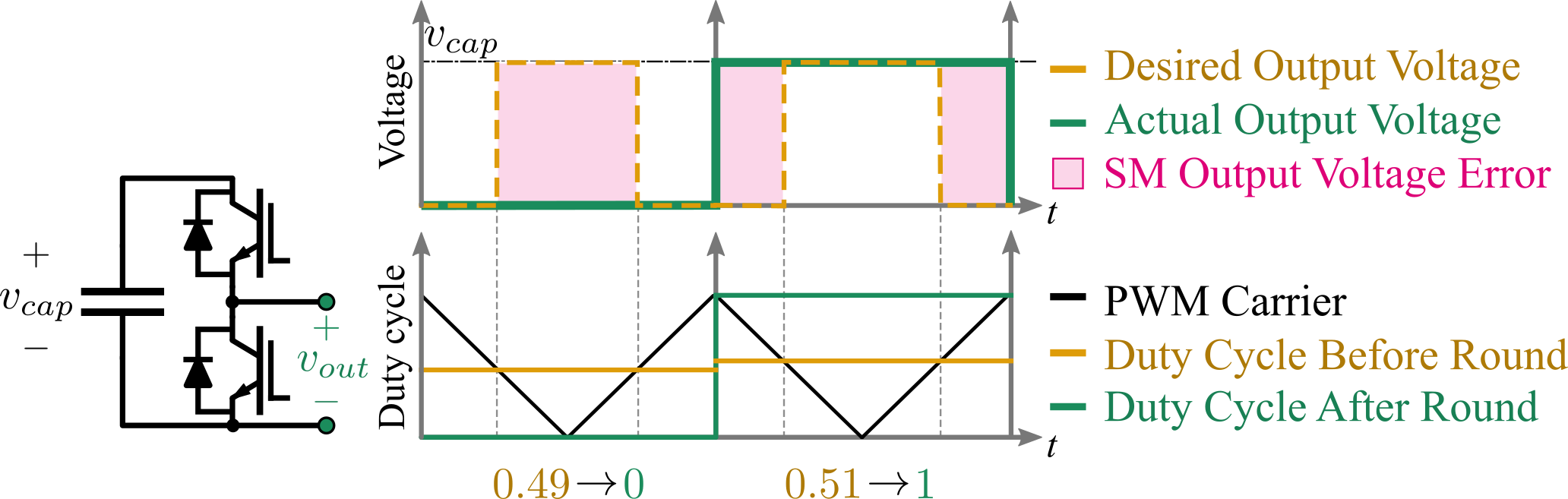}
        \caption{}
        \vspace{2em}
        \label{fig:mmc_nlm_distortion_sm}
    \end{subfigure}
    \hfill
    \begin{subfigure}[b]{1.0\linewidth}
        \includegraphics[width=1.0\linewidth,center]{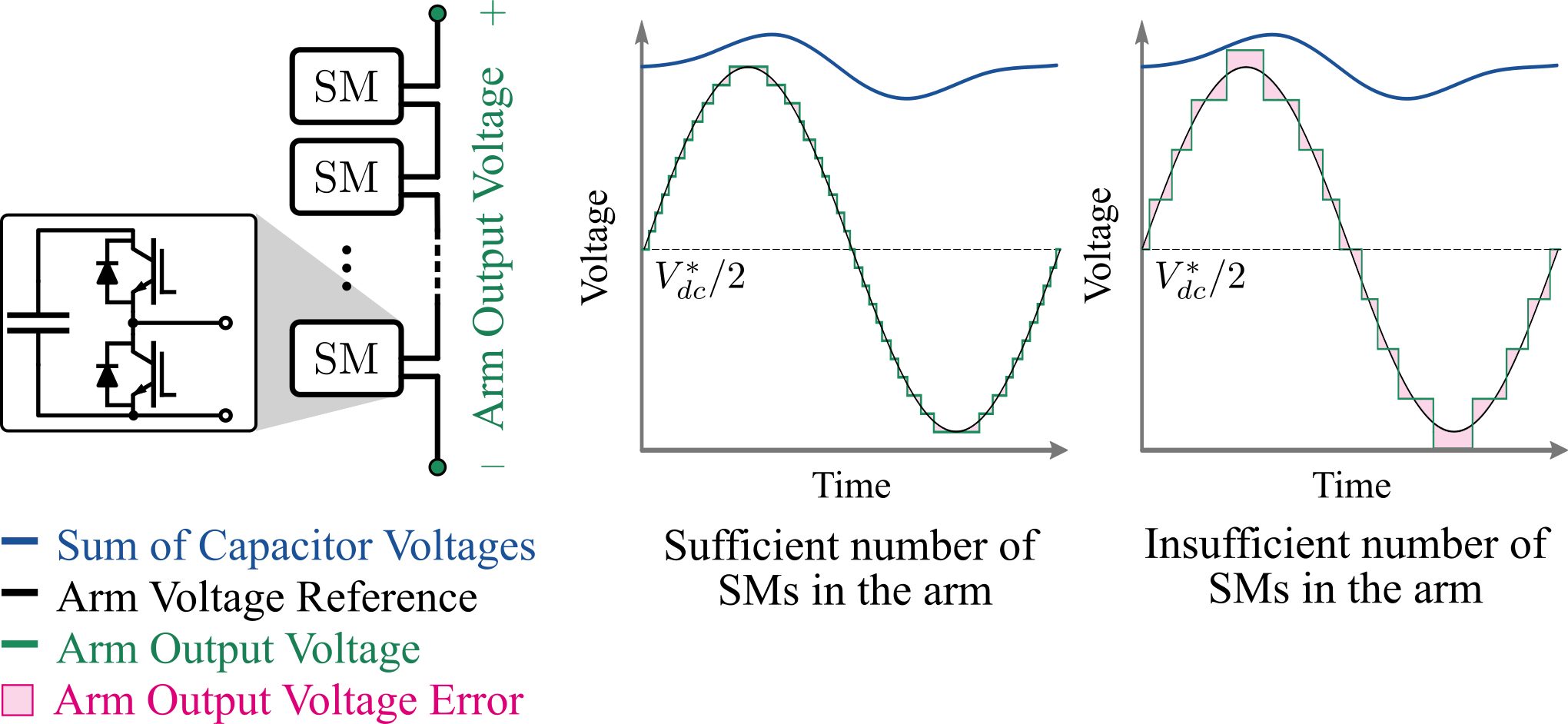}
        \caption{}
        \vspace{1em}
        \label{fig:mmc_nlm_distortion_arm}
    \end{subfigure}
    \caption{Conceptual diagram of voltage distortion caused by NLM.
    (a) Output voltage distortion of boundary SM,
    (b) Arm voltage distortion caused by NLM depending on the number of SMs in the arm.}
    \label{fig:mmc_nlm_distortion}
   \vspace{0.0em}  
\end{figure}

NLM uses the nearest available voltage level 
to the desired arm voltage reference \cite{Meshram2015simplified}.
In the case of direct-modulated MMC,
the number of inserted SMs is determined based on the nominal capacitor voltage,
which does not consider the actual capacitor voltages of the SMs.
However, in the case of indirect-modulated MMC,
the number of inserted SMs is determined based on the instantaneously measured capacitor voltages of the SMs.

\figurename{\ref{fig:mmc_nlm_flowchart}} illustrates the operation flowchart of NLM for indirect-modulated MMC.
The duty cycle for each SM in the arm is determined 
by sequentially summing the measured capacitor voltages of the SMs 
and comparing this accumulated value to the arm voltage reference.
If the PWM operation is combined with NLM, as described in \cite{Rohner2010modulation,Yi2018nearest},
an SM at the boundary between inserted and bypassed SMs 
can have a fractional duty cycle between 0 and 1.
This boundary SM operates as a PWM switch, to ensure the accurate synthesis of the arm voltage reference.
However, in the case of NLM, the insertion of this boundary SM is determined by rounding its duty cycle.
If the duty cycle is 0.5 or above, the SM is inserted; otherwise, it is bypassed.
Therefore, from the perspective of the arm voltage reference,
NLM in  indirect-modulated MMC introduces an arm voltage synthesis error
of up to half of the capacitor voltage of the boundary SM.
\figurename{\ref{fig:mmc_nlm_distortion_sm}}
shows the conceptual diagram of the output voltage distortion of the corresponding boundary SM.

In HVDC applications, where the number of SMs in each arm is sufficiently large,
NLM can be conveniently used for the indirect-modulated MMC since
the arm voltage can be synthesized with a sufficient number of levels,
therefore, the aforementioned voltage synthesis errors are negligible.
For example, in \cite{Leon2017energy},
NLM based on the measured capacitor voltages is employed
for 400 SMs in the arm of the MMC for HVDC applications,
while the decoupled control of the indirect-modulated MMC is employed.
However, in the case of small number of SMs per arm, such as those used in MVDC applications,
the number of SMs in each arm is limited,
and the arm voltage reference cannot be synthesized with a sufficient number of levels.
\figurename{\ref{fig:mmc_nlm_distortion_arm}}
illustrates the conceptual diagram of the arm voltage distortion caused by NLM.
Without a sufficient number of SMs or PWM functionality, 
the arm voltage synthesis error caused by NLM can be significant.
Such voltage synthesis error not
only degrades the ac-side current quality but also the 
control performance of circulating current and dc-side current, 
which are all critical control targets in indirect modulation based control.

\subsection{Principles of Disturbance Observer}\label{subsec:PrinciplesOfDisturbanceObserver}

\begin{figure}[t]
    \centering
    \includegraphics[width=0.90\linewidth]{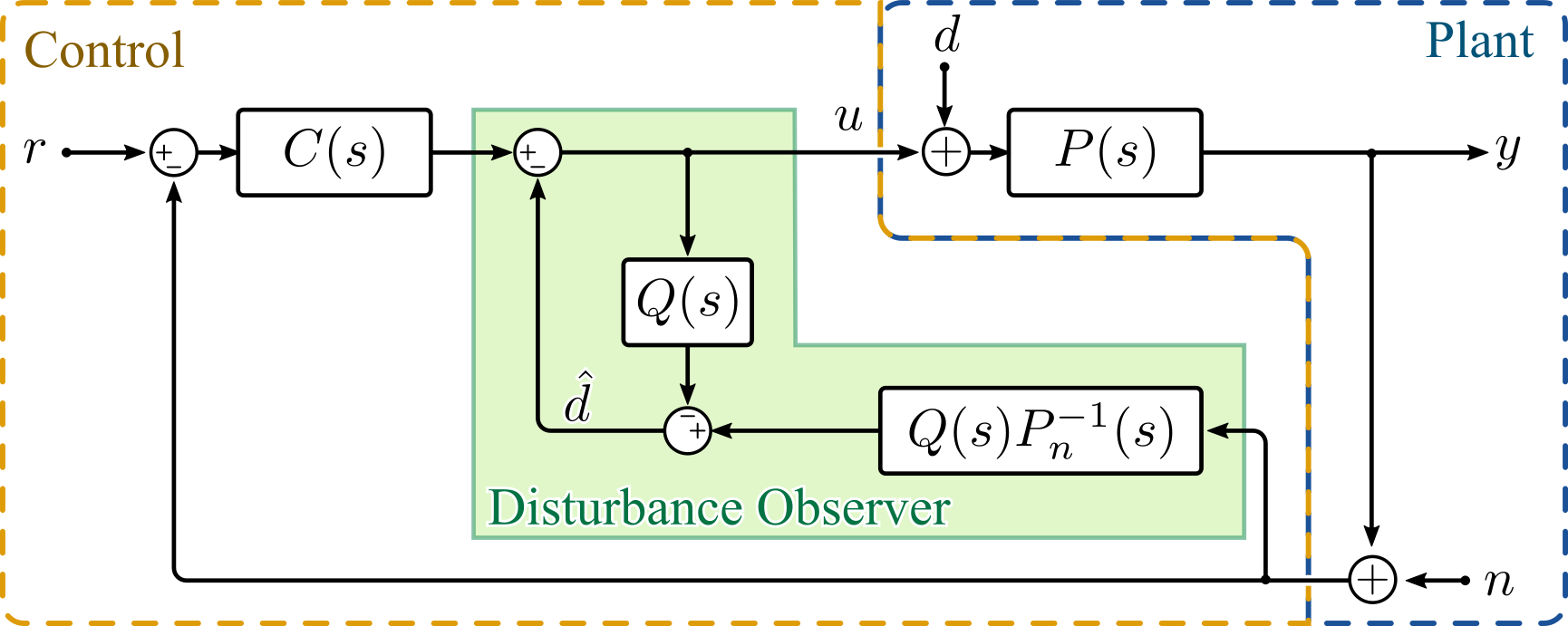}
    \caption{Block diagram of DOB for a linear plant.
    $C(s)$, $P(s)$, and $Q(s)$ represent the transfer function for 
    controller, plant, and $Q$-filter of DOB, respectively.
    $r$ stands for the reference signal,
    $u$ for the controller output,
    $y$ for the output of the plant,
    $n$ for the noise signal, and
    $d$ for the disturbance.}
    \label{fig:dob_structure}
   \vspace{0.0em}  
\end{figure}

\figurename{\ref{fig:dob_structure}} illustrates 
the structure of a DOB for a linear plant.
$C(s)$, $P(s)$, and $Q(s)$ represent the transfer function for
the controller, plant, and $Q$-filter of DOB, respectively.
$r$ stands for the reference signal,
$u$ for the controller output,
$y$ for the output of the plant,
$n$ for the noise signal, and
$d$ for the disturbance.

In order to consider the parameter uncertainty of the plant,
the set of plant transfer functions, $\mathcal{P}$, 
to which $P(s)$ belongs,
can be defined as follows:
\begin{align}
    \mathcal{P} &= \{g\frac{s^{n-\nu}+\beta_{n-\nu-1}s^{n-\nu-1}+\cdots+\beta_0}{s^n+\alpha_{n-1}s^{n-1}+\cdots+\alpha_0} \nonumber\\
    &:\alpha_k\in[\underline{\alpha}_k,\overline{\alpha}_k],\, \beta_k\in[\underline{\beta}_k,\overline{\beta}_k],\, g\in[\underline{g},\overline{g}] \},\, \underline{g}>0,\label{eq:plant_set}
\end{align}
where the interval $[\underline{\alpha}_k,\overline{\alpha}_k]$, $[\underline{\beta}_k,\overline{\beta}_k]$, and $[\underline{g},\overline{g}]$ 
represent the uncertainty of the coefficients of the denominator polynomial,
the numerator polynomial, and the static gain, respectively.
The nominal-parameter plant transfer function 
is denoted as $P_n(s)$, and it also belongs to $\mathcal{P}$ as $P(s)$ belongs to $\mathcal{P}$.

The DOB is designed to estimate the disturbance $d(s)$
and compensate it to the output of the controller $u(s)$.
The disturbance is assumed to be an unknown input to the plant, $P(s)$.
The DOB operates 
by comparing the estimated plant input, $\hat{u}(s)$,
with the controller output, $u(s)$,
to estimate the disturbance, $\hat{d}(s)$,
and this estimate is then used to compensate for the disturbance in the control loop.

To estimate the plant input,
the inverse transfer function of the nominal-parameter plant, $P_n^{-1}(s)$, is used.
However, if the plant model, $P(s)$, is a strictly proper transfer function, 
whose degree of numerator is less than that of denominator,
the inverse transfer function of the nominal-parameter plant, $P_n^{-1}(s)$, becomes improper and thus unrealizable.
In this case, a $Q$-filter, $Q(s)$, 
is used to ensure that $Q(s)P_{n}^{-1}(s)$ becomes a proper transfer function,
thereby making the DOB realizable.
$Q(s)$ is implemented as a low-pass filter (LPF) whose denominator degree equals the relative degree of $P_n(s)$, $\nu\in\mathbb{N}_0$, 
as follows:
\begin{equation}
    Q(s) = \frac{a_0}{(\tau s)^\nu+a_{\nu-1}(\tau s)^{\nu-1}+\cdots+a_1(\tau s)+a_0}, \label{eq:q_filter}
\end{equation}
where $a_0$ and $a_k$ are real-valued coefficients, 
and $\tau>0$ determines the cut-off frequency of the $Q$-filter.
With these considerations, 
the output of the plant, $y(s)$, in \figurename{\ref{fig:dob_structure}}
is expressed as follows:
\begin{align}
    y(s) =&\, \frac{P_n(s)P(s)C(s)}{P_n(s)[1+P(s)C(s)]+Q(s)[P(s)-P_n(s)]}r(s) \nonumber\\
    &+\frac{P_n(s)P(s)[1-Q(s)]}{P_n(s)[1+P(s)C(s)]+Q(s)[P-P_n(s)]}d(s) \nonumber\\
    &-\frac{P(s)[Q(s)+P_n(s)C(s)]}{P_n(s)[1+P(s)C(s)]+Q(s)[P(s)-P_n(s)]}n(s), \label{eq:dob_output}
\end{align}
where $r(s)$ and $n(s)$ represent the reference signal and noise signal, respectively.

Following the approach in \cite{Shim2020disturbance},
let $\omega_{Q}$ be the cutoff frequency of $Q(s)$.
By approximating  $Q(j\omega)\approx 1$ for $|\omega|\ll \omega_{Q}$
and $Q(j\omega)\approx 0$ for $|\omega|\gg \omega_{Q}$,
output of the plant, $y(s)$, in \eqref{eq:dob_output}
can be approximated as follows:
\begin{equation}
    y(j\omega) \approx \frac{P_n(j\omega)C(j\omega)}{1+P_n(j\omega)C(j\omega)}r(j\omega)-n(j\omega),\,  |\omega|\ll\omega_Q 
    \label{eq:dob_output_approx}
\end{equation}
under the assumption that all transfer functions in \eqref{eq:dob_output} are stable.
This input-output relation in \eqref{eq:dob_output_approx} indicates that, 
within the frequency range below the cut-off frequency of the $Q$-filter, 
the DOB enables 
the closed-loop system to follow 
the nominal input-output behavior 
while effectively rejecting the disturbance $d(s)$.

\subsection{Stability Conditions for Disturbance Observer}\label{subsec:StabilityForDisturbanceObserver}
In \cite{Shim2009almost}, stability conditions for the control loops employing DOB
are derived.
To properly design the DOB while ensuring the stability of the control loop,
following conditions are assumed:
\begin{enumerate}
    \item $C(s)$ internally stabilizes $P_{n}(s)\in \mathcal{P}$.
    \item $\forall P(s)\in\mathcal{P}$ are minimum phase.
    \item Coefficients of $Q(s)$ in \eqref{eq:q_filter} are chosen such that following polynomial, $p_f(s)$,
    is a Hurwitz polynomial for all $g\in[\underline{g},\overline{g}]$.
\end{enumerate}
\begin{equation}
    p_f(s) \coloneqq s^\nu+a_{\nu-1}s^{\nu-1}+\cdots+a_1s+\frac{g}{g^*}a_0, \label{eq:hurwitz_poly}
\end{equation}
where $g^*$ is the nominal static gain of the plant, $P_n(s)$.
Under these conditions,
there exists $\tau^*>0$ such that
all transfer functions in \eqref{eq:dob_output} are stable
for $\tau\in(0,\tau^*]$.
It means that the cut-off frequency of $Q$-filter has a lower bound,
since $\tau$ must not exceed $\tau^*$ in order to guarantee the stability.

The first condition implies that 
the controller, $C(s)$, should be designed to 
stabilize $P_{n}(s)$.
Additionally, 
the numerator and denominator degrees of $P_{n}(s)$ 
should match those of the actual plant $P(s)$, 
even in the presence of parameter uncertainties,
because both of $P(s)$ and $P_{n}(s)$ belong to $\mathcal{P}$.
The second condition implies that
both zeros and poles of $P(s)$, which is a rational transfer function for the linear system,
should be located in the left half-plane (LHP) of the complex plane.
Once the first and second assumptions are satisfied,
the DOB for the stable control loop 
can be implemented by 
determining coefficients of $Q(s)$ to satisfy the third assumption.
The third assumption means that all roots of $p_f(s)=0$ are 
located in the LHP or on the imaginary axis.
Then, the $\tau$ is chosen to be within the range $(0,\tau^*]$.

Based on the principles and stability conditions of the DOB 
discussed in this section, 
a DOB design to improve 
the decoupled control of 
the indirect-modulated MMC 
under NLM
is presented in the following sections.

\section{Proposed Disturbance Observer Design 
for Decoupled Control of MMC under NLM} \label{sec:proposed}
In this section,
the arm voltage distortion caused by NLM is modeled and decomposed
into disturbances affecting the ac-side current, dc-side current, and circulating current.
Based on this disturbance model, 
the proposed DOB design is presented. 
For practical applications, a simple implementation is also presented.
To ensure the stability of the control loops with the designed DOBs,
the proposed method is analyzed
based on the stability criterion discussed in Section \ref{subsec:PrinciplesOfDisturbanceObserver}.

\subsection{Modeling of Disturbances Induced by NLM} \label{sec:disturbance_analysis}

\begin{figure}[t]
    \vspace{0em}  
    \centering
    \begin{subfigure}[b]{0.325\linewidth}
        \includegraphics[width=1\linewidth,center]{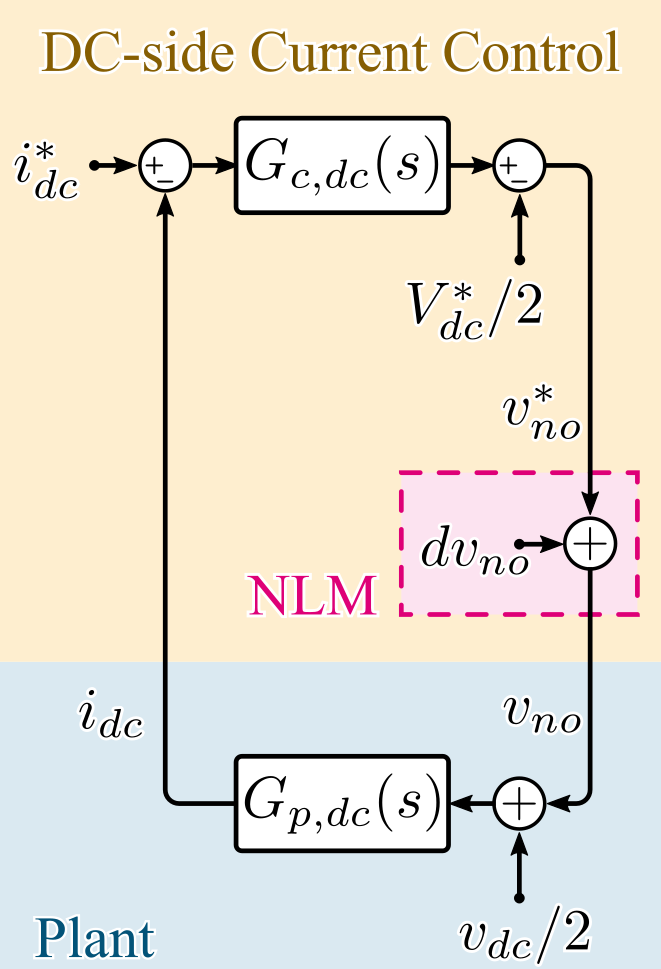}
        \caption{}
        \label{fig:mmc_inner_loop_dc_cc_nlm}
    \end{subfigure}
    \hfill
    \begin{subfigure}[b]{0.325\linewidth}
        \includegraphics[width=1\linewidth,center]{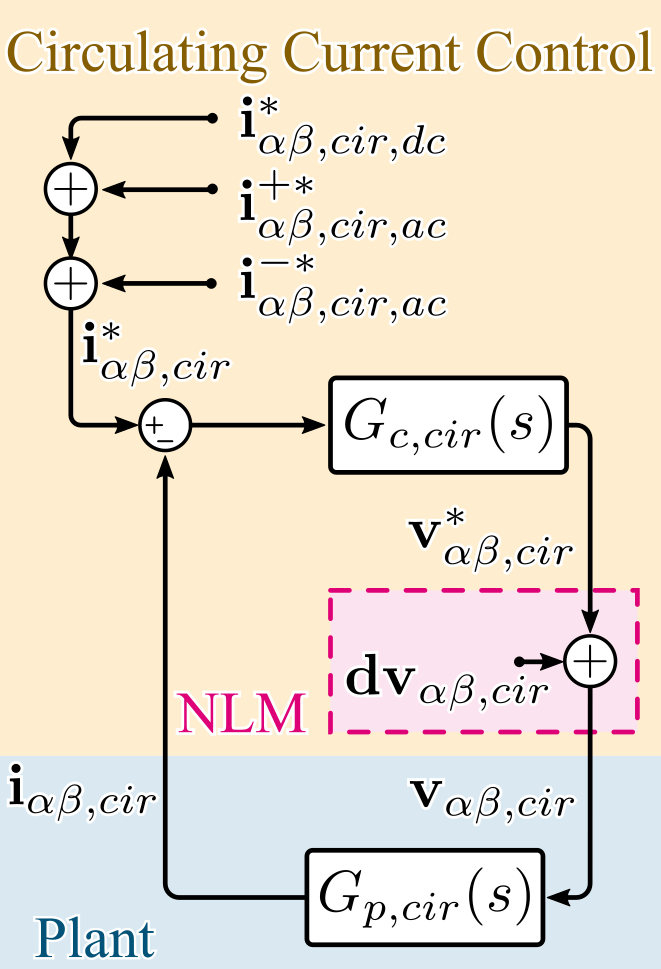}
        \caption{}
        \label{fig:mmc_inner_loop_cir_cc_nlm}
    \end{subfigure}
    \hfill
    \begin{subfigure}[b]{0.325\linewidth}
        \includegraphics[width=1\linewidth,center]{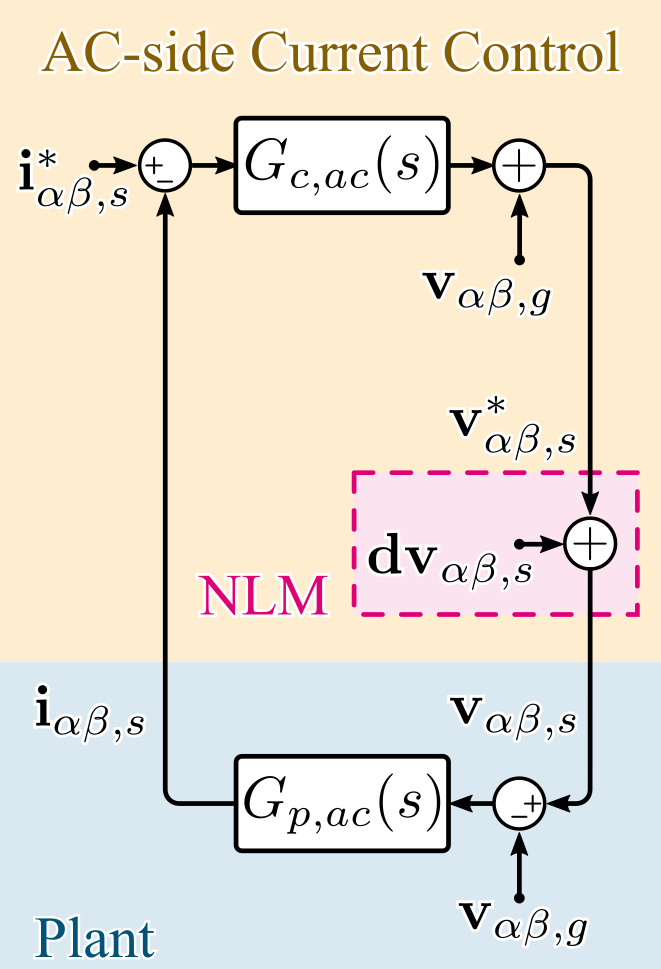}
        \caption{}
        \label{fig:mmc_inner_loop_ac_cc_nlm}
    \end{subfigure}
    \caption{Equivalent disturbances induced by NLM in decoupled current controls for indirect-modulated MMC.
    (a) dc-side current, (b) circulating current, (c) ac-side current control loops.}
    \label{fig:mmc_inner_loop_nlm}
   \vspace{0.0em}  
\end{figure}

For the indirect-modulated MMC,
NLM determines the number of inserted SMs in an arm
based on the capacitor voltages of each SM and the arm voltage reference.
As shown in \figurename{\ref{fig:mmc_nlm_flowchart}} and \figurename{\ref{fig:mmc_nlm_distortion}},
the duty cycle of the boundary SM between the inserted and bypassed SMs
is rounded to either 0 or 1.
Therefore, from the perspective of arm-voltage synthesis, 
NLM in the indirect-modulated MMC introduces the arm voltage error of
up to half the capacitor voltage of the boundary SM.
As a result, the actual arm voltage is synthesized as follows:
\begin{equation}
    v_{xu}=\sum_{k=1}^N D_{xu}[k]v_{cap,xu}[k],\quad  v_{xl}=\sum_{k=1}^N D_{xl}[k]v_{cap,xl}[k],\label{eq:upper_arm_voltage}
\end{equation}
where $D_{xu}[k],\, D_{xl}[k] \in \{0,1\}$ are the duty cycles.
$v_{cap,xu}[k]$ and $v_{cap,xl}[k]$ are the capacitor voltages 
of the $k$-th SM in the $x$-phase upper arm and lower arm, respectively.

The arm voltage error caused by NLM can be 
interpreted as voltage disturbances formulated as follows:
\begin{equation}
    \mathbf{dv}_{u}=\mathbf{v}_{u}-\mathbf{v}_{u}^*,\quad \mathbf{dv}_{l}=\mathbf{v}_{l}-\mathbf{v}_{l}^*, 
    \label{eq:arm_voltage_disturbance} 
\end{equation}
where $\mathbf{v}_{u}^*$ and $\mathbf{v}_{l}^*$ are 
the arm voltage references
from the current controllers in \figurename{\ref{fig:mmc_ctrl}}, 
while $\mathbf{v}_{u}$ and $\mathbf{v}_{l}$ are the actual arm voltages of the upper and lower arms, respectively.
$\mathbf{dv}_{u}=[dv_{au}\;\;dv_{bu}\;\;dv_{cu}]^\top$ 
and $\mathbf{dv}_{l}=[dv_{al}\;\;dv_{bl}\;\;dv_{cl}]^\top$
are the arm voltage disturbances caused by NLM 
at the upper and lower arms, respectively.

Following definitions of the ac-side output voltage and
the leg internal voltage  in  \eqref{eq:mmc-voltage},
$\mathbf{dv}_{u}$ and $\mathbf{dv}_{l}$ 
also can be equivalently mapped 
into the ac-side output voltage disturbance, $\mathbf{dv}_{s}=[dv_{as}\;\;dv_{bs}\;\;dv_{cs}]^\top$,
and the leg internal voltage disturbance, $\mathbf{dv}_{o}=[dv_{ao}\;\;dv_{bo}\;\;dv_{co}]^\top$, as follows:
\begin{equation}
    \mathbf{dv}_{s}=-\frac{1}{2}(\mathbf{dv}_{u}-\mathbf{dv}_{l}),\quad
    \mathbf{dv}_{o}=-\frac{1}{2}(\mathbf{dv}_{u}+\mathbf{dv}_{l}).
    \label{eq:mmc_disturb_voltage}
\end{equation}

From the perspective of the decoupled MMC model, 
the arm voltage error can be equivalently interpreted as a distortion in the output voltage of each decoupled circuit, 
leading to disturbances in the dc-side current, circulating current, and ac-side current, individually.
The output voltages of each circuit can be formulated 
with the voltage references and disturbances as follows:
\begin{align}
    v_{no}&=v_{no}^*+dv_{no},
    \label{eq:zero-sequence-leg-internal-voltage-disturbance}\\
    \mathbf{v}_{\alpha\beta,cir}&= \mathbf{v}_{\alpha\beta,cir}^*+ \mathbf{dv}_{\alpha\beta,cir},
    \label{eq:leg-internal-voltage-voltage-disturbance}\\
    \mathbf{v}_{\alpha\beta,s}&= \mathbf{v}_{\alpha\beta,s}^*+ \mathbf{dv}_{\alpha\beta,s},
    \label{eq:ac-side-voltage-disturbance}
\end{align}
where $\mathbf{dv}_{\alpha\beta,cir}=\mathbf{T}\mathbf{dv}_{o}$, 
$\mathbf{dv}_{\alpha\beta,s}=\mathbf{T}\mathbf{dv}_{s}$,
and $dv_{no}=(dv_{ao}+dv_{bo}+dv_{co})/3$.

By substituting
\eqref{eq:zero-sequence-leg-internal-voltage-disturbance},
\eqref{eq:leg-internal-voltage-voltage-disturbance},
\eqref{eq:ac-side-voltage-disturbance}
into \eqref{eq:dc-side-circuit-s},
\eqref{eq:circulating-current-circuit-alpha-beta-s},
\eqref{eq:ac-side-circuit-alpha-beta-s}, respectively,
the decoupled current control loops depicted in \figurename{\ref{fig:mmc_inner_loop}}
can be redrawn with the corresponding voltage disturbances caused by NLM as shown in \figurename{\ref{fig:mmc_inner_loop_nlm}}.
It highlights that the effects of NLM
can be equivalently mapped to the voltage disturbances in decoupled models,
although NLM operation is conducted for the arm voltage synthesis.

In the following section,
the proposed DOB design is presented
to increase the rejection capability for the disturbances caused by NLM,
consequently improving the decoupled control performance of the MMC under NLM.

\subsection{Design of Disturbance Observer for NLM} \label{sec:dob_design}

\begin{figure}[t]
    \vspace{-0.0em}  
    \centering
    \begin{subfigure}[b]{1\linewidth}
        \includegraphics[width=1.0\linewidth,center]{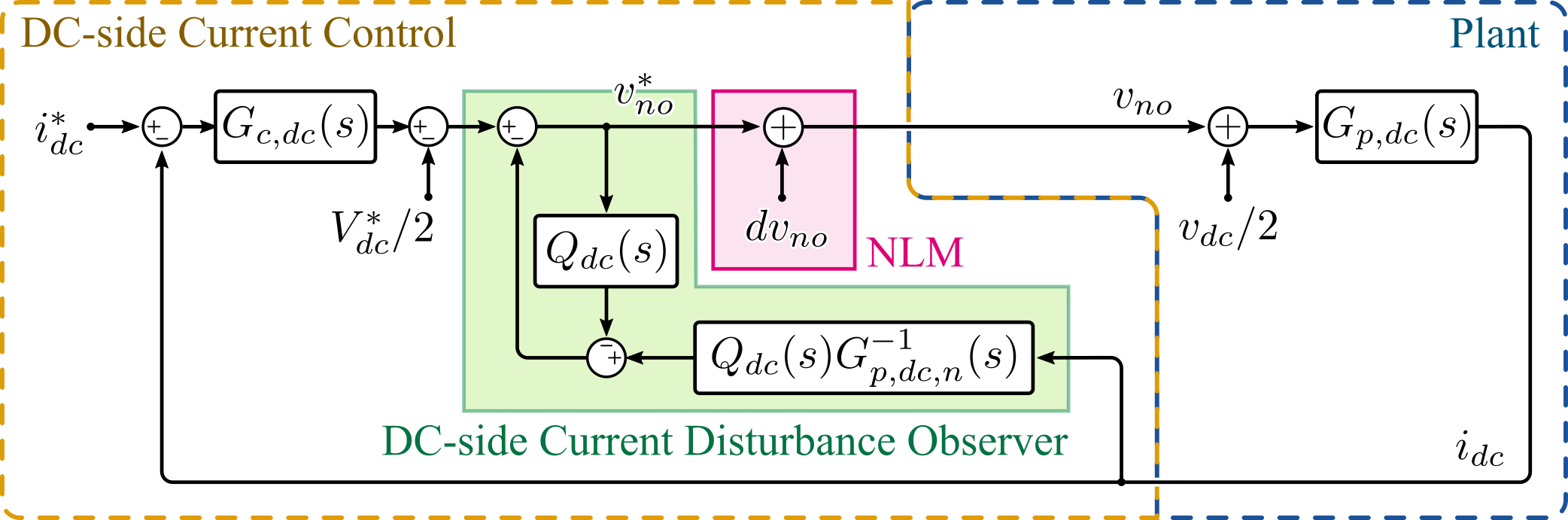}
        \caption{}
        \vspace{2.0em}
        \label{fig:mmc_dob_dc}
    \end{subfigure}
    \hfill
    \begin{subfigure}[b]{1\linewidth}
        \includegraphics[width=1.0\linewidth,center]{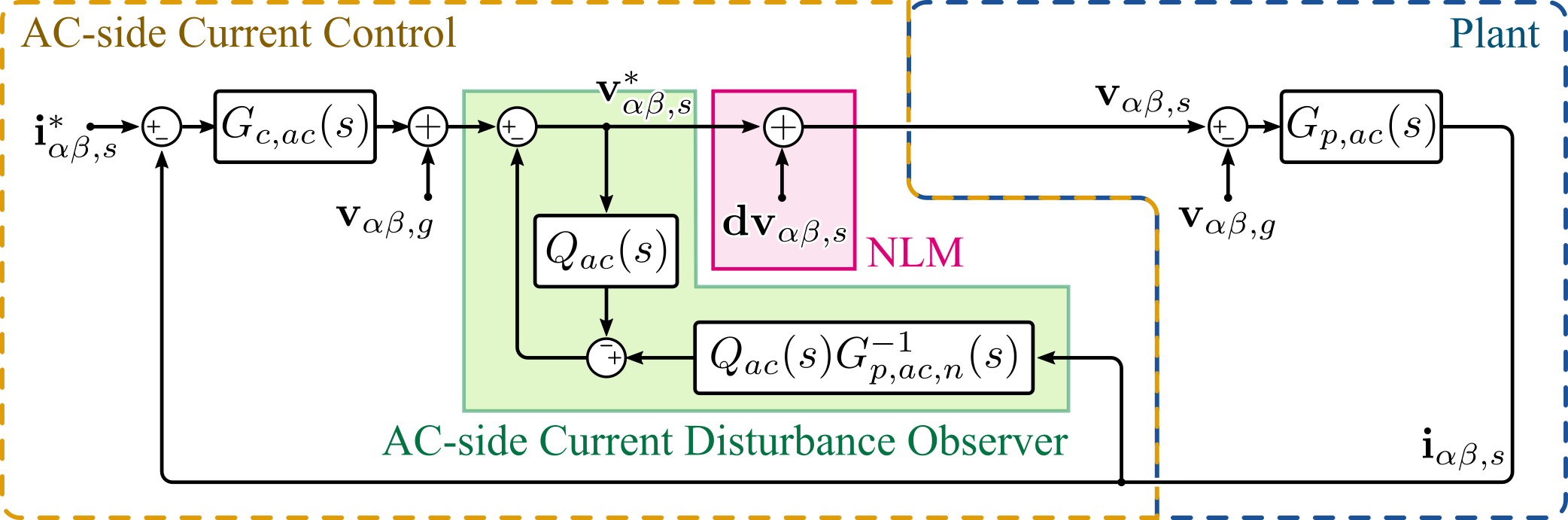}
        \caption{}
        \vspace{2.0em}
        \label{fig:mmc_dob_ac}
    \end{subfigure}
    \hfill
    \begin{subfigure}[b]{1\linewidth}
        \includegraphics[width=1.0\linewidth,center]{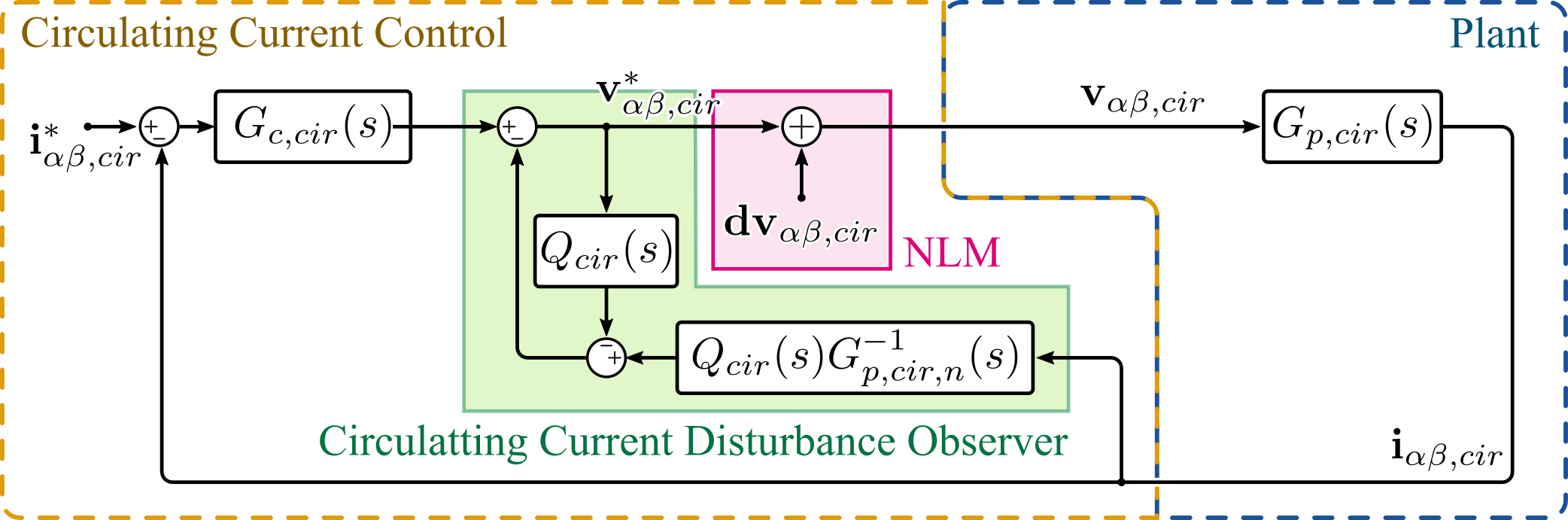}
        \caption{}
        \vspace{1em}
        \label{fig:mmc_dob_cir}
    \end{subfigure}
    \caption{Proposed DOB design for (a) dc-side current, (b) ac-side current, (c) circulating current control loops.
    $G_{p,x,n}^{-1}(s),\,\forall x\in\{dc,ac,cir\}$ denotes the inverse transfer function of the nominal resistive-inductive load for each decoupled model.
    }
    \label{fig:mmc_dob}
   \vspace{0.0em}  
\end{figure}

\figurename{\ref{fig:mmc_dob}}
illustrates the proposed DOB design 
to improve the decoupled control performance of the MMC under NLM.
The DOB is designed for each decoupled equivalent model of the MMC
to estimate and compensate the disturbances caused by NLM.
For each decoupled model, $Q$-filters are designed as follows:
\begin{equation}
    Q_{x}(s)=\frac{a_{0,x}}{\tau_x s+a_{0,x}},\label{eq:q-filter-mmc}
\end{equation}
where $x\in\{dc,ac,cir\}$.
The subscript `$x$' denotes the decoupled model for the dc-side, ac-side, and circulating currents, respectively.
$a_{0,x}$ and $\tau_x$ stand for the coefficient and time constant of each $Q$-filter, respectively.  

Since the plant for each decoupled model is resistive-inductive (RL) load,
whose relative degree is 1,
the first order LPF is used as the $Q$-filter in the DOB
based on \eqref{eq:q_filter}.
As shown in \figurename{\ref{fig:mmc_dob}},
the estimated input of the RL load 
is obtained by the transfer function 
combining the $Q$-filter and the inverse transfer function of the plant.
The voltage error is estimated 
by subtracting the $Q$-filtered voltage reference
from the estimated input to the RL load.
This estimated voltage error 
is then used to compensate 
in each decoupled control loop.

\begin{figure}[t]
    \vspace{-0.0em}  
    \centering
    \begin{subfigure}[b]{1\linewidth}
        \includegraphics[width=1.0\linewidth,center]{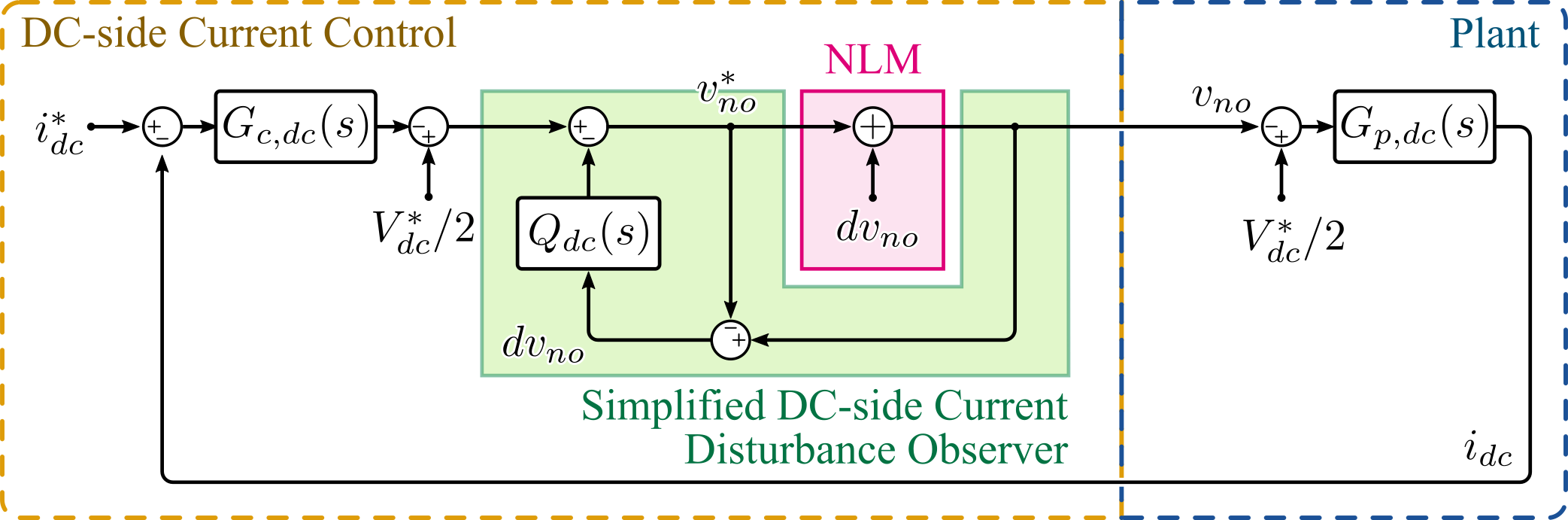}
        \caption{}
        \vspace{2.0em}
        \label{fig:mmc_dob_dc_simple}
    \end{subfigure}
    \hfill
    \begin{subfigure}[b]{1\linewidth}
        \includegraphics[width=1.0\linewidth,center]{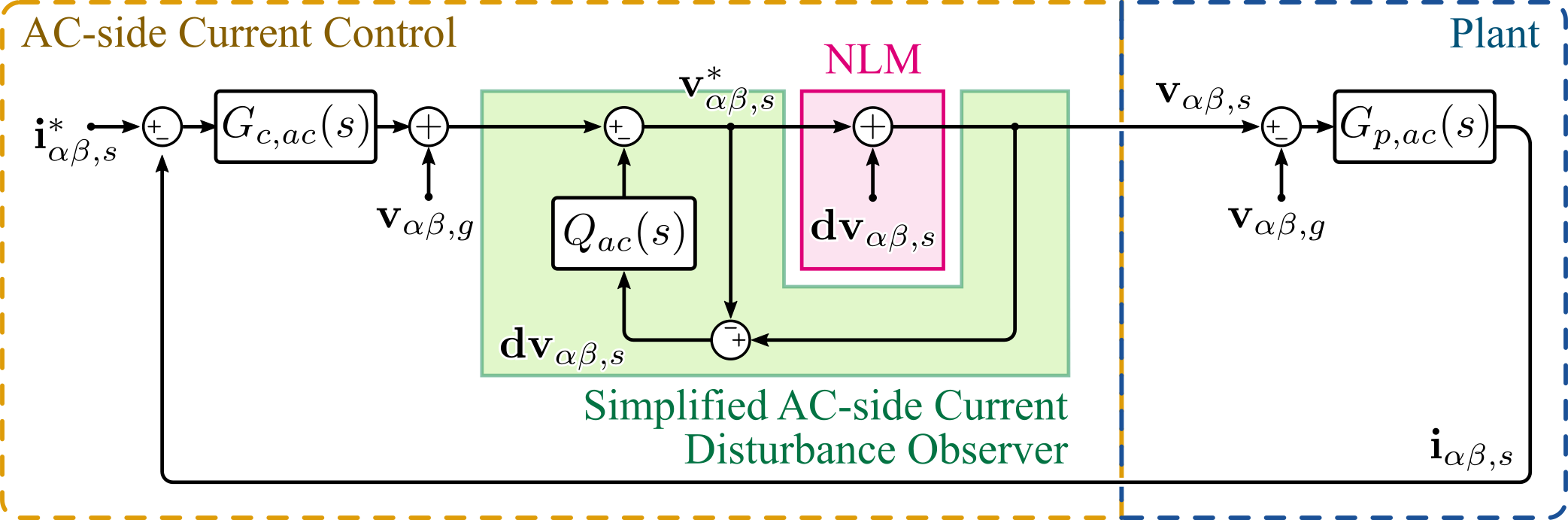}
        \caption{}
        \vspace{2.0em}
        \label{fig:mmc_dob_ac_simple}
    \end{subfigure}
    \hfill
    \begin{subfigure}[b]{1\linewidth}
        \includegraphics[width=1.0\linewidth,center]{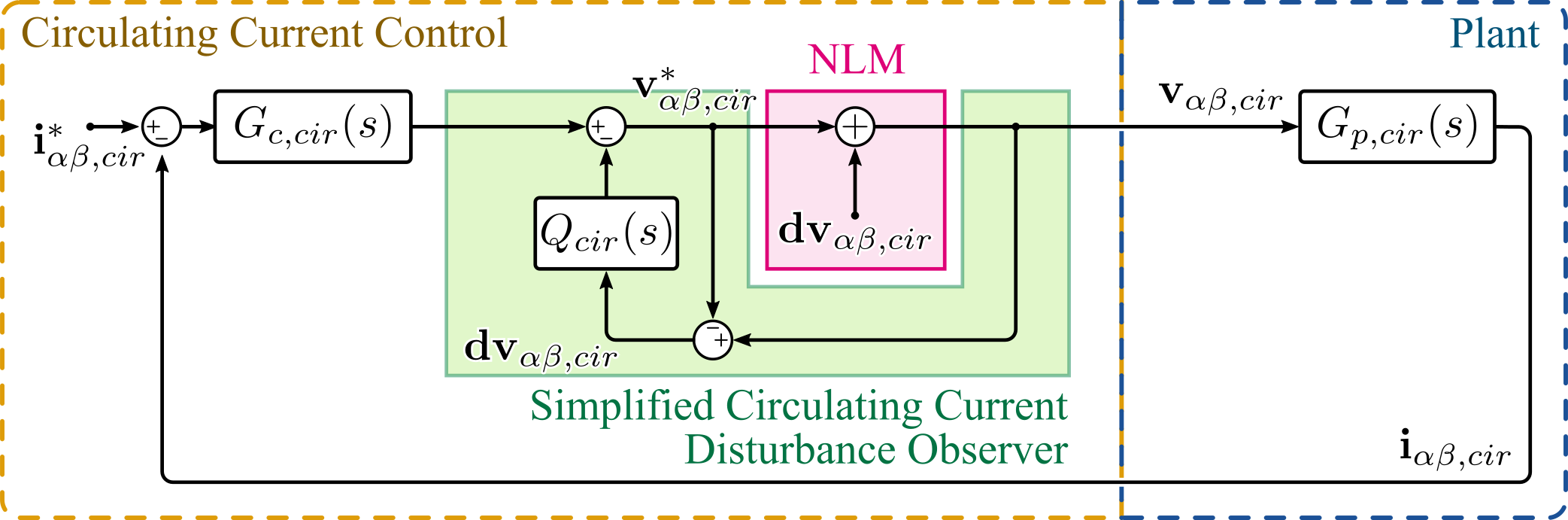}
        \caption{}
        \vspace{1em}
        \label{fig:mmc_dob_cir_simple}
    \end{subfigure}
    \caption{Simplified implementation of the proposed DOB for 
    (a) dc-side current, 
    (b) ac-side current, 
    (c) circulating current control loops.}
    \label{fig:mmc_dob_simple}
   \vspace{0.0em}  
\end{figure}

\begin{figure}[t!]
    \centering
    \includegraphics[width=1\linewidth]{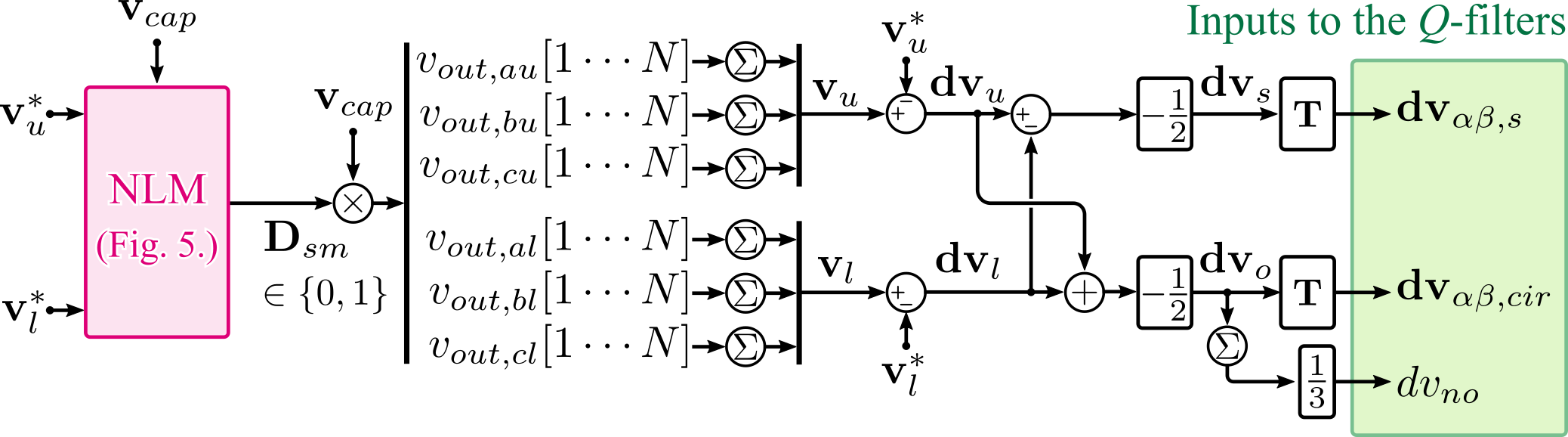}
    \caption{
        Implementation of disturbance extraction for the proposed DOB.
        Extracted voltage disturbances for each decoupled model,
        $\mathbf{dv}_{\alpha\beta,s}$, $\mathbf{dv}_{\alpha\beta,cir}$, and $dv_{no}$,
        are used as the inputs to the $Q$-filters.
    }
    \label{fig:mmc_dob_extraction}
   \vspace{0.0em}  
\end{figure}

In principle, the proposed DOB
can be implemented as described above.
However, for practical purposes,
the DOBs in \figurename{\ref{fig:mmc_dob}}
can be simplified as shown in 
\figurename{\ref{fig:mmc_dob_simple}}.
Since the arm voltage error 
is caused by rounding duty cycles of SMs in NLM,
it can be directly obtained from the capacitor voltages and duty cycles of the SMs.
Based on 
\eqref{eq:arm_voltage_disturbance} and \eqref{eq:mmc_disturb_voltage},
$\mathbf{dv}_s$ and $\mathbf{dv}_o$ can be directly calculated.
Consequently, 
as shown in \figurename{\ref{fig:mmc_dob_extraction}}, 
the voltage disturbances,
$\mathbf{dv}_{\alpha\beta,s}$, $\mathbf{dv}_{\alpha\beta,cir}$, 
and $dv_{no}$, 
which are used as inputs to the $Q$-filters, 
can be directly extracted without the need for the inverse transfer function of the RL load in each decoupled model.
Therefore, 
the proposed DOB can be efficiently implemented 
by directly extracting the voltage disturbances for each decoupled model 
and applying a single LPF to each current-control loop.

By adopting the proposed DOBs depicted in \figurename{\ref{fig:mmc_dob_simple}},
the responses of the dc-side, ac-side, 
and circulating currents to both the current references 
and the disturbances caused by NLM can be formulated as follows:
\begin{equation}
    i_{dc}(s) = T_{r,dc}(s)i_{dc}^*(s)+T_{d,dc}(s)dv_{no}(s),\label{eq:idc-w-dob}
\end{equation}
\begin{equation}
    \mathbf{i}_{\alpha\beta,ac}(s) = T_{r,ac}(s)i_{cir}^*(s)+T_{d,ac}(s)\mathbf{dv}_{\alpha\beta,ac}(s),\label{eq:iac-w-dob}
\end{equation}
\begin{equation}
    \mathbf{i}_{\alpha\beta,cir}(s) = T_{r,cir}(s)\mathbf{i}_{\alpha\beta,cir}^*(s)+T_{d,cir}(s)\mathbf{dv}_{\alpha\beta,cir}(s),\label{eq:icir-w-dob}
\end{equation}
where $T_{r,dc}(s)$, $T_{r,ac}(s)$, and $T_{r,cir}(s)$ 
denote the closed-loop transfer functions from the reference to the current output,
while $T_{d,dc}(s)$, $T_{d,ac}(s)$, and $T_{d,cir}(s)$ represent the closed-loop transfer functions 
from the disturbance input to the current output, for each decoupled model.
The closed-loop transfer functions from the disturbance and the reference input to the current output
are formulated as follows:
\begin{equation}
    T_{d,x}(s) = \frac{G_{p,x}(s)\left(1-Q_{x}(s)\right)}{1+G_{p,x}(s)G_{c,x}(s)},\label{eq:T_dx}
\end{equation}
\begin{equation}
    T_{r,x}(s) = \frac{G_{p,x}(s)G_{c,x}(s)}{1+G_{p,x}(s)G_{c,x}(s)},\label{eq:T_rx}
\end{equation}
where $x\in\{dc,ac,cir\}$.

As shown in \eqref{eq:T_dx},
the closed-loop system with respect to the disturbance input
shows the disturbance rejection capability depending on the $Q$-filter.
By approximating $Q_{x}(j\omega)\approx 1$ for $|\omega|\ll\omega_{Q,x}$,
and $Q_{x}(j\omega)\approx 0$ for $|\omega|\gg\omega_{Q,x}$, 
the current responses induced by the disturbance input
can be approximated as follows:
\begin{equation}
    T_{d,x}(j\omega) \approx 0,\,  |\omega|\ll\omega_{Q,x}.\label{eq:T_dx_approx}
\end{equation}

On the other hand, 
the closed-loop system with respect to the reference input in \eqref{eq:T_rx}
exhibits the response characteristics of the designed current controller, 
independently of NLM.
Therefore, 
the current controller can be designed regardless of NLM operation,
while the disturbance effects caused by NLM can be suppressed by the proposed DOB.

By increasing the cut-off frequency of the $Q$-filter in each decoupled current-control loop, 
the current quality of each decoupled model can be improved 
when NLM is applied to the indirect-modulated MMC.
Consequently,
the cut-off frequency of each $Q$-filter 
can be independently adjusted 
to improve the control quality of the dc-side, ac-side, and circulating currents
without interference among them.

\subsection{Stability Analysis of the Proposed Method} \label{sec:dob_stability}
Although a larger cut-off frequency of $Q$-filter leads to the 
better current-control performance
due to the augmented disturbance rejection,
the stability of the control loop with the DOB
should be considered to ensure the stable operation.
For simplicity of analysis,
it is assumed that
the sampling frequency of the MMC controller
is sufficiently high to neglect discretization effects.
In addition,
all current controllers to regulate the dc-side, ac-side and circulating currents
are assumed to be properly designed to stabilize the nominal RL load
in corresponding decoupled model.
Therefore, 
the stability of the control loop employing the proposed DOB 
is analyzed with respect to the second and third conditions discussed in Section \ref{subsec:StabilityForDisturbanceObserver},
for both the principle-based implementation shown in \figurename{\ref{fig:mmc_dob}} and 
the simplified implementation shown in \figurename{\ref{fig:mmc_dob_simple}}.

\subsubsection{Principle-based Implementation}\label{subsub:PincipleBasedImplementation}
The principle-based implementation of the proposed DOB 
is shown in \figurename{\ref{fig:mmc_dob}}.
To satisfy the second condition,
the transfer function of RL load needs to be the minimum phase.
Since $R_o$, $L_o$, $R_s$, and $L_s$
are positive real value,
the plant transfer function for each decoupled model
has a pole in LHP and no zero,
as shown in \eqref{eq:dc-side-circuit-s}, \eqref{eq:circulating-current-circuit-alpha-beta-s}, and \eqref{eq:ac-side-circuit-alpha-beta-s}.
Therefore, the plant transfer functions for each current controller
are the minimum phase.

To satisfy the third condition,
the coefficient and time constant for $Q_{dc}(s)$,
$Q_{cir}(s)$, and $Q_{ac}(s)$
should be chosen such that following polynomials are Hurwitz polynomials,
meaning that all roots of polynomials are located in LHP.
\begin{align}
    p_{f,dc}(s)&=s+\frac{L_o}{L_{o}^*}a_{0,dc}, \label{eq:poly-dc}\\
    p_{f,cir}(s)&=s+\frac{L_o}{L_{o}^*}a_{0,cir}, \label{eq:poly-cir}\\
    p_{f,ac}(s)&=s+\frac{L_o+\frac{1}{2}L_s}{L_o^*+\frac{1}{2}L_s^*}a_{0,ac}, \label{eq:poly-ac}
\end{align}
where the superscript `*' denotes the 
nominal parameter value.
Under the assumption that 
$L_o$, $L_o^*$, $L_s$, and $L_s^*$
are positive values,
the roots of $p_{f,dc}$, $p_{f,cir}$, and $p_{f,ac}$
are located in LHP for all $ a_{0,dc},\, a_{0,cir},\, a_{0,ac}>0$.
This is consistent with practical systems, 
as the inductance of a physical inductor, 
being a passive component, is always positive.
Note that 
arbitrary positive values for 
$a_{0,dc}$, $a_{0,cir}$, and $a_{0,ac}$
make $p_{f,dc}$, $p_{f,cir}$, and $p_{f,ac}$
Hurwitz polynomials,
which is guaranteed by the physical property that the inductance of a physical inductor is always positive.

By satisfying the first, second and third conditions in Section \ref{subsec:StabilityForDisturbanceObserver},
there exist positive constants $\tau_{dc}^*$, $\tau_{cir}^*$, and $\tau_{ac}^*$ 
such that the control loops employing the DOB are stable 
for $\tau_{dc} \in (0, \tau_{dc}^*]$, $\tau_{cir} \in (0, \tau_{cir}^*]$, and $\tau_{ac} \in (0, \tau_{ac}^*]$, respectively.

To properly design $\tau_{x}$ of $Q_x(s)$ in \eqref{eq:q-filter-mmc},
it appears necessary to know 
the upper bounds of the time constant, $\tau_{x}^*$,
in order to ensure  $\tau_{x} \in (0,\tau_{x}^*]$.
However, instead of directly determining $\tau_x$ and $a_{0,x}$ in $Q$-filter,
$Q$-filter for stable control loop 
of each decoupled model 
can be indirectly realized by parameterizing $a_{0,x}$ in $Q_{x}(s)$ as follows:
\begin{equation}
    a_{0,x}=\omega_{Q,x}\tau_{x}^*,\,x\in\{dc,cir,ac\}.  \label{eq:t_x^*}
\end{equation}

Since $p_{f,dc}(s)$, $p_{f,cir}(s)$, and $p_{f,ac}(s)$ 
are Hurwitz polynomials for arbitrary positive 
$a_{0,dc}$, $a_{0,cir}$, and $a_{0,ac}$,
there exists positive $\tau_{x}^*$ satisfying \eqref{eq:t_x^*} for $\omega_{Q,x}>0$.
In summary, 
by substituting \eqref{eq:t_x^*} into \eqref{eq:q-filter-mmc}, 
the $Q$-filters for the decoupled models can be flexibly designed 
with arbitrary cut-off frequency, $\omega_{Q,x}$, as shown below, while maintaining the stability of the control loops.
\begin{equation}
    Q_{x}(s)=\frac{\omega_{Q,x}}{s+\omega_{Q,x}},\quad x\in\{dc,cir,ac\}. \label{eq:q-filter-mmc2}
\end{equation}

\subsubsection{Simplified Implementation}\label{subsub:SimplifiedImplementation}
The simplified implementation of the proposed DOB
is shown in \figurename{\ref{fig:mmc_dob_simple}}.
This realization is the equivalent 
to the principle-based implementation depicted in \figurename{\ref{fig:mmc_dob}},
under the assumption that 
the nominal parameters --- $L_o^*$, $L_s^*$, $R_o^*$, and $R_s^*$ ---
are identical to the actual parameters of the plant, 
namely $L_o$, $L_s$, $R_o$, and $R_s$.
Therefore, 
the simplified implementation is a special case of the principle-based implementation.
Accordingly,
the stability analysis in Section \ref{subsub:PincipleBasedImplementation}
remains valid.

For the intuitive understanding,
the stability analysis can be simplified 
based on the transfer functions in \eqref{eq:T_dx} and \eqref{eq:T_rx}.
Since the current controllers are assumed 
to be designed to stabilize the nominal RL load in each model,
all poles of $T_{r,x}(s)$ in \eqref{eq:T_rx} are located in LHP.
It means that solutions of following equation are located in LHP.
\begin{gather}
    D_{p,x}(s)D_{c,x}(s)+N_{p,x}(s)N_{c,x}(s)=0,\nonumber\\
     x\in\{dc,cir,ac\},\label{eq:char_eq}
\end{gather}
where $D_{p,x}(s)$ and $D_{c,x}(s)$
denote the denominator polynomials of $G_{p,x}(s)$ and $G_{c,x}(s)$, 
while $N_{p,x}(s)$ and $N_{c,x}(s)$
denote the numerator polynomials of $G_{p,x}(s)$ and $G_{c,x}(s)$, respectively.

By substituting $D_{p,x}(s)$ and $N_{p,x}(s)$, 
as well as $D_{c,x}(s)$ and $N_{c,x}(s)$, 
into $G_{p,x}(s)$ and $G_{c,x}(s)$ in \eqref{eq:T_dx}, respectively,
and further substituting $Q_x(s)$ in \eqref{eq:q-filter-mmc2},
$T_{d,x}(s)$
can be rewritten as follows:
\begin{align}
    T_{d,x}(s)
    &=\frac{N_{p,x}(s)D_{c,x}(s)}{D_{p,x}(s)D_{c,x}(s)+N_{p,x}(s)N_{c,x}(s)}\frac{s}{s+\omega_{Q,x}}.\label{eq:T_dx2}
\end{align}

Since $\omega_{Q,x}$ is positive 
and the solutions of \eqref{eq:char_eq} are located 
in the LHP, all poles of $T_{d,x(s)}$ are also located in the LHP.
Therefore, 
the closed-loop systems described in \eqref{eq:idc-w-dob}, \eqref{eq:iac-w-dob}, and \eqref{eq:icir-w-dob}
are stable for arbitrary cut-off frequency of the $Q$-filter in each decoupled model.

From a control perspective, 
increasing the cut-off frequency of the $Q$-filter 
in each decoupled current-control loop 
is desirable 
for enhancing the decoupled control performance 
of the MMC under NLM operation, 
as it does not compromise the stability of the control loops. 
However, in practice, there exists a trade-off 
between the cut-off frequency of the $Q$-filter 
and the average switching frequency of the SMs. 
In the following simulation and experimental results, 
this trade-off is examined, 
and the effectiveness of the proposed DOB design 
is validated. 

\begin{table}[t]
        \begin{threeparttable}
            \caption{\textsc{Parameters for Decoupled Current Control}}
            \label{tab:params_ctrl}
            \centering
            {\renewcommand{\arraystretch}{1.2}
        
            \begin{tabular*}{\columnwidth}{@{\extracolsep{\fill}}ll}
                \toprule
                \multicolumn{2}{c}{DC-side Current Control} \\ 
                \midrule
                Controller, $G_{c,dc}(s)$ & $K_{p,dc}+K_{i,dc}/s$    \\
                Bandwidth, $\omega_{bw,dc}$ & $2\pi\times 100\,$ rad/s      \\
                Proportional Gain, $K_{p,dc}$   &$\left(L_o/3\right)\times \omega_{bw,dc}$\\
                Integral Gain, $K_{i,dc}$   &$\left(R_o/3\right)\times \omega_{bw,dc}$\\
                \toprule
                \multicolumn{2}{c}{AC-side Current Control \tnote{*} } \\ 
                \midrule
                Controller, $G_{c,ac}(s)$ & $K_{p,ac}+K_{i,ac}/s++K_{r,ac}s/(s^2+\omega_r^2)$     \\
                Reference frame & Stationary Reference Frame ($\alpha\beta$-frame)    \\
                Bandwidth, $\omega_{bw,ac}$ & $2\pi\times 100\,$ rad/s      \\
                Proportional Gain, $K_{p,dc}$   &$\left(L_o/2\right)\times \omega_{bw,ac}$\\
                Integral Gain, $K_{i,dc}$   &$\left(R_o/2\right)\times \omega_{bw,ac}$\\
                Resonant Gain, $K_{r,cir}$   &$(R_o/2)\times \omega_{bw,ac}$\\
                \toprule
                \multicolumn{2}{c}{Circulating Current Control \tnote{*} } \\ 
                \midrule
                Controller, $G_{c,cir}(s)$ & $K_{p,cir}+K_{i,cir}/s+K_{r,cir}s/(s^2+\omega_r^2)$     \\
                Reference frame & Stationary Reference Frame ($\alpha\beta$-frame)    \\
                Bandwidth, $\omega_{bw,cir}$ & $2\pi\times 100\,$ rad/s      \\
                Proportional Gain, $K_{p,cir}$   &$L_o\times \omega_{bw,cir}$\\
                Integral Gain, $K_{i,cir}$   &$(R_o)\times \omega_{bw,cir}$\\
                Resonant Gain, $K_{r,cir}$   &$(R_o)\times \omega_{bw,cir}$\\
                \bottomrule
            \end{tabular*}%
            \scriptsize
            \begin{tablenotes}
            \item[*]{Resonant frequency, $\omega_r$, is set to the ac-side grid frequency obtained by phase-locked loop (PLL) }
            \end{tablenotes}
            \vspace{-0.0em}
        }
        \end{threeparttable}
        \vspace{0em}
\end{table}
\begin{table}[t]
        \begin{threeparttable}
            \caption{\textsc{System Parameters for Simulation}}
            \label{tab:params_sys}
            \centering
            {\renewcommand{\arraystretch}{1.2}
        
            \begin{tabular*}{\columnwidth}{@{\extracolsep{\fill}}llll}
                \toprule
                \multicolumn{4}{c}{Modular Multilevel Converter} \\ 
                \midrule
                Rated power & 1 MW         & Number of SM per arm & 10 \\
                Rated dc voltage, $V_{dc}^*$     & 10 kV         & SM nominal dc voltage  & 1 kV\\
                Arm inductance, $L_o$       & 16 mH   & SM capacitance & 1 mF\\
                Arm resistance, $R_o$ & 8 m$\Omega$ & Sampling frequency & 10 kHz\\
                \toprule
                \multicolumn{4}{c}{AC Grid} \\ 
                \midrule
                Grid line-to-line voltage\tnote{*} & 6.6 kV & Nominal grid frequency & 60 Hz\\
                \bottomrule
            \end{tabular*}%
            \scriptsize
            \begin{tablenotes}
            \item[*]{Voltage is expressed as its root-mean-square (RMS) value.}
            \end{tablenotes}
            \vspace{-0.0em}
        }
        \end{threeparttable}
        \vspace{0em}
\end{table}

\section{Simulation Results} \label{sec:simulation}

To verify the validity of the proposed DOB design,
a simulation of MMC-based DC/AC power conversion is conducted
by using PLECS\textsuperscript{\textregistered}.
The control and system parameters are summarized in \tablename{\ref{tab:params_ctrl}} and \tablename{\ref{tab:params_sys}}, respectively.

In this section, the following aspects are described:
1)$\,$ It is demonstrated that,
for each decoupled current control loop,
the voltage disturbance caused by NLM is
independently rejected by the proposed DOB.
2)$\,$ It is demonstrated that,
the capacitor voltage balancing is achieved by 
employing the proposed DOB for dc-side and circulating currents.
3)$\,$ The effects of the $Q$-filter cut-off frequency 
on the current quality and the average switching frequency of SMs 
are numerically investigated.
 
\subsection{Improvement of Decoupled Current Control} \label{sec:sims_improvement_of_decoupled_current_control}

\begin{figure*}[t!]
    \vspace{-0.0em}  
    \centering
    \begin{subfigure}[b]{0.325\linewidth}
        \includegraphics[width=1.0\linewidth,center]{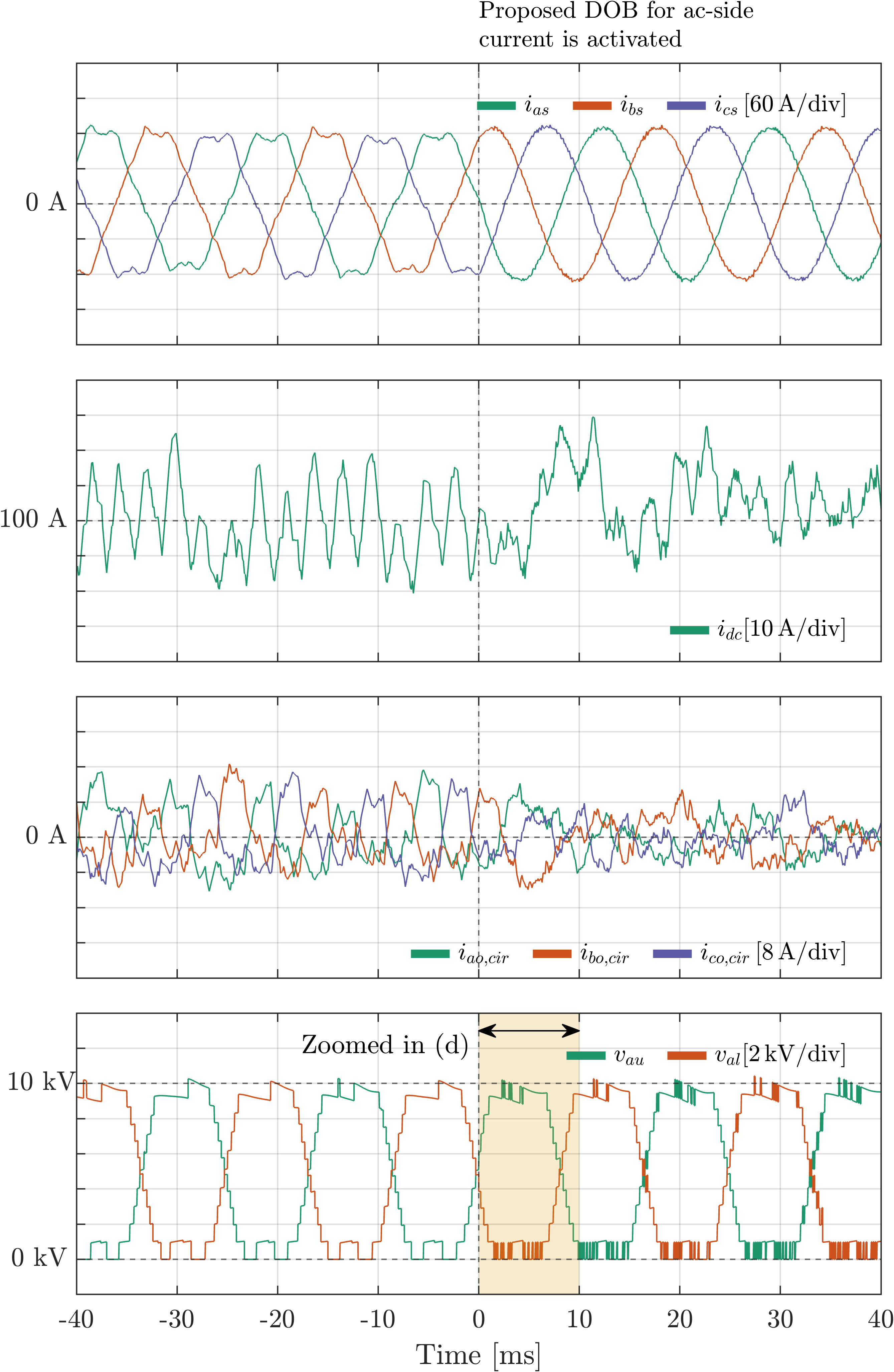}
        \caption{}
        \label{fig:mmc_dob_ac_enable}
    \end{subfigure}
    \hfill
    \begin{subfigure}[b]{0.325\linewidth}
        \includegraphics[width=1.0\linewidth,center]{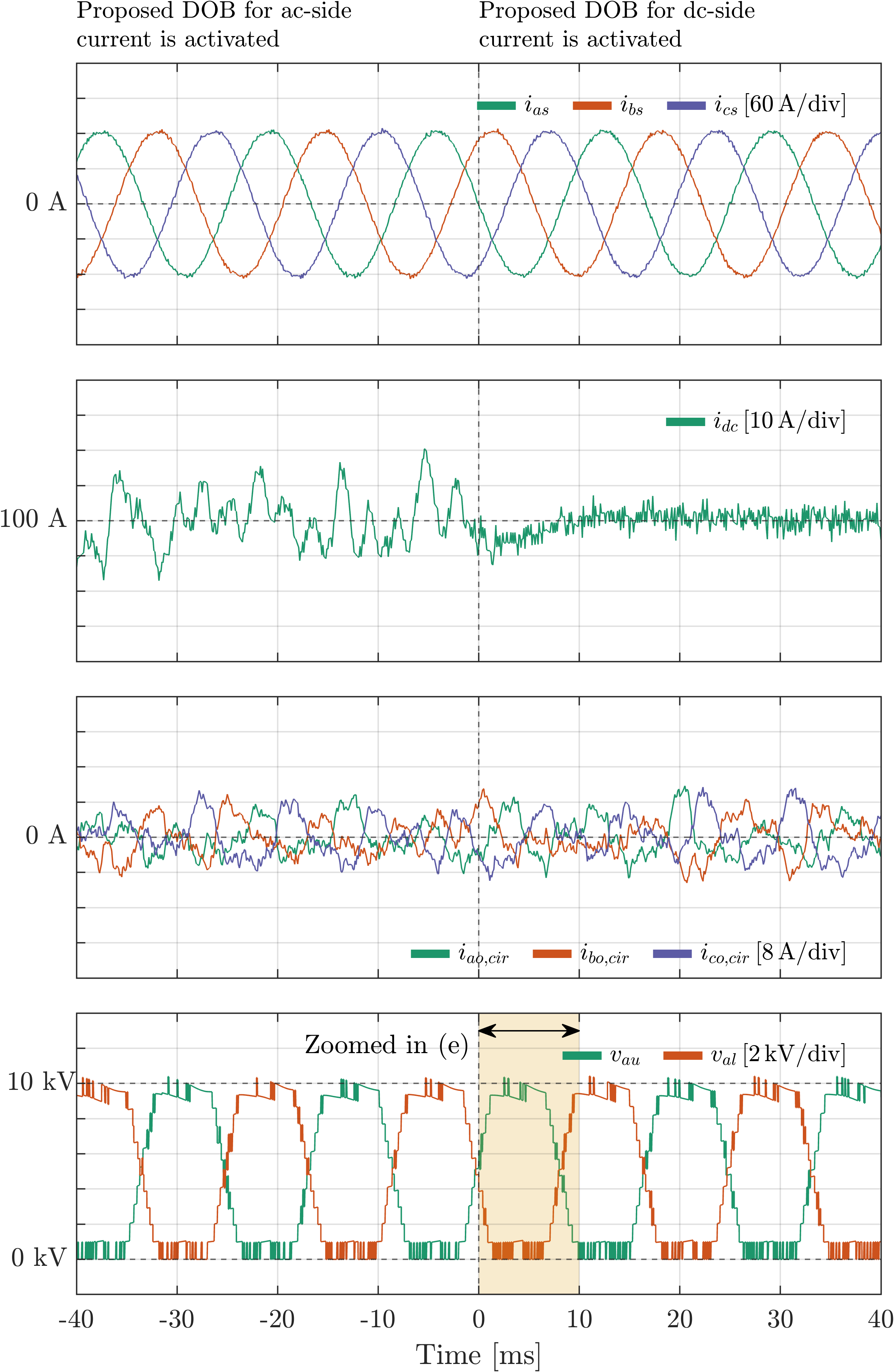}
        \caption{}
        \label{fig:mmc_dob_dc_enable}
    \end{subfigure}
    \hfill
    \begin{subfigure}[b]{0.325\linewidth}
        \includegraphics[width=1.0\linewidth,center]{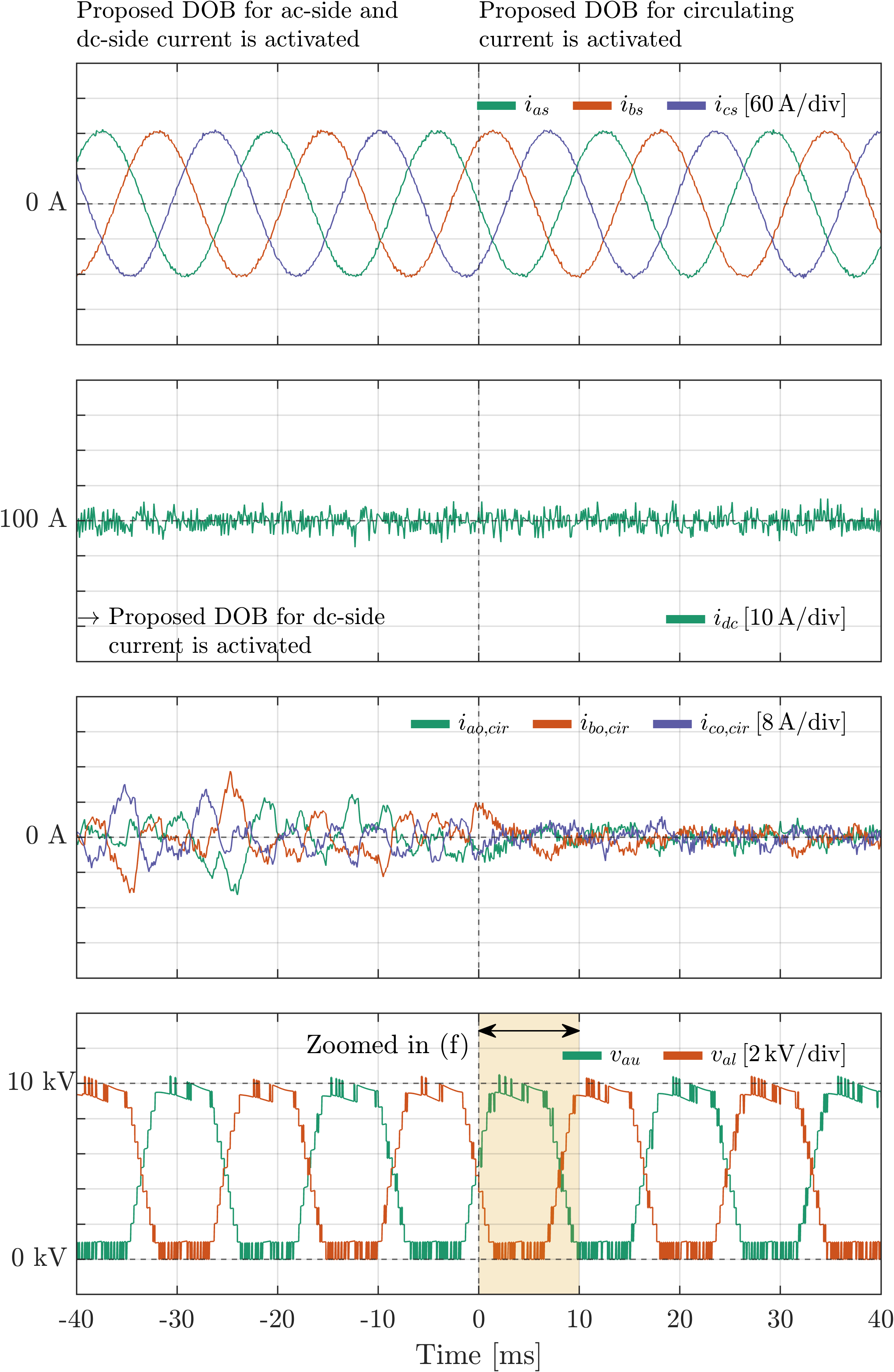}
        \caption{}
        \label{fig:mmc_dob_cir_enable}
    \end{subfigure}

    \begin{subfigure}[b]{0.325\linewidth}
        \vspace{1em}
        \includegraphics[width=1.0\linewidth,center]{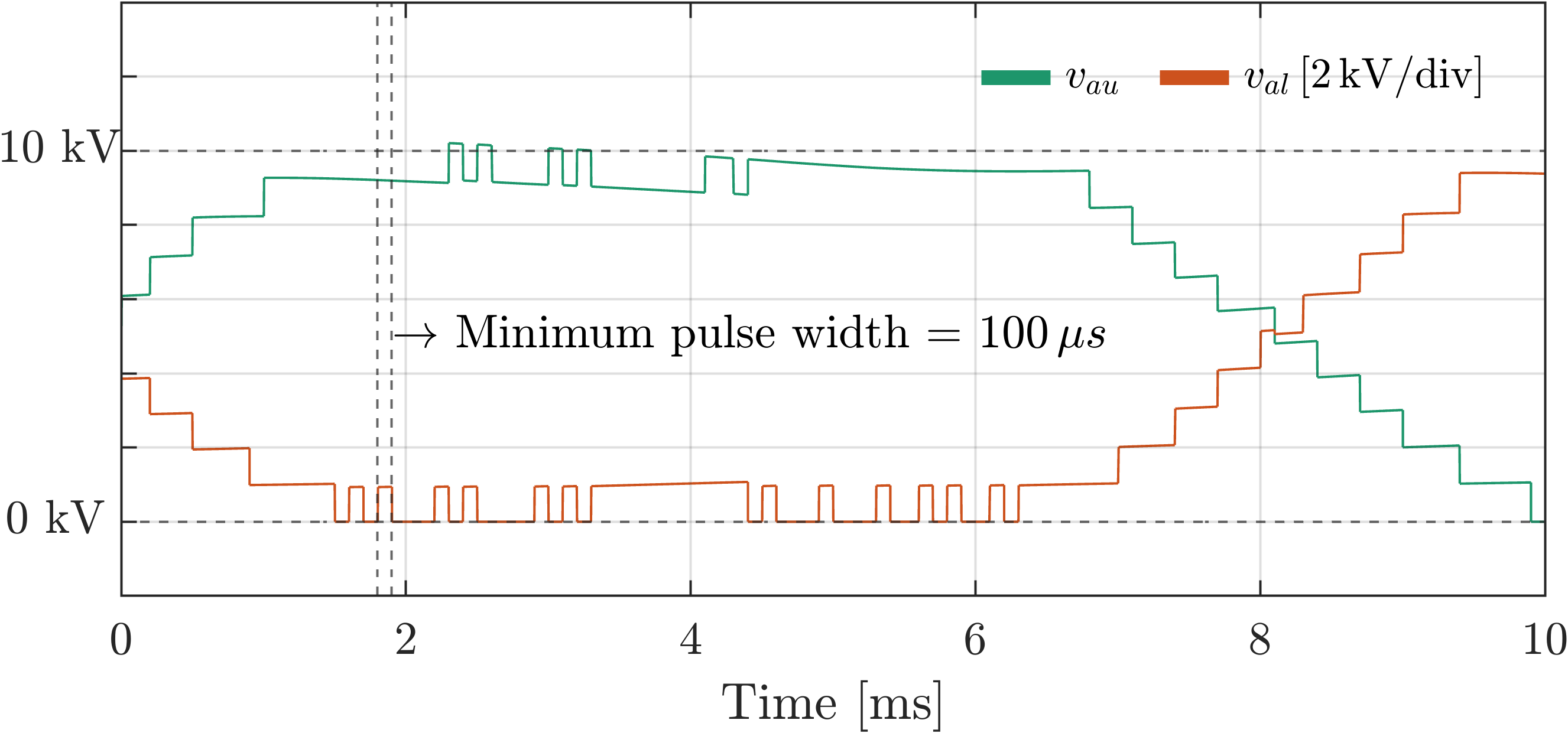}
        \caption{}
        \label{fig:mmc_dob_ac_enable_zoom}
    \end{subfigure}
    \hfill
    \begin{subfigure}[b]{0.325\linewidth}
        \vspace{1em}
        \includegraphics[width=1.0\linewidth,center]{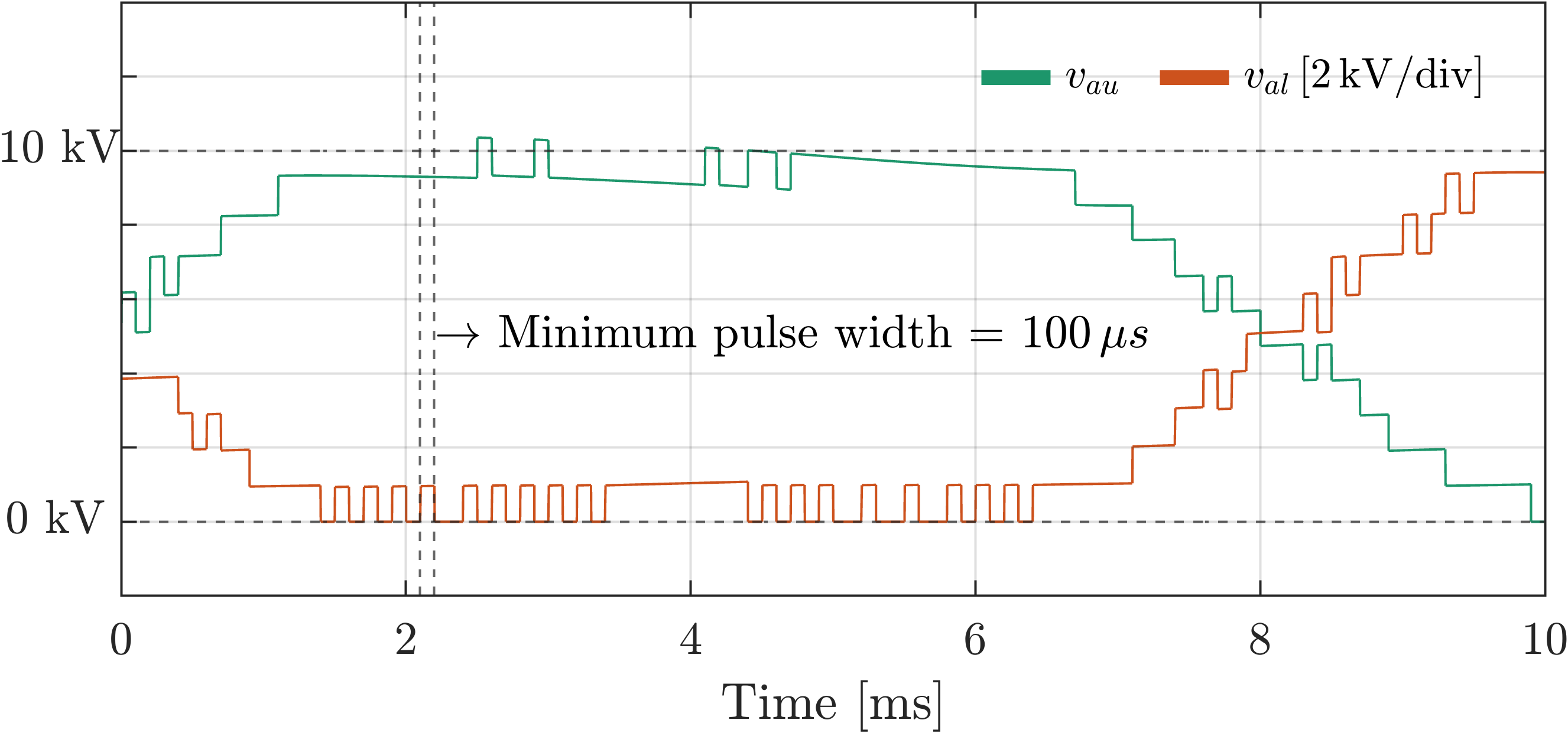}
        \caption{}
        \label{fig:mmc_dob_dc_enable_zoom}
    \end{subfigure}
    \hfill
    \begin{subfigure}[b]{0.325\linewidth}
        \vspace{1em}
        \includegraphics[width=1.0\linewidth,center]{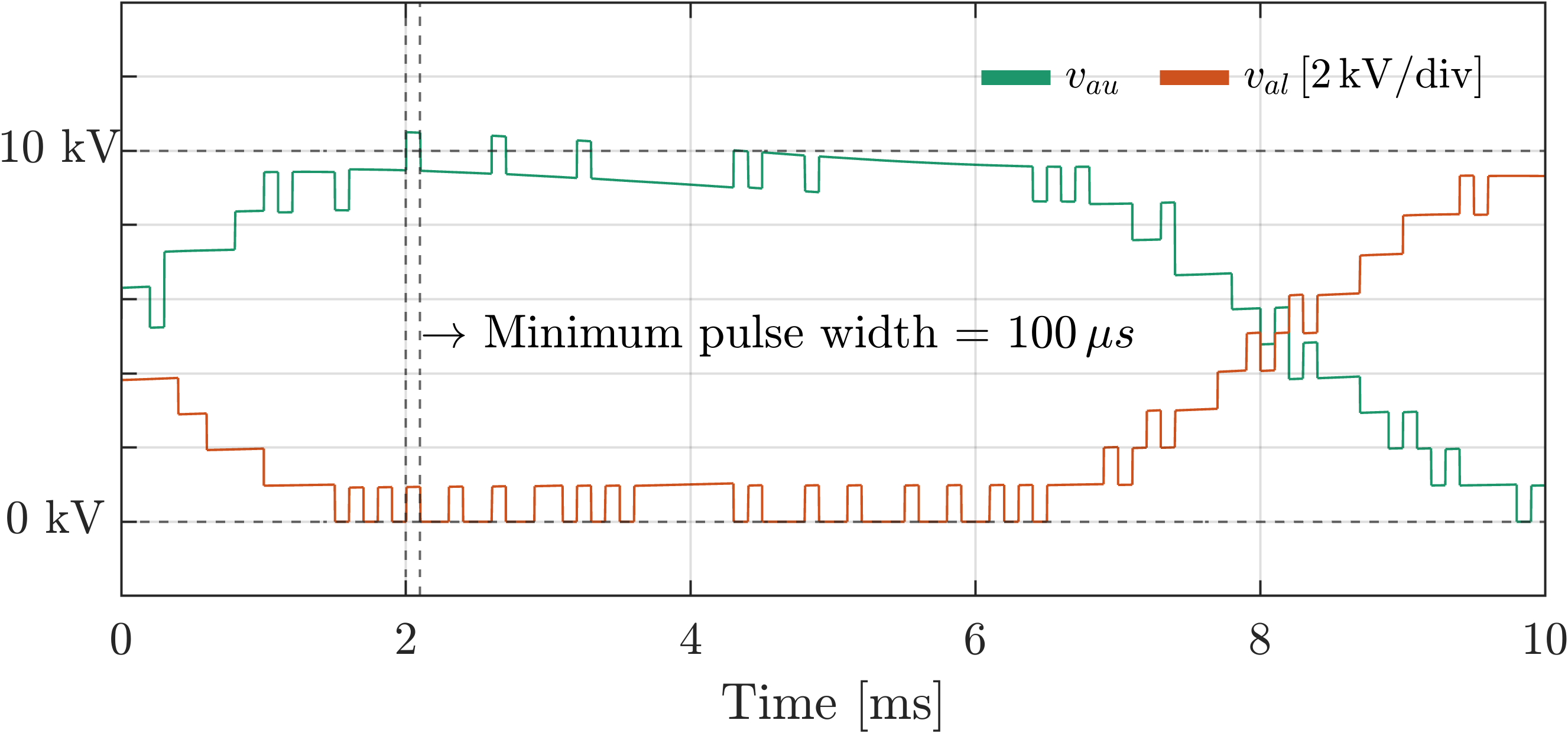}
        \caption{}
        \label{fig:mmc_dob_cir_enable_zoom}
    \end{subfigure}

    \caption{
    Simulation results representing the effects of the DOB
    on (a) ac-side current, (b) dc-side current, and (c) circulating current.
    For case (a), only the DOB for ac-side current is activated at $t=0\,$s.
    For case (b), the DOB for dc-side current is activated at $t=0\,$s
    while only the DOB for ac-side current is initially activated.
    For case (c), the DOB for circulating current is activated at $t=0\,$s
    while the DOBs for ac-side and dc-side currents are initially activated.
    Arm voltages shown in (d), (e), and (f) present magnified views of the arm voltages 
    from 0 ms to 10 ms, corresponding to (a), (b), and (c), respectively.}
    \label{fig:mmc_dob_enable}
   \vspace{0.0em}  
\end{figure*}

\begin{figure}[t]
    \vspace{-0.0em}  
    \centering
    \begin{subfigure}[b]{0.493\linewidth}
        \includegraphics[width=1.0\linewidth,center]{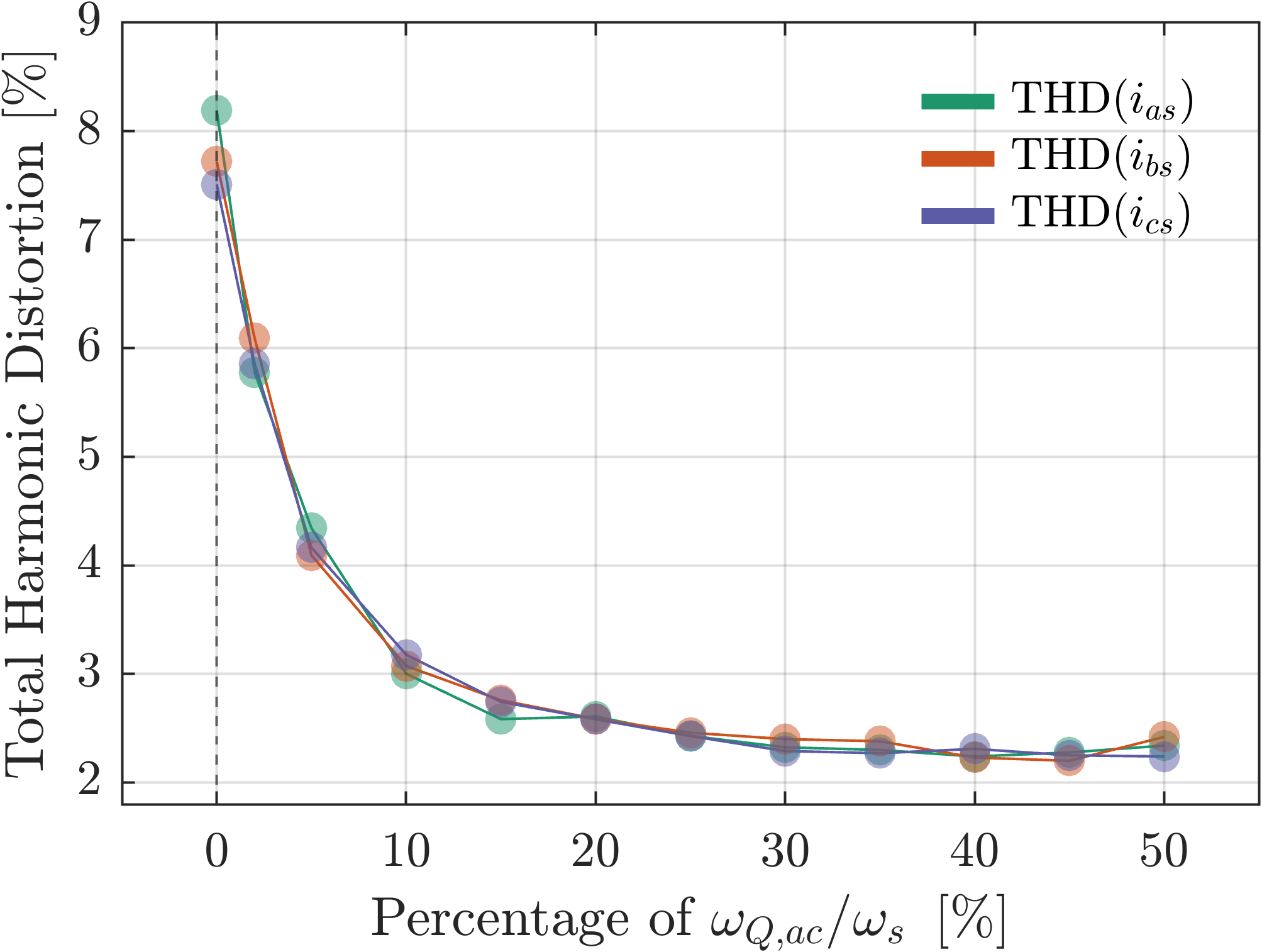}
        \caption{}
        \label{fig:dob_thd}
    \end{subfigure}
    \hfill
    \begin{subfigure}[b]{0.493\linewidth}
        \includegraphics[width=1.0\linewidth,center]{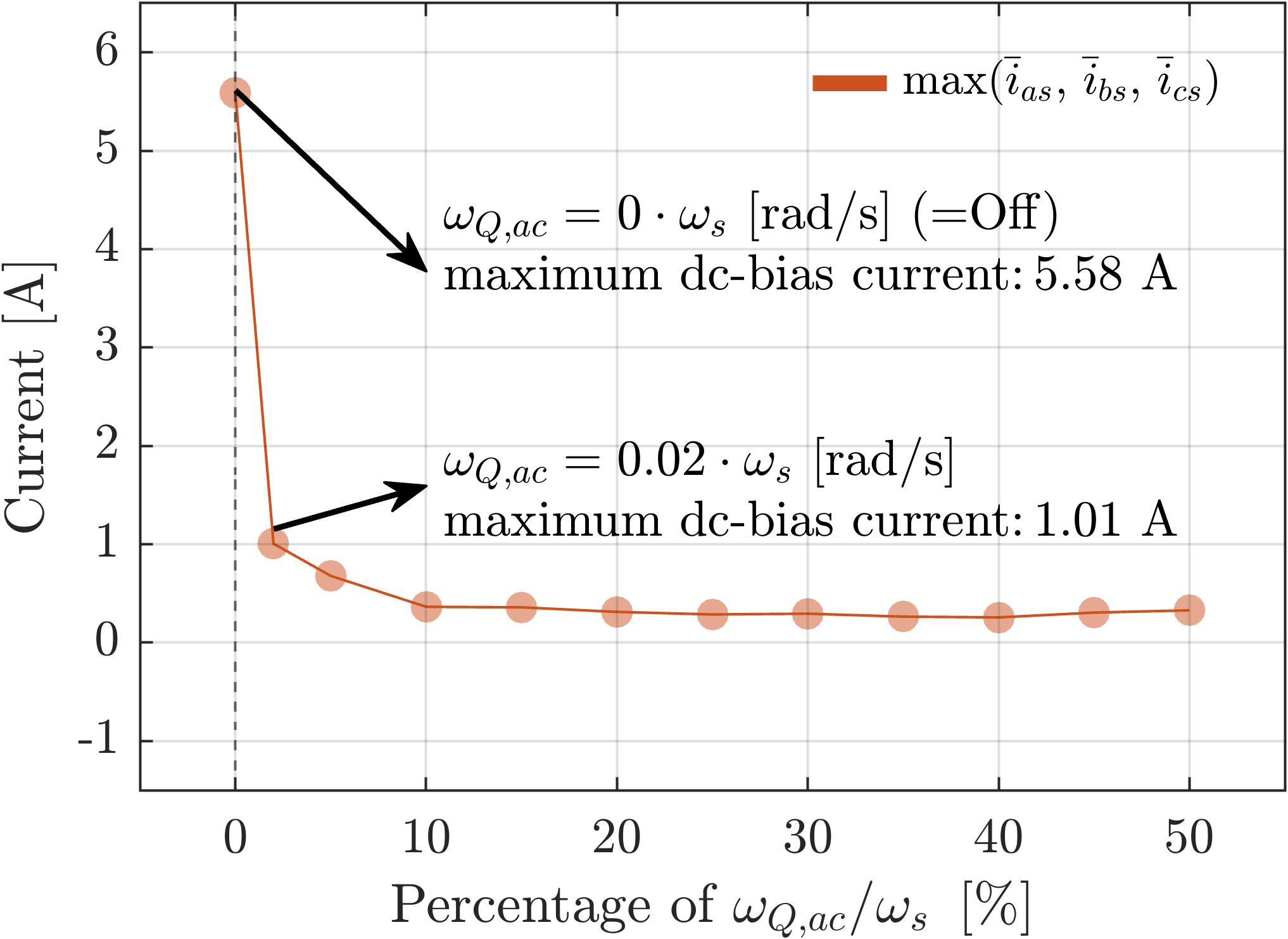}
        \caption{}
        \label{fig:dob_dc_bias}
    \end{subfigure}
    \caption{Simulation results showing the effects of the DOB cut-off frequency for ac-side current on 
    (a) total harmonic distortion (THD) and 
    (b) dc bias of the ac-side current.
    $\bar{i}_{x}$, indicates the average value of $i_{x\in\{as,bs,cs\}}$ over a fundamental period.
    max$(\cdot)$ represents the maximum value among three phases.
     $\omega_{s}$ represents the angular sampling frequency in rad/s.}
    \label{fig:dob_ac_cutoff_effect}
   \vspace{0.0em}  
\end{figure}

\figurename{\ref{fig:mmc_dob_enable}} illustrates effects of DOBs 
for each decoupled current component, $i_{dc}$, $\mathbf{i}_{s}$, and $\mathbf{i}_{o}$.
The cut-off frequency of the $Q$-filter for each DOB
is set to half of the sampling frequency.

In \figurename{\ref{fig:mmc_dob_ac_enable}}, 
all proposed DOBs are initially deactivated for $t<0\,$s.
At $t=0\,$s, only the DOB for ac-side current is activated.
As a result, the ac-side current restores three-phase balance
and the harmonic distortion is conspicuously reduced,
while the dc-side current is barely affected.
It is observed that the circulating current is slightly improved. 
This is because the ac-side output power is stably regulated 
by the proposed DOB for ac-side current for $t>0\,$s, 
which indirectly helps mitigate the irregular energy imbalance between the arms.
Consequently, the circulating current, carrying the power between arms, is improved to some extent.
However, its improvement is limited
since the disturbance caused by NLM operation for circulating current
is not directly addressed.

Similarly, \figurename{\ref{fig:mmc_dob_dc_enable}}
demonstrates that the DOB for the dc-side current effectively
suppresses the fluctuation in $i_{dc}$.
\figurename{\ref{fig:mmc_dob_cir_enable}} shows 
that the DOB for the circulating current 
mitigates its irregular fluctuations.
The improvement in circulating current is not as notable as that in the dc or ac-side currents;
however, as will be discussed later, it brings significant enhancement in capacitor energy control.
These results in \figurename{\ref{fig:mmc_dob_enable}} validate that the proposed DOB can be independently
applied to each decoupled current component, 
effectively rejecting the corresponding NLM-induced disturbance.

The arm voltages illustrated in 
\figurename{\ref{fig:mmc_dob_ac_enable_zoom}, \subref{fig:mmc_dob_dc_enable_zoom}, and \subref{fig:mmc_dob_cir_enable_zoom}},
show magnified views of the arm voltages 
from 0 ms to 10 ms, corresponding to
\figurename{\ref{fig:mmc_dob_ac_enable}, \subref{fig:mmc_dob_dc_enable}, and \subref{fig:mmc_dob_cir_enable}}, respectively.
The minimum pulse width of the arm voltage is the same with the sampling period, 100 $\mu s$,
because NLM operation is employed.

\figurename{\ref{fig:dob_ac_cutoff_effect}}
illustrates the effects of the DOB cut-off frequency on the ac-side current quality.
Total harmonic distortion (THD) and dc bias of the ac-side current 
are observed under various cut-off frequencies of the DOB for ac-side current,
ranging from 0\% to 50\% of the sampling frequency,
while cut-off frequencies of the other DOBs are fixed to half of the sampling frequency.
As the cut-off frequency increases,
THD of the ac-side current decreases,
as shown in \figurename{\ref{fig:dob_thd}}.
Furthermore, the proposed DOB mitigates the dc bias of the ac-side current, 
as depicted in \figurename{\ref{fig:dob_dc_bias}}. 
This is a significant advantage because such a dc component 
can cause asymmetric magnetic core saturation 
in the ac-side transformer \cite{Buticchi2013detection}.

\subsection{Improvement of Capacitor Voltage Balancing}\label{sec:sims_improvement_of_capacitor_voltage_balancing}

\begin{figure}[t]
    \vspace{-0.0em}  
    \centering
    \begin{subfigure}[b]{0.493\linewidth}
        \includegraphics[width=1.0\linewidth,center]{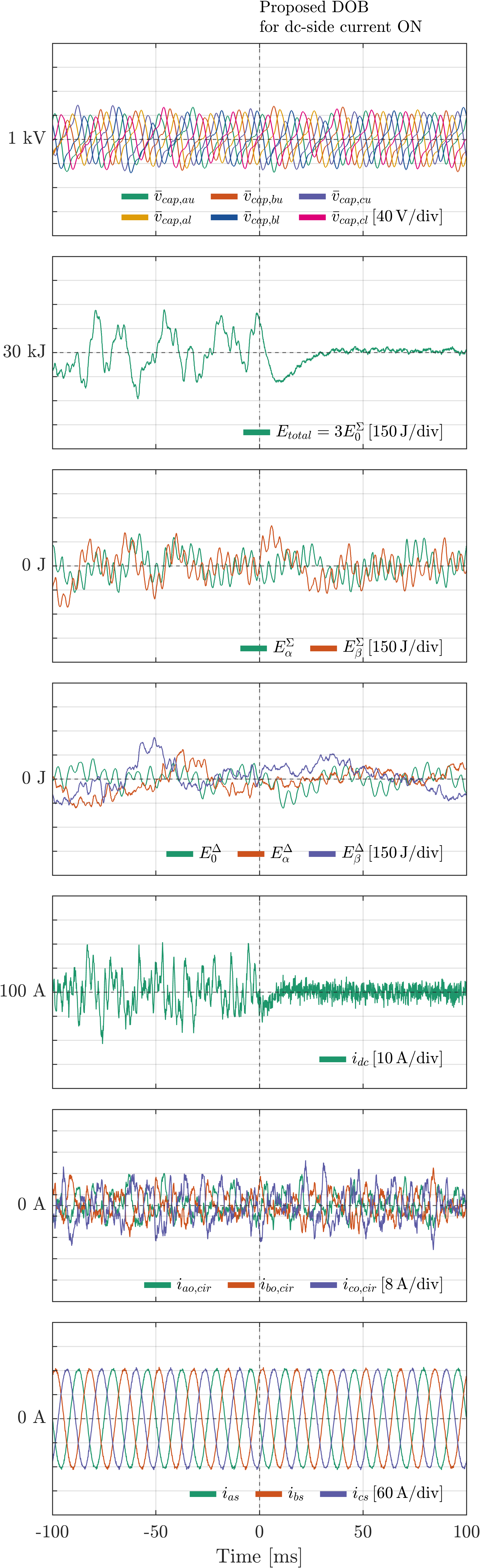}
        \caption{}
        \label{fig:dob_capvoltage_effect_dc}
    \end{subfigure}
    \hfill
    \begin{subfigure}[b]{0.493\linewidth}
        \includegraphics[width=1.0\linewidth,center]{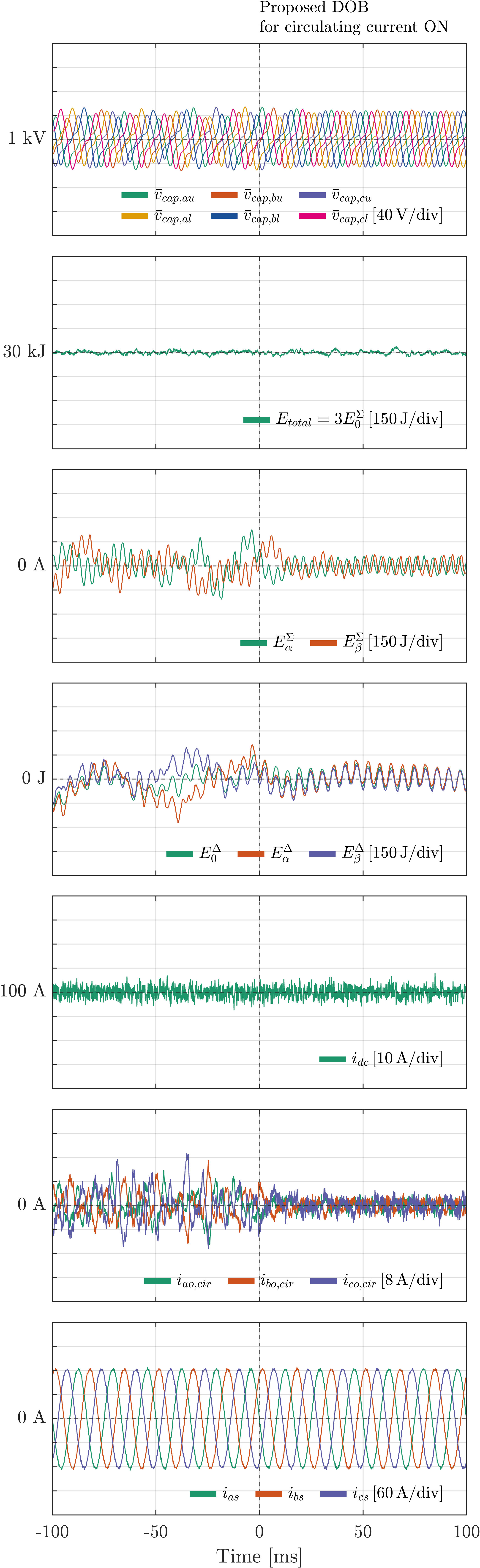}
        \caption{}
        \label{fig:dob_capvoltage_effect_cir}
    \end{subfigure}
    \caption{Simulation results showing the effects of DOBs on capacitor voltage balancing.
        (a) DOB for dc-side current is activated at $t=0\,$s
        while only the DOB for ac-side current is initially activated.
        (b) DOB for circulating current is activated at $t=0\,$s,
        while the DOBs for the ac-side and dc-side currents are initially activated.}
    \label{fig:dob_capvoltage_effect}
   \vspace{0.0em}  
\end{figure}

\figurename{\ref{fig:dob_capvoltage_effect}} illustrates
the effects of the proposed DOBs on capacitor voltage balancing.
$\mathbf{E}_{abc}^\Sigma$ and $\mathbf{E}_{abc}^\Delta$
are represented as $\mathbf{E}_{\alpha\beta0}^\Sigma=[E_{\alpha}^\Sigma\quad E_{\beta}^\Sigma\quad E_{0}^\Sigma]^\top$ 
and $\mathbf{E}_{\alpha\beta0}^\Delta=[E_{\alpha}^\Delta\quad E_{\beta}^\Delta\quad E_{0}^\Delta]^\top$
in the $\alpha\beta0$-frame,
respectively, by using Clarke transformation.
In \figurename{\ref{fig:dob_capvoltage_effect_dc}},
only the DOB for ac-side current is activated for $t<0\,$s.
At $t=0\,$s, the DOB for dc-side current is activated.
As a result, the total capacitor energy in MMC, $E_{total}$,
is properly regulated to its nominal value,
by achieving the proper dc-side current reference tracking
via the proposed DOB. However,
since the circulating current is not well regulated,
$\mathbf{E}_{\alpha\beta}^\Sigma$ and $\mathbf{E}_{\alpha\beta 0}^\Delta$
are irregularly fluctuated, 
as shown in the third and fourth plots of \figurename{\ref{fig:dob_capvoltage_effect_dc}}.

In \figurename{\ref{fig:dob_capvoltage_effect_cir}},
the DOB for circulating current is additionally activated.
By suppressing the disturbance in the circulating current,
$\mathbf{E}_{\alpha\beta}^\Sigma$ and $\mathbf{E}_{\alpha\beta 0}^\Delta$
are effectively controlled
to converge to zero.
Consequently, the capacitor voltages of all SMs are balanced,
as shown in the first plot of \figurename{\ref{fig:dob_capvoltage_effect_cir}}.

\subsection{Trade-off in DOB Cut-off Frequency} \label{sec:sims_trade_off_dob}

\begin{figure}[t]
    \vspace{-0.0em}  
    \centering
    \begin{subfigure}[b]{0.493\linewidth}
        \includegraphics[width=1.0\linewidth,center]{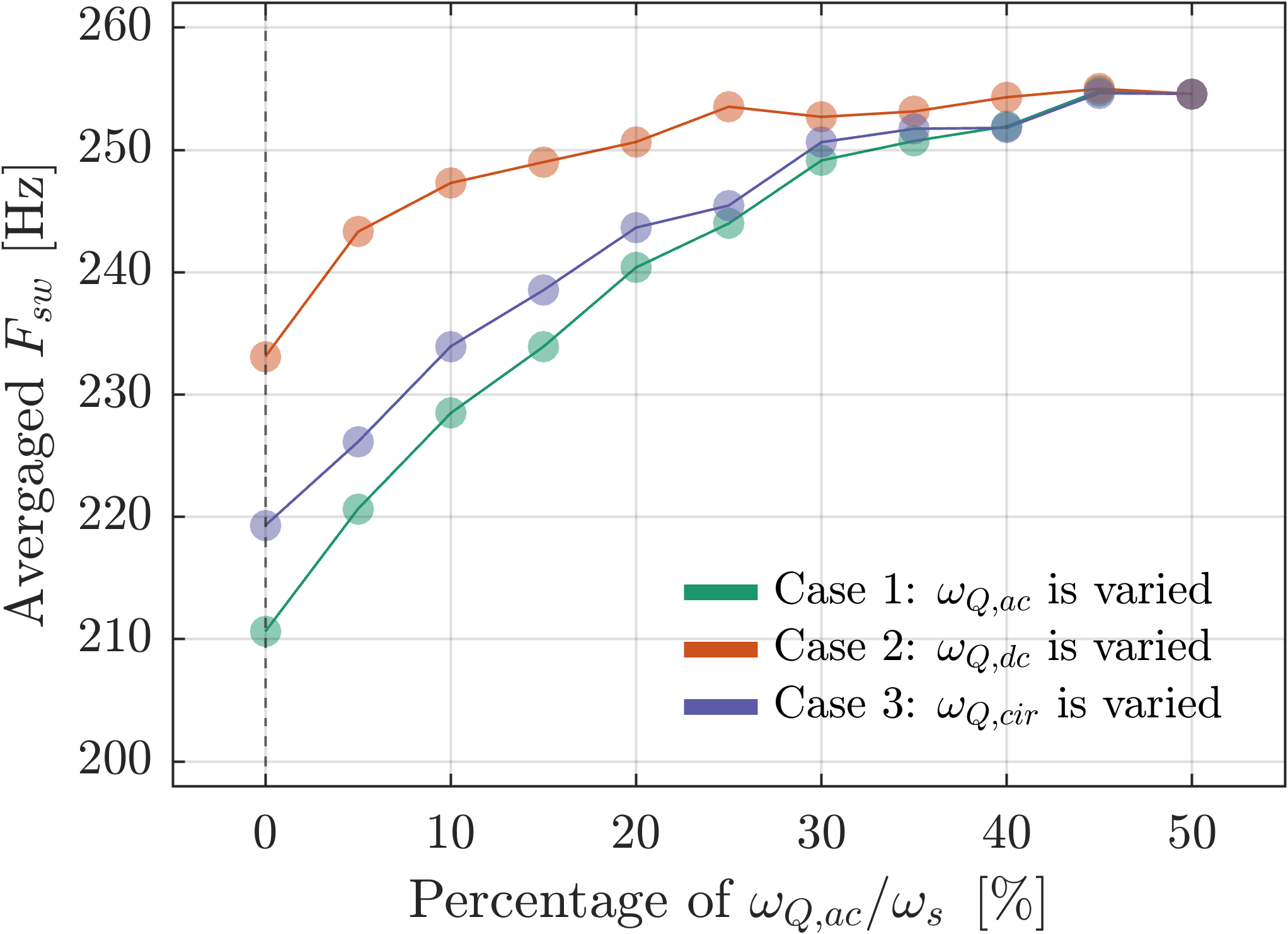}
        \caption{}
        \label{fig:dob_fsw_cutoff}
    \end{subfigure}
    \hfill
    \begin{subfigure}[b]{0.492\linewidth}
        \includegraphics[width=1.0\linewidth,center]{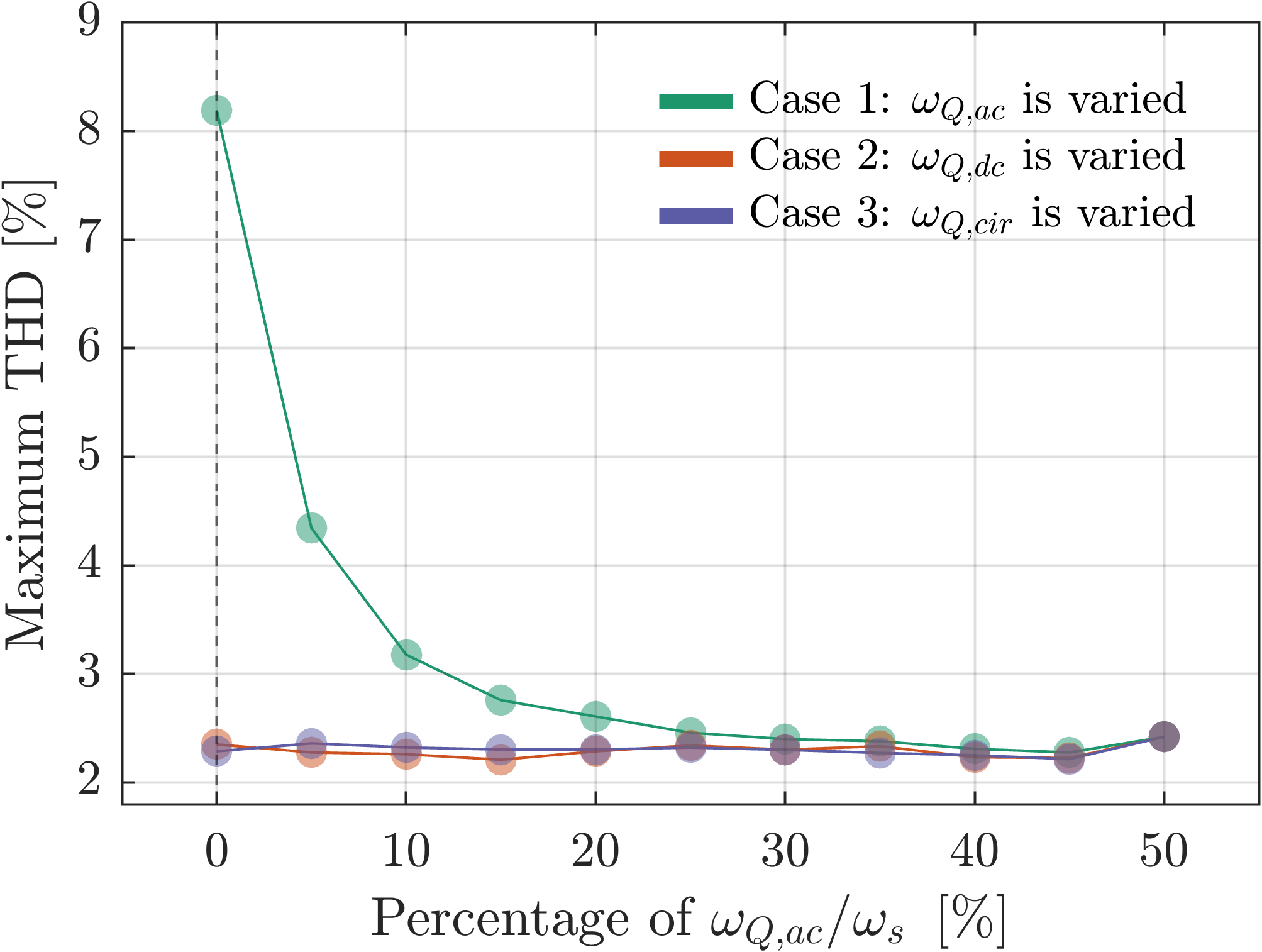}
        \caption{}
        \label{fig:dob_thd_cutoff}
    \end{subfigure}
    
    \caption{Simulation results showing the effects of the DOB cut-off frequency on 
    (a) SM-averaged switching frequency and
    (b) THD variation of the ac-side current.
    Case 1, 2, and 3 represent that only the cut-off frequency of the DOB for
    ac-side, dc-side, and circulating current is varied, respectively,
    while the cut-off frequencies of the other DOBs are fixed to half of the sampling frequency.
    $\omega_{s}$ represents the angular sampling frequency in rad/s.}
    \label{fig:dob_cutoff_effect}
   \vspace{1.0em}  
\end{figure}

Despite the benefits of the proposed DOB as aforementioned,
there exists a trade-off in selecting 
the cut-off frequency of the $Q$-filter.
As shown in \figurename{\ref{fig:mmc_dob_ac_enable_zoom}},
\ref{sub@fig:mmc_dob_dc_enable_zoom},
and \ref{sub@fig:mmc_dob_cir_enable_zoom}, 
enabling DOB causes the arm voltage 
to exhibit pseudo-PWM behavior, even under NLM operation.
This behavior increases the averaged switching frequency of SMs,
which can lead to higher switching losses.
This trade-off is numerically investigated in this section,
by varying the cut-off frequency of each DOB.
An analytical approach needs to be further studied.

\figurename{\ref{fig:dob_fsw_cutoff}} illustrates
the effects of the DOB cut-off frequency on
the SM-averaged switching frequency.
As the cut-off frequency of DOB increases,
the switching frequency of SMs also increases.
On the other hand,
THD of the ac-side is affected only by the cut-off frequency of the DOB for ac-side current,
as shown in \figurename{\ref{fig:dob_thd_cutoff}}.
Therefore, a practical tuning approach is to set a low cut-off frequency 
for the circulating current DOB$\,$—$\,$addressing switching loss 
concerns$\,$—$\,$while employing high cut-off frequencies for the ac- and dc-side current DOBs 
to ensure high-quality current for grid interaction. 
The waveforms in Fig.~\ref{fig:dob_capvoltage_effect_cir} for $t < 0\,$s 
exemplify this tuning approach. In this interval, 
the ac- and dc-side DOBs are active, while the circulating current DOB is disabled, 
representing the extreme case of setting its cut-off frequency to zero to minimize its impact on switching frequency.
This strategy is particularly effective 
when precise capacitor voltage balancing is not a primary requirement.


\begin{figure}[t!]
    \centering
    \includegraphics[width=1\linewidth]{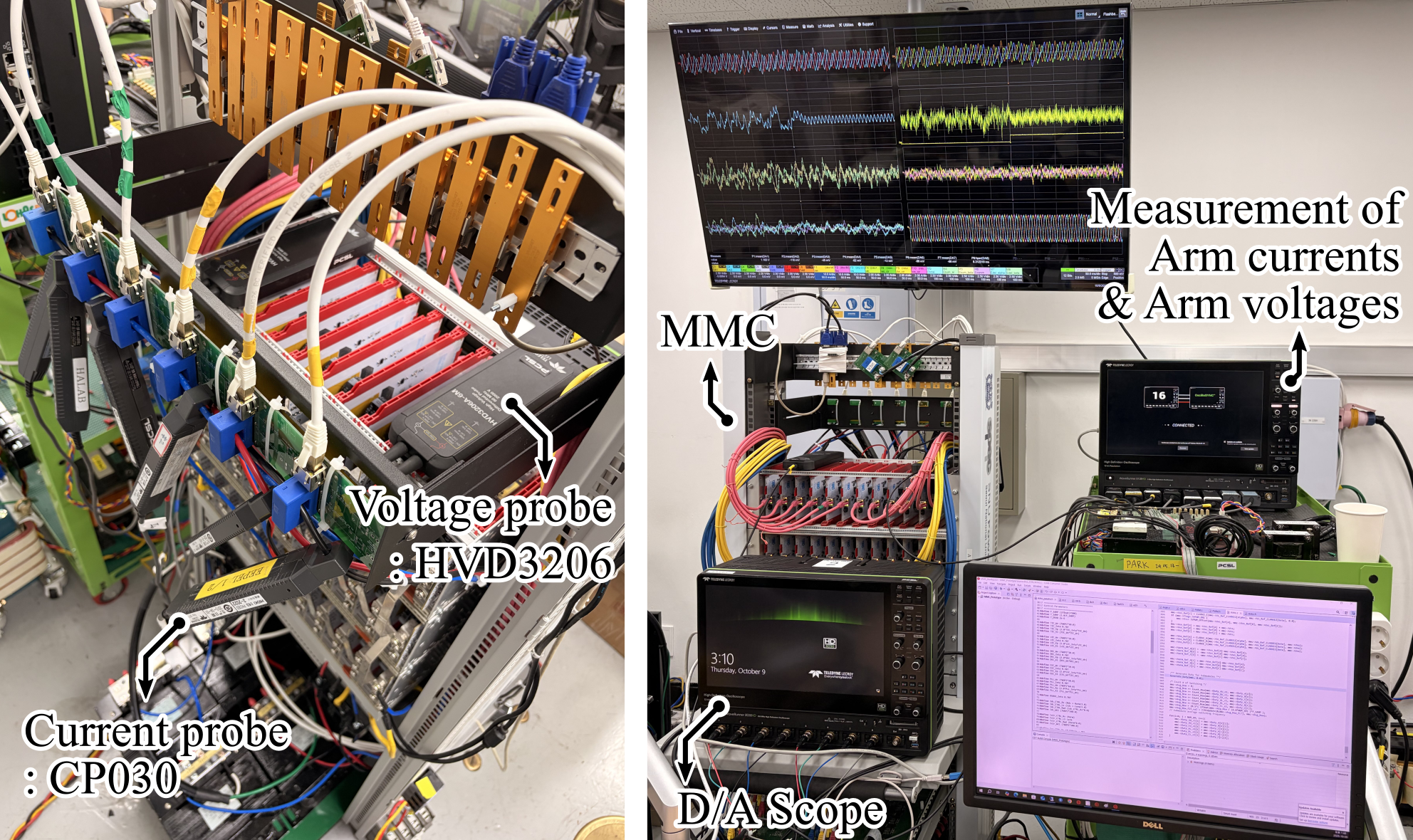}
    \caption{
        Experimental setup.
    }
    \label{fig:experimental_setup}
   \vspace{0.0em}  
\end{figure}

\begin{table}[t!]
        \begin{threeparttable}
            \caption{\textsc{System Parameters for Experiment}}
            \label{tab:params_sys_exp}
            \centering
            {\renewcommand{\arraystretch}{1.2}
        
            \begin{tabular*}{\columnwidth}{@{\extracolsep{\fill}}llll}
                \toprule
                \multicolumn{4}{c}{Modular Multilevel Converter} \\ 
                \midrule
                Rated power & 5 kW         & Number of SM per arm & 6 \\
                Rated dc voltage, $V_{dc}^*$     & 360 V         & SM nominal dc voltage  & 60 V\\
                Arm inductance, $L_o$       & 4 mH   & SM capacitance & 2.39 mF\\
                Arm resistance, $R_o$ & 23 m$\Omega$ & Sampling frequency & 7.2 kHz\\
                \toprule
                \multicolumn{4}{c}{AC Grid} \\ 
                \midrule
                Grid line-to-line voltage\tnote{*} & 200 V & Nominal grid frequency & 60 Hz\\
                \bottomrule
            \end{tabular*}%
            \scriptsize
            \begin{tablenotes}
            \item[*]{Voltage are expressed as its root-mean-square (RMS) value.}
            \end{tablenotes}
            \vspace{-0.0em}
        }
        \end{threeparttable}
        \vspace{-0em}
\end{table}

\section{Experimental Results} \label{sec:experiment}

\begin{figure*}[t]
    \vspace{-0.0em}  
    \centering
    \begin{subfigure}[b]{0.325\linewidth}
        \includegraphics[width=1.0\linewidth,center]{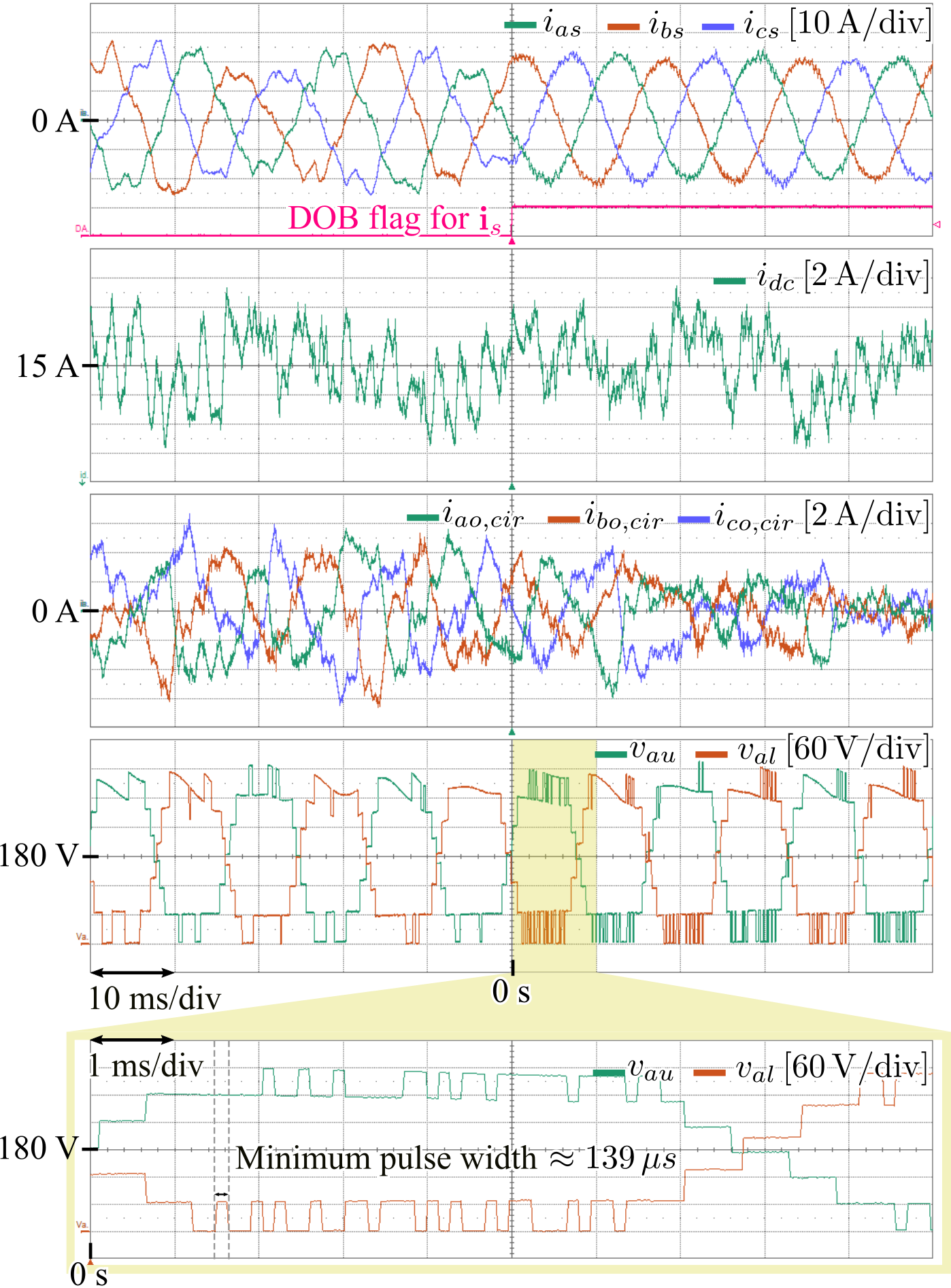}
        \caption{}
        \label{fig:exp_dob_ac_enable}
    \end{subfigure}
    \hfill
    \begin{subfigure}[b]{0.325\linewidth}
        \includegraphics[width=1.0\linewidth,center]{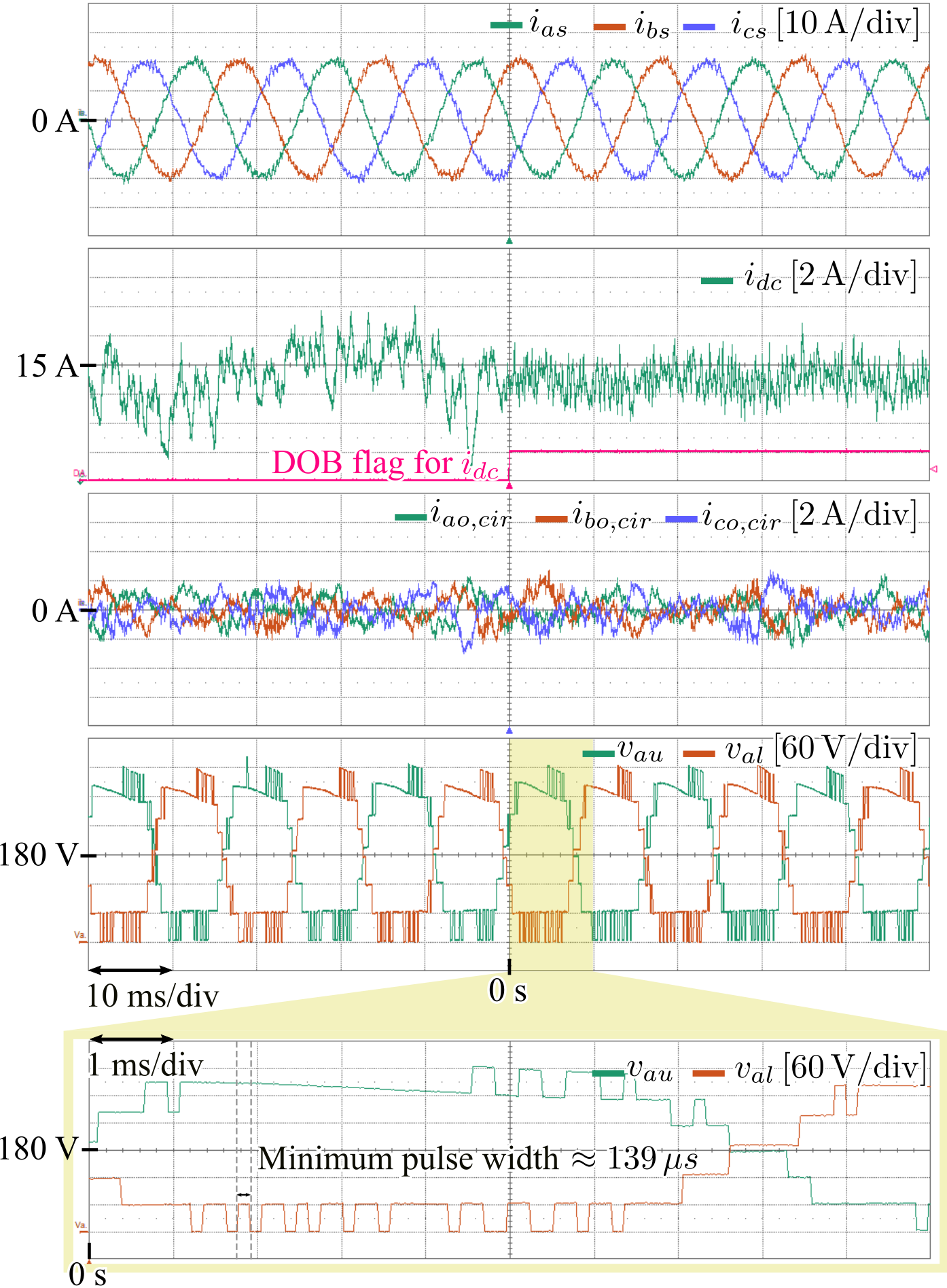}
        \caption{}
        \label{fig:exp_dob_dc_enable}
    \end{subfigure}
    \hfill
    \begin{subfigure}[b]{0.325\linewidth}
        \includegraphics[width=1.0\linewidth,center]{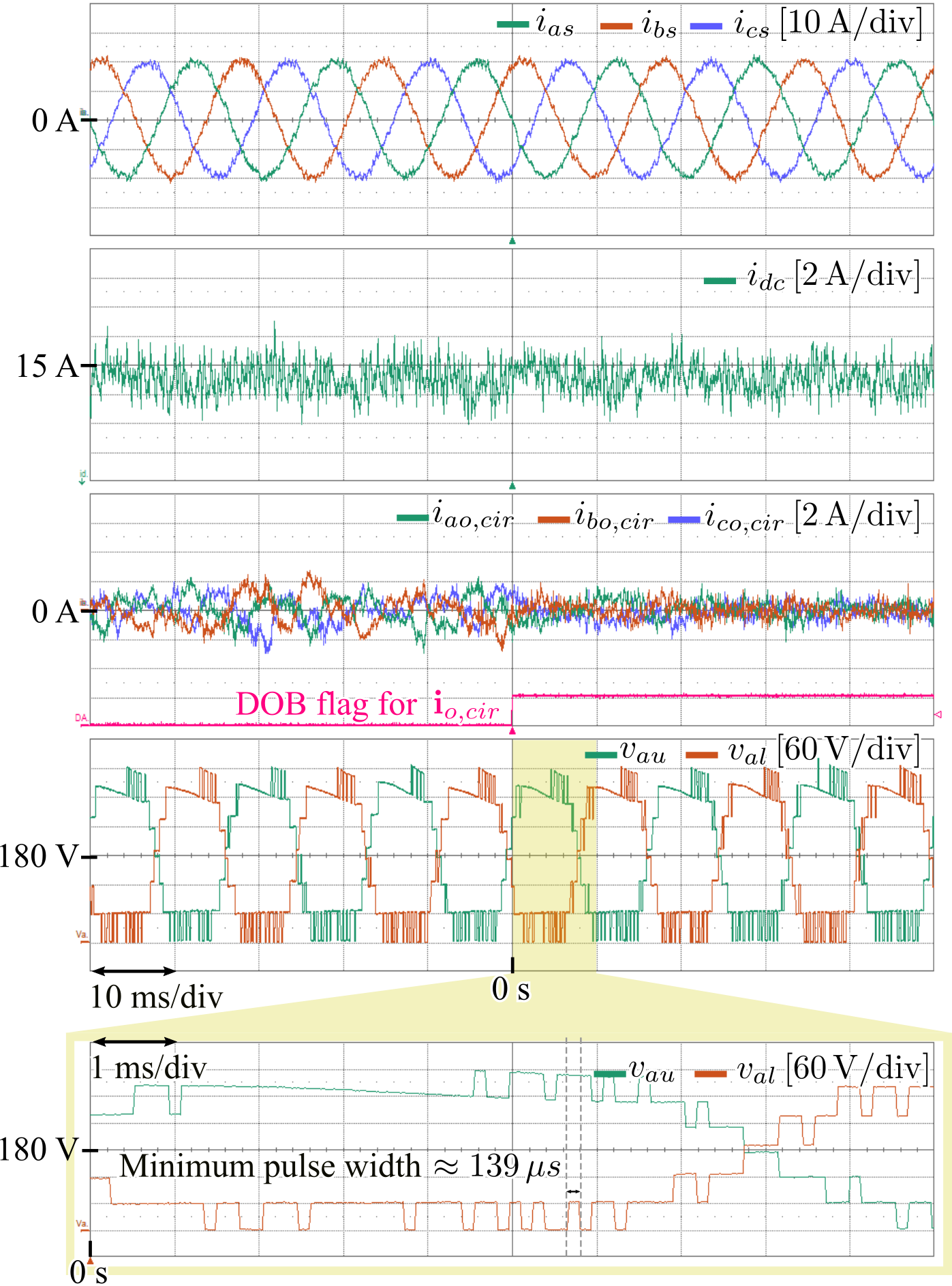}
        \caption{}
        \label{fig:exp_dob_cir_enable}
    \end{subfigure}

    \caption{
    Experimental results showing the
    effects of the DOB
    on (a) ac-side current, (b) dc-side current, and (c) circulating current.
    For (a) case, only the DOB for ac-side current is activated at $t=0\,$s.
    For (b) case, the DOB for dc-side current is activated at $t=0\,$s 
    while only the DOB for ac-side current is initially activated.
    For (c) case, the DOB for circulating current is activated at $t=0\,$s
    while the DOBs for ac-side and dc-side currents are initially activated.}
    \label{fig:exp_dob_enable}
   \vspace{0.0em}  
\end{figure*}

\begin{figure}[t!]
    \vspace{-0.0em}  
    \centering
    \begin{subfigure}[b]{0.493\linewidth}
        \includegraphics[width=1.0\linewidth,center]{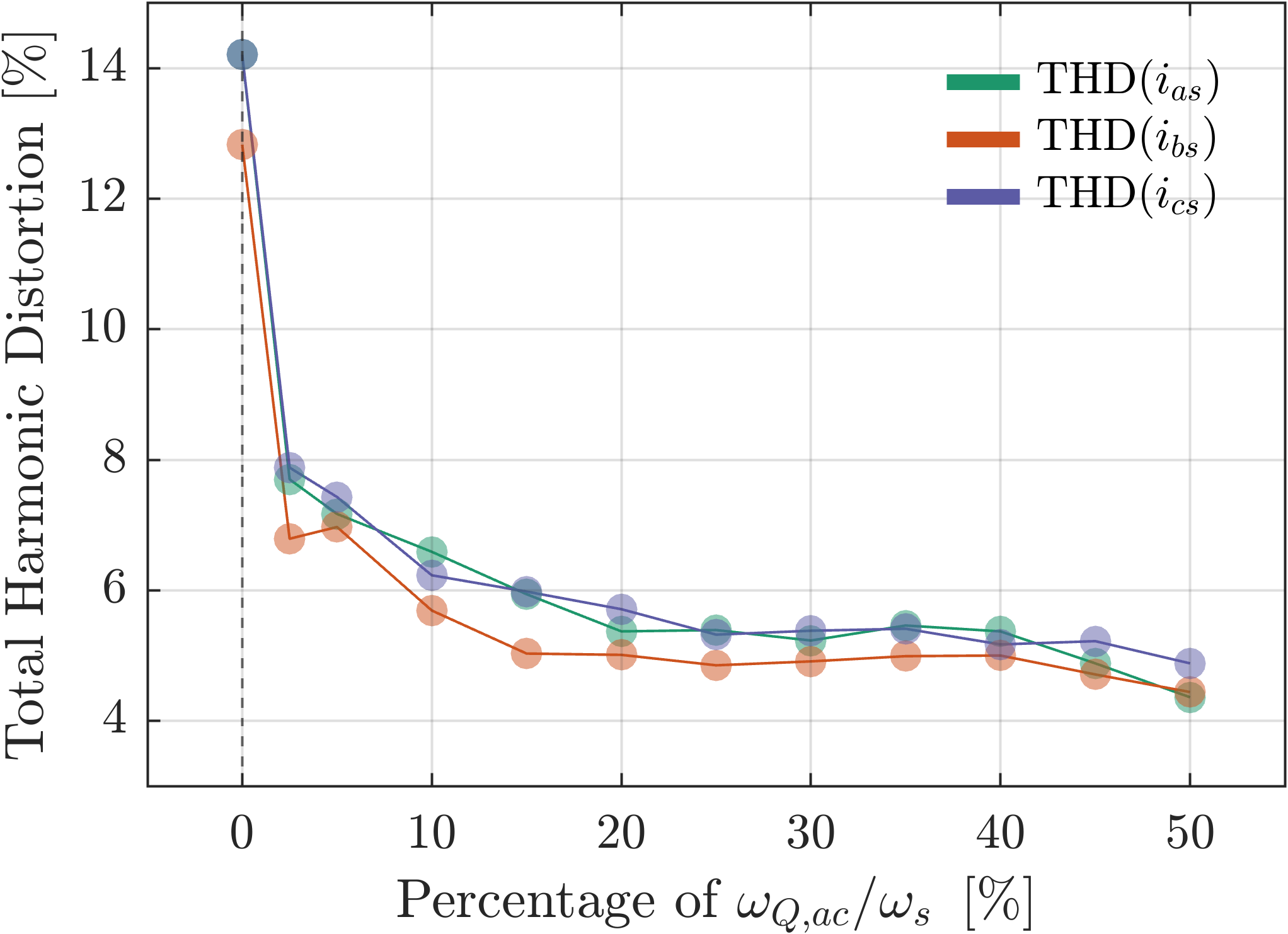}
        \caption{}
        \label{fig:exp_dob_thd}
    \end{subfigure}
    \hfill
    \begin{subfigure}[b]{0.493\linewidth}
        \includegraphics[width=1.0\linewidth,center]{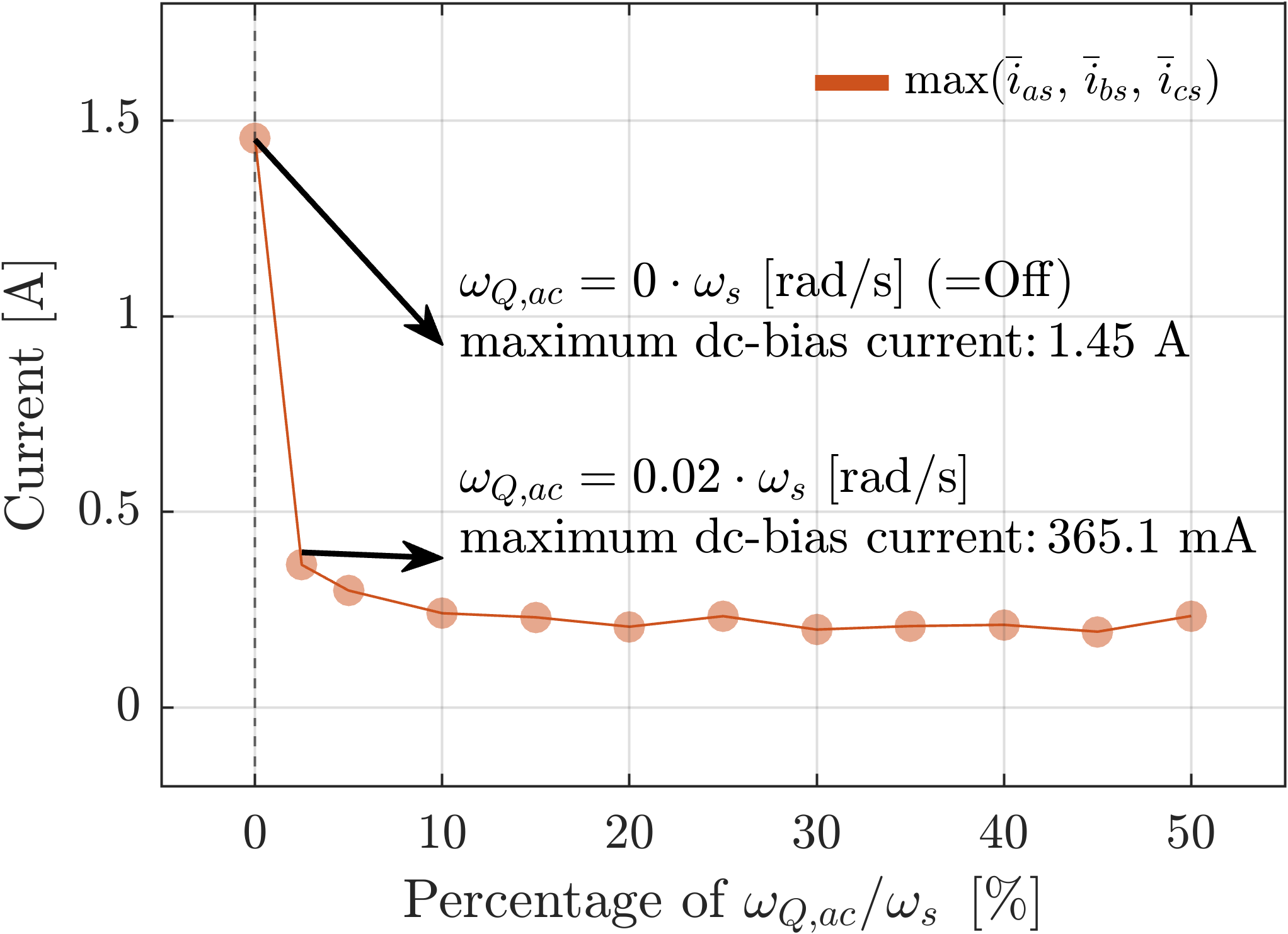}
        \caption{}
        \label{fig:exp_dob_dc_bias}
    \end{subfigure}
    \caption{
    Experimental results showing the effects of the ac-side DOB's cut-off frequency 
    on (a) total harmonic distortion and (b) dc bias of the ac-side current. 
    The cut-off frequency is normalized to the angular sampling frequency, $\omega_s$.
    $\bar{i}_{x}$ indicates the average value of $i_{x\in\{as,bs,cs\}}$ over a fundamental period.
    max$(\cdot)$ represents the maximum value among three phases.}
    \label{fig:exp_dob_ac_cutoff_effect}
   \vspace{-0.0em}  
\end{figure}

\begin{figure*}[t]
    \vspace{-0.0em}  
    \centering
    \begin{subfigure}[b]{0.48\linewidth}
        \includegraphics[width=1.0\linewidth,center]{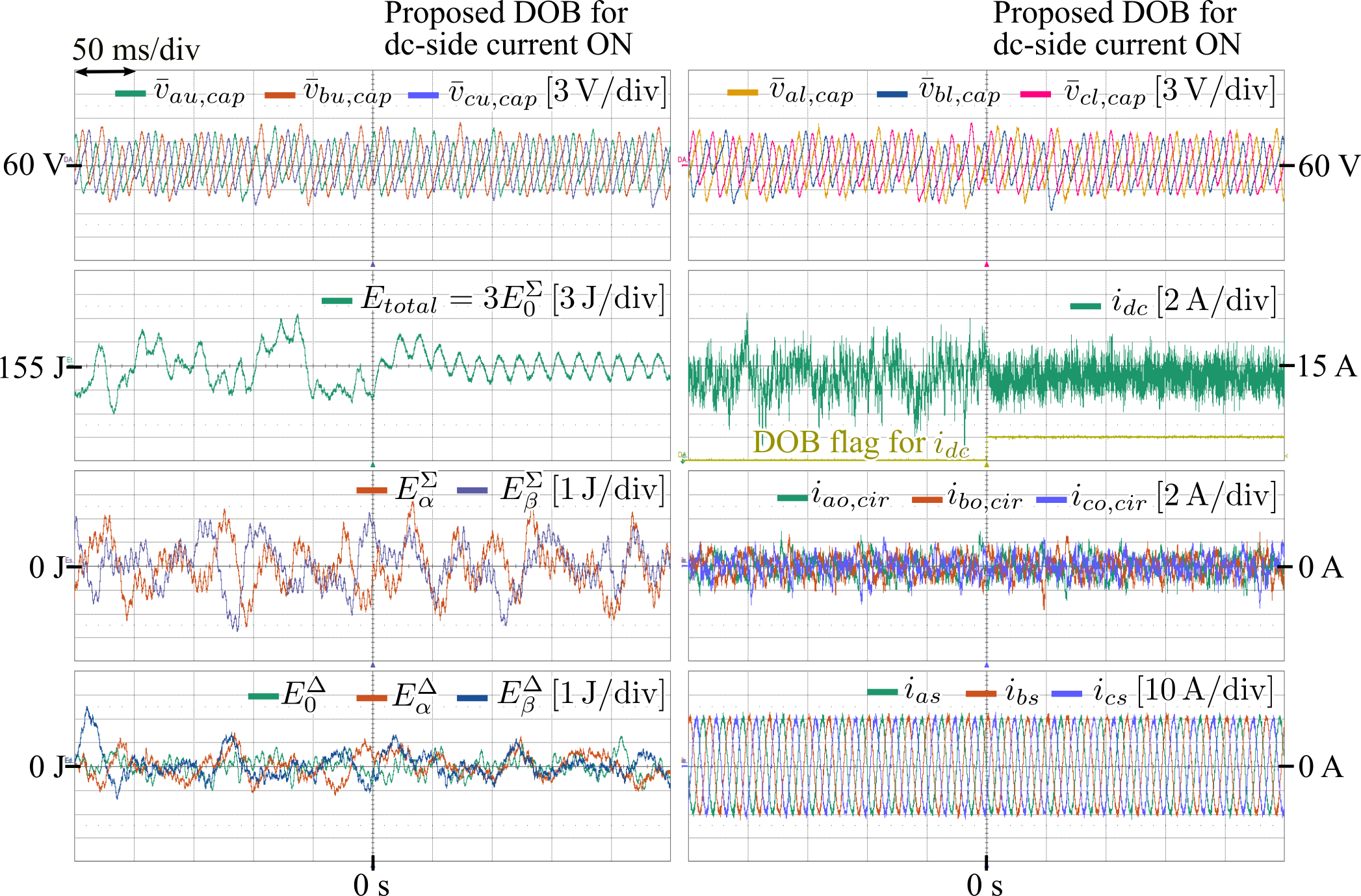}
        \caption{}
        \label{fig:exp_dob_dc_on_cap}
    \end{subfigure}
    \hfill
    \begin{subfigure}[b]{0.48\linewidth}
        \includegraphics[width=1.0\linewidth,center]{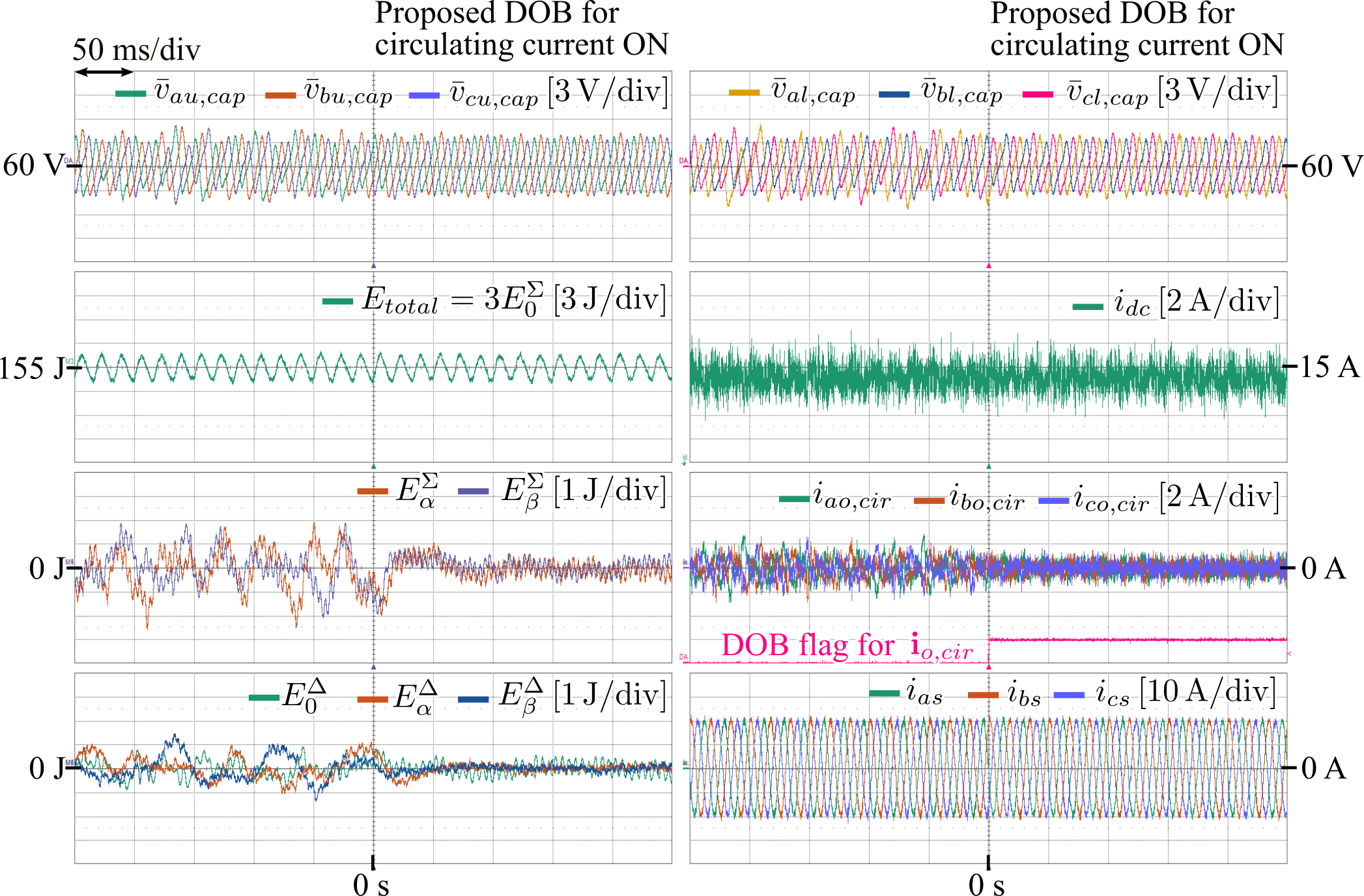}
        \caption{}
        \label{fig:exp_dob_cir_on_cap}
    \end{subfigure}
    \caption{
        Experimental results representing the effects of DOB on the 
        capacitor energy (voltage) balancing.
        (a) DOB for dc-side current is activated at $t=0\,$s
        while only the DOB for ac-side current is initially activated.
        (b) DOB for circulating current is activated at $t=0\,$s,
        while the DOBs for ac-side and dc-side currents are initially activated.
    }
    \label{fig:exp_dob_cap_effect}
   \vspace{0.0em}  
\end{figure*}

\begin{figure*}[t!]
    \vspace{-0.0em}  
    \centering
    \begin{subfigure}[b]{0.495\linewidth}
        \includegraphics[width=1.0\linewidth,center]{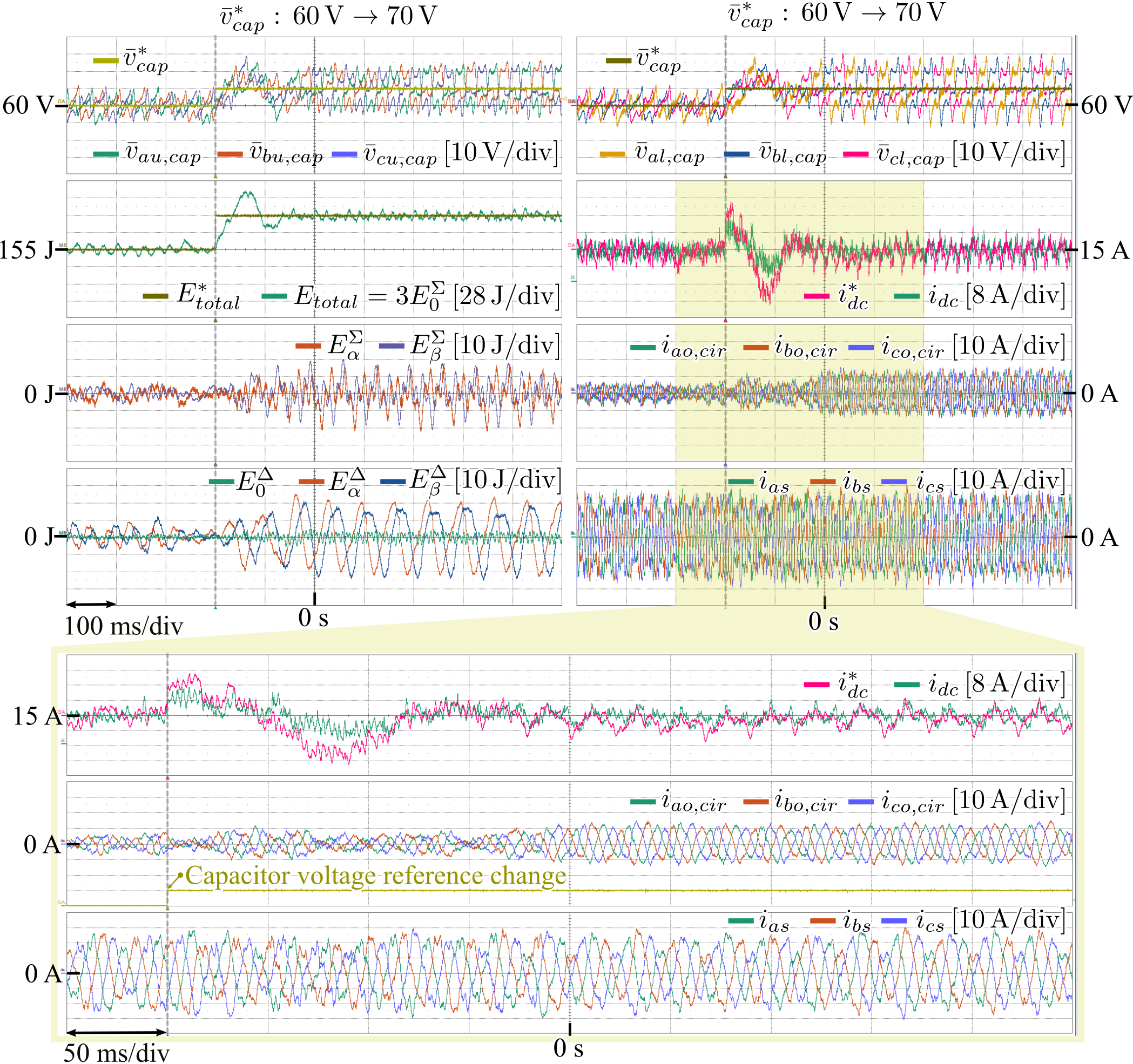}
        \caption{}
        \label{fig:exp_vcap_change_nlm}
    \end{subfigure}
    \hfill
    \begin{subfigure}[b]{0.495\linewidth}
        \includegraphics[width=1.0\linewidth,center]{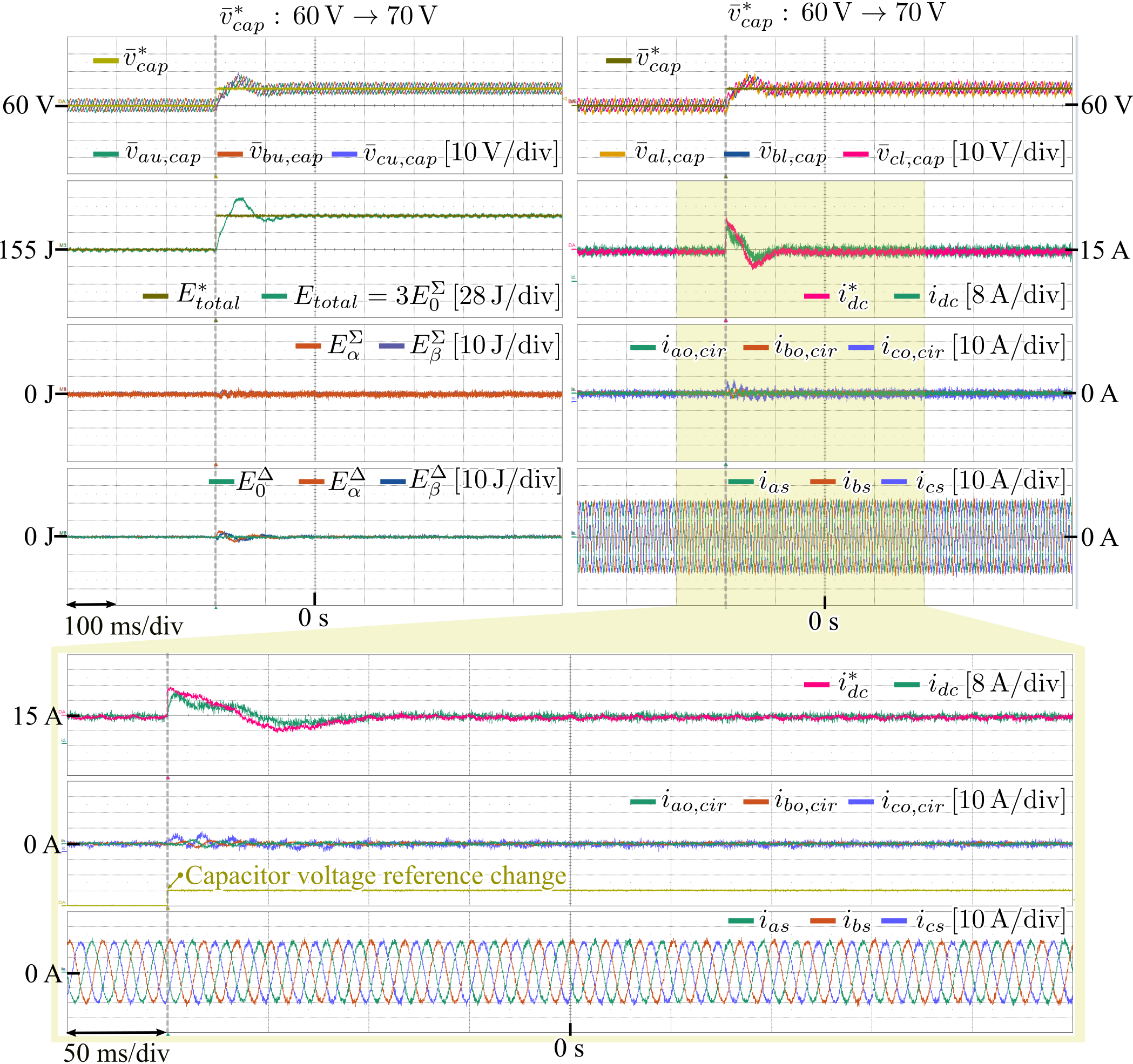}
        \caption{}
        \label{fig:exp_vcap_change_dob}
    \end{subfigure}

    \caption{
    Experimental results showing the
    effects of the DOB on decoupled capacitor energy control.
    The capacitor voltage reference is changed from 60 V to 70 V at $t=-0.2$ s
    (a) without DOB and (b) with DOB for each decoupled current control.
    The superscript `*' indicates the reference value.}
    \label{fig:exp_vcap_change}
   \vspace{0.0em}  
\end{figure*}

To validate the effectiveness and feasibility of the proposed DOB design,
a small-scale hardware experimental setup is implemented,
which is shown in \figurename{\ref{fig:experimental_setup}}.
The system parameters for experiment are summarized in
\tablename{\ref{tab:params_sys_exp}}. Control parameters 
are the same as those in the simulation, as shown in \tablename{\ref{tab:params_ctrl}}.
The six arm currents are measured using 
CP030\textsuperscript{\textregistered} current probes, 
while the $a$-phase upper and lower arm voltages are measured with
HVD3206\textsuperscript{\textregistered} voltage probes. 
All signals are sampled at 5 MS/s.
Control variables, such as 
capacitor energy, are captured through 8-channel digital-to-analog converter.
Two 8-channel oscilloscopes are used and synchronized via OscilloSYNC\textsuperscript{\texttrademark} 
to simultaneously capture all signals.

\subsection{Improvement of Decoupled Current Control} \label{sec:exp_improvement_of_decoupled_current_control}

\figurename{\ref{fig:exp_dob_enable}} 
shows the effects of the proposed DOB on the
decoupled current control in experiment.
The cut-off frequency of the $Q$-filter for each DOB
is set to half of the sampling frequency, $2\pi \times 3600\,$ rad/sec.

In \figurename{\ref{fig:exp_dob_ac_enable}},
the DOB for ac-side current is activated
while the DOBs for dc-side and circulating currents are deactivated.
As a result, the ac-side current restores three-phase balanced sinusoidal waveforms,
while the dc-side current is left unaffected.
    As aforementioned in Section~\ref{sec:sims_improvement_of_decoupled_current_control},
    it is observed that
    irregular fluctuation in circulating current $\mathbf{i}_{o,cir}$ is reduced
    to some extent,
    because the ac-side output power is stably regulated
    by the proposed DOB for ac-side current for $t>0\,$s.
    However, since 
    the DOB for circulating current is not activated,
    its improvement is limited 
    compared to the cases when the DOB for circulating current is activated,
    which is shown in \figurename{\ref{fig:exp_dob_cir_enable}}.

\figurename{\ref{fig:exp_dob_dc_enable}} shows the results when
the DOB for dc-side current is activated
while the DOB for ac-side current is initially activated.
Similarly, the DOB for dc-side current effectively suppresses the current distortion in $i_{dc}$
without affecting the circulating current.
\figurename{\ref{fig:exp_dob_cir_enable}} shows the results when
the DOB for circulating current is activated
while the DOBs for ac-side and dc-side currents are initially activated.
As a result, the irregular fluctuation in $\mathbf{i}_{o,cir}$ is effectively regulated.
As seen in the magnified arm voltage waveform,
the minimum pulse width is observed as one sampling period due to NLM operation.
Notably, these results in \figurename{\ref{fig:exp_dob_enable}}
represent that the proposed DOB can be individually
applied to each decoupled current control
to improve the disturbance rejection capability from NLM operation.

\figurename{\ref{fig:exp_dob_ac_cutoff_effect}} shows the relationship 
between the cut-off frequency of DOB for the ac-side current and the quality of the ac-side current, specifically its THD and dc bias.
Similar to the simulation results depicted in \figurename{\ref{fig:dob_ac_cutoff_effect}},
the THD and dc bias current decrease as the cut-off frequency increases.

\subsection{Improvement of Capacitor Voltage Balancing} \label{sec:exp_improvement_of_capacitor_voltage_balancing}

\figurename{\ref{fig:exp_dob_cap_effect}} shows the effects of the proposed DOB
on the capacitor voltage balancing in experiment.
Differential components of $\mathbf{E}_{abc}^\Sigma$
are represented as $\mathbf{E}_{\alpha\beta}^\Sigma=[E_{\alpha}^\Sigma\quad E_{\beta}^\Sigma]^\top$in the $\alpha\beta$ frame,
while those of $\mathbf{E}_{abc}^\Delta$ are represented as 
$\mathbf{E}_{\alpha\beta}^\Delta=[E_{\alpha}^\Delta\quad E_{\beta}^\Delta]^\top$
in the $\alpha\beta$-frame.
Common components of $\mathbf{E}_{abc}^\Sigma$ and $\mathbf{E}_{abc}^\Delta$
are corresponded as $E_{0}^\Sigma$ and $E_{0}^\Delta$, respectively.

In \figurename{\ref{fig:exp_dob_dc_on_cap}},
the DOB for dc-side current is activated at $t=0\,$s
while only the DOB for ac-side current is initially activated.
As a result,
the total capacitor energy, $E_{total}$, is well regulated to its rated value
by improving the dc-side current quality via the proposed DOB.
However, $\mathbf{E}_{\alpha\beta}^\Sigma$ and $\mathbf{E}_{\alpha\beta 0}^\Delta$
are not improved as shown in the third and fourth plots of \figurename{\ref{fig:exp_dob_dc_on_cap}},
because the NLM-induced disturbance for circulating current is still not addressed.

In \figurename{\ref{fig:exp_dob_cir_on_cap}},
the DOB for circulating current is activated at $t=0\,$s,
while the DOBs for ac-side and dc-side currents are initially activated.
As a result, not only $E_{total}$ but also
$\mathbf{E}_{\alpha\beta}^\Sigma$ and $\mathbf{E}_{\alpha\beta 0}^\Delta$
are well regulated to zero.
Consistent with the simulation results, the improved circulating current quality directly contributes to effective capacitor energy balancing.
As shown in the first plot of \figurename{\ref{fig:exp_dob_cir_on_cap}}, 
the capacitor voltage balancing is enhanced.

\subsection{Effectiveness in Decoupled Capacitor Energy Control} \label{sec:applicability_in_decoupled_control}

\figurename{\ref{fig:exp_vcap_change}} shows the effects of the proposed DOB
on the decoupled capacitor energy control in experiment.
Since the indirect-modulated MMC can offer
a control capability for SM capacitor energy decoupled from the ac-side current control
as well as the dc-link voltage,
the SM capacitor voltage reference is boosted up from 60 V to 70 V at $t=-0.2$ s
by drawing dc-link current instantaneously,
while maintaining the ac-side 
active power and the dc-link voltage at 360$\,$V unchanged.

In \figurename{\ref{fig:exp_vcap_change_nlm}},
the capacitor voltage reference is changed without the proposed DOB.
As a result, the capacitor energy components, 
$\mathbf{E}_{\alpha\beta}^\Sigma$ and $\mathbf{E}_{\alpha\beta 0}^\Delta$, 
are not regulated to zero, causing large fluctuations in the capacitor voltages. 
These voltage fluctuations exacerbate the NLM-induced arm voltage synthesis error. 
Consequently, the ac-side current quality is significantly degraded, 
and a large circulating current is induced.

In \figurename{\ref{fig:exp_vcap_change_dob}},
the capacitor voltage reference is changed with the proposed DOB for each decoupled current control all activated.
The cut-off bandwidth of $Q$-filter for each DOB is set to half of the sampling frequency.
As a result, the capacitor energy components, 
$\mathbf{E}_{\alpha\beta}^\Sigma$ and $\mathbf{E}_{\alpha\beta0}^\Delta$,
are clearly regulated to zero.
During the transient period following the reference change, 
a slight deviation is observed in $\mathbf{E}_{\alpha\beta0}^\Delta$. 
However, the system promptly restores the balance by temporarily injecting a corresponding circulating current.
Once the capacitor energy reaches a balanced state at the new reference, 
the circulating current is regulated back to zero, which is the nature of indirect modulation based control.
These results demonstrate that the proposed DOB enables 
decoupled control of capacitor energy for an indirect-modulated MMC 
with a low number of SMs, even under the constraint of NLM operation.

\subsection{Trade-off in DOB Cut-off Frequency} \label{sec:exp_trade_off_dob}

\begin{figure}[t]
    \vspace{-0.0em}  
    \centering
    \begin{subfigure}[b]{0.493\linewidth}
        \includegraphics[width=1.0\linewidth,center]{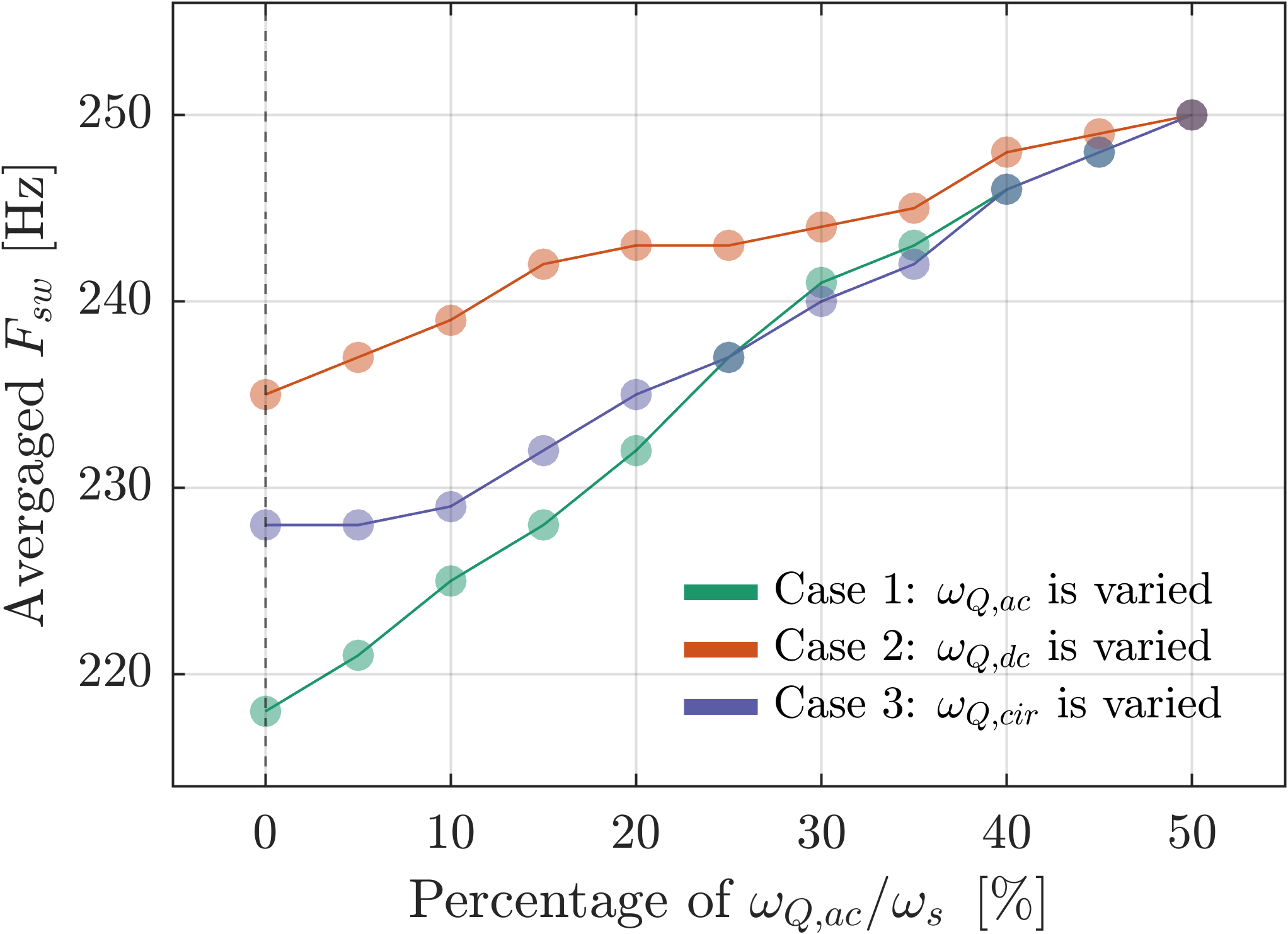}
        \caption{}
        \label{fig:exp_dob_fsw_cutoff}
    \end{subfigure}
    \hfill
    \begin{subfigure}[b]{0.492\linewidth}
        \includegraphics[width=1.0\linewidth,center]{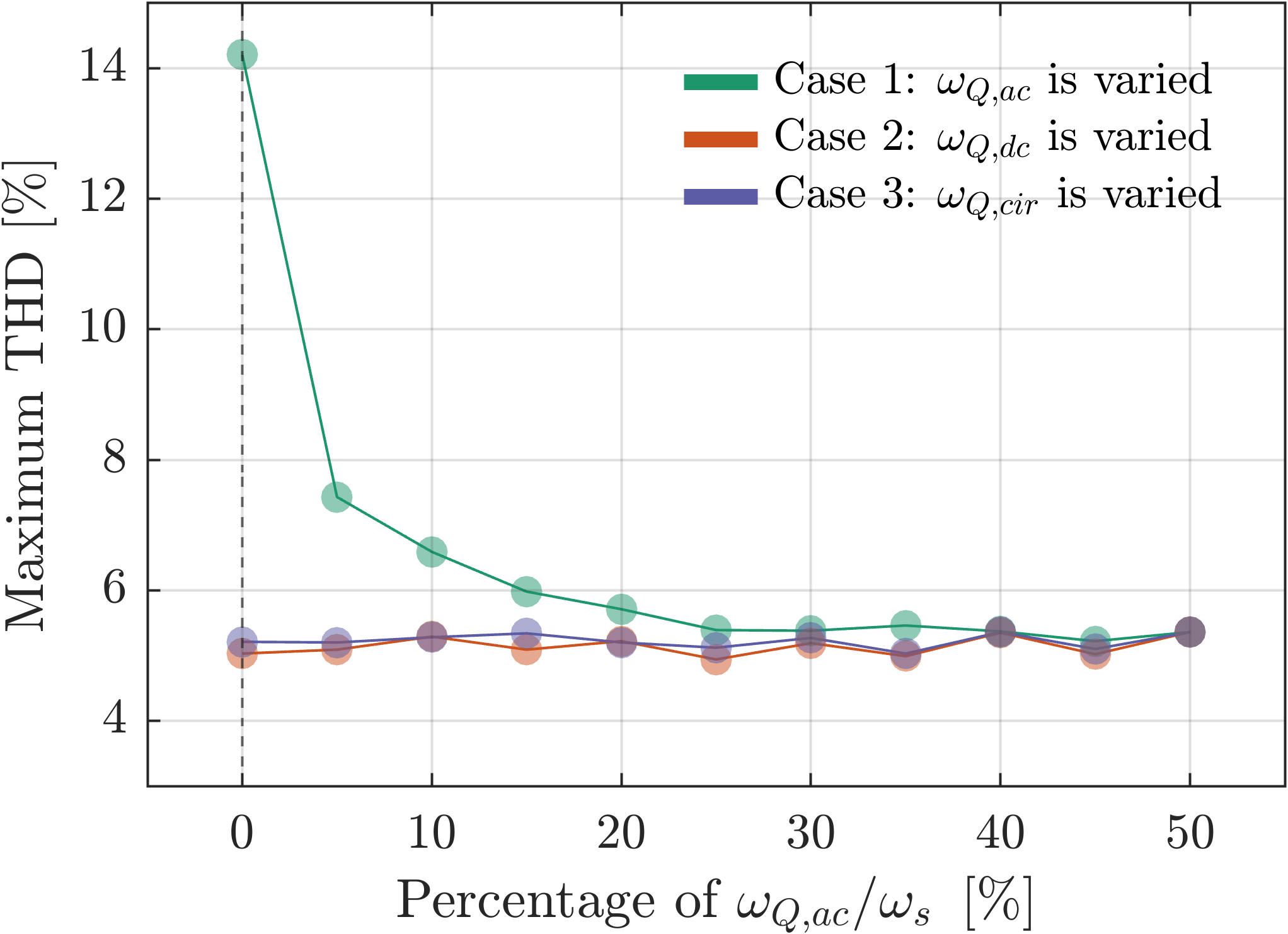}
        \caption{}
        \label{fig:exp_dob_thd_cutoff}
    \end{subfigure}
    
    \caption{Experimental results showing the effects of the DOB cut-off frequency on 
    (a) SM-averaged switching frequency and
    (b) THD variation of the ac-side current.
    Case 1, 2, and 3 represent that only the cut-off frequency of the DOB for
    ac-side, dc-side, and circulating current is varied, respectively,
    while the cut-off frequencies of the other DOBs are fixed to half of the sampling frequency.}
    \label{fig:exp_dob_cutoff_effect}
   \vspace{0.0em}  
\end{figure}

As discussed in Section~\ref{sec:sims_trade_off_dob}, 
a trade-off exists between the disturbance rejection performance of the DOB 
and the resulting switching frequency. 

This relationship is experimentally investigated in \figurename{\ref{fig:exp_dob_cutoff_effect}}, 
which illustrates the SM-averaged switching frequency and the ac-side current THD as the cut-off frequency of each DOB is individually varied.
\figurename{\ref{fig:exp_dob_fsw_cutoff}} confirms 
that increasing the cut-off frequency of any DOB leads to a higher average switching frequency. 
This is because, in principle, a wider DOB bandwidth 
allows for the compensation of higher-frequency disturbances, 
which in turn requires more frequent switching actions.

\figurename{\ref{fig:exp_dob_thd_cutoff}} shows that 
the THD of the ac-side current is predominantly influenced by its own DOB (Case 1). 
Increasing the cut-off frequency for the ac-side DOB significantly improves the current quality by reducing the THD. 
In contrast, varying the cut-off frequencies of DOBs for the dc-side and circulating currents 
has a negligible effect on the ac-side current THD.
These experimental findings are consistent with the simulation results.

\section{Conclusion}\label{sec:conclusion}
In this paper,
a new control method based on DOB
is proposed to enable fully decoupled control of
indirect-modulated MMC under the constraint of NLM operation with a small number of SMs per arm.
The proposed method estimates the arm voltage synthesis error inherent in NLM and 
compensates for it as a disturbance in the ac-side, dc-side, and circulating current control loops.
A key advantage of the proposed method is its ease of implementation,
as it requires only LPF for the proposed method 
and no modifications to the conventional NLM and decoupled control structure.
Furthermore, since the proposed method is designed to 
selectively improve the ac-side, dc-side, and circulating current dynamics,
it can offer flexible performance tuning without interference among the respective currents.
The validity and effectiveness of the proposed method
in improving current quality and decoupled SM energy control
are verified through both simulation and experimental results. 

%
\IEEEpeerreviewmaketitle

\ifCLASSOPTIONcaptionsoff
  \newpage
\fi



\bibliographystyle{IEEEtran}
\bibliography{References.bib}

@inbook{Shim2020disturbance,
   title={Disturbance Observers},
   ISBN={9781447151029},
   url={http://dx.doi.org/10.1007/978-1-4471-5102-9_100068-1},
   DOI={10.1007/978-1-4471-5102-9_100068-1},
   booktitle={Encyclopedia of Systems and Control},
   publisher={Springer London},
   author={Shim, Hyungbo},
   year={2020},
   pages={1–8} }

@article{Shim2009almost,
title = {An almost necessary and sufficient condition for robust stability of closed-loop systems with disturbance observer},
journal = {Automatica},
volume = {45},
number = {1},
pages = {296-299},
year = {2009},
issn = {0005-1098},
doi = {https://doi.org/10.1016/j.automatica.2008.10.009},
url = {https://www.sciencedirect.com/science/article/pii/S0005109808003749},
author = {Hyungbo Shim and Nam H. Jo}
}

@INPROCEEDINGS{Perez2011generalized,
  author={Pérez, Marcelo A. and Rodríguez, José},
  booktitle={2011 IEEE International Symposium on Industrial Electronics}, 
  title={Generalized modeling and simulation of a modular multilevel converter}, 
  year={2011},
  volume={},
  number={},
  pages={1863-1868},
  doi={10.1109/ISIE.2011.5984441}}

@ARTICLE{Buticchi2013detection,
  author={Buticchi, Giampaolo and Lorenzani, Emilio},
  journal={IEEE Transactions on Industrial Electronics}, 
  title={Detection Method of the DC Bias in Distribution Power Transformers}, 
  year={2013},
  volume={60},
  number={8},
  pages={3539-3549},
  doi={10.1109/TIE.2012.2226418}}

@INPROCEEDINGS{Lizana2012capacitor,
  author={Lizana, Ricardo and Castillo, Cristian and Perez, Marcelo A. and Rodriguez, Jose},
  booktitle={IECON 2012 - 38th Annual Conference on IEEE Industrial Electronics Society}, 
  title={Capacitor voltage balance of MMC converters in bidirectional power flow operation}, 
  year={2012},
  volume={},
  number={},
  pages={4935-4940},
  doi={10.1109/IECON.2012.6389573}}

@ARTICLE{Perez2021modular,
  author={Perez, Marcelo A. and Ceballos, Salvador and Konstantinou, Georgios and Pou, Josep and Aguilera, Ricardo P.},
  journal={IEEE Open Journal of the Industrial Electronics Society}, 
  title={Modular Multilevel Converters: Recent Achievements and Challenges}, 
  year={2021},
  volume={2},
  number={},
  pages={224-239},
  doi={10.1109/OJIES.2021.3060791}}

@ARTICLE{Cui2019modular,
  author={Cui, Shenghui and Lee, Joon-Hee and Hu, Jingxin and De Doncker, Rik W. and Sul, Seung-Ki},
  journal={IEEE Transactions on Power Electronics}, 
  title={A Modular Multilevel Converter with a Zigzag Transformer for Bipolar MVDC Distribution Systems}, 
  year={2019},
  volume={34},
  number={2},
  pages={1038-1043},
  doi={10.1109/TPEL.2018.2855082}}

@ARTICLE{Cui2020fault,
  author={Cui, Shenghui and Hu, Jingxin and De Doncker, Rik},
  journal={IEEE Transactions on Power Electronics}, 
  title={Fault-Tolerant Operation of a TLC-MMC Hybrid DC-DC Converter for Interconnection of MVDC and HVdc Grids}, 
  year={2020},
  volume={35},
  number={1},
  pages={83-93},
  doi={10.1109/TPEL.2019.2911853}}

@INPROCEEDINGS{Cui2014comprehensive,
  author={Cui, Shenghui and Kim, Sungmin and Jung, Jae-Jung and Sul, Seung-Ki},
  booktitle={2014 IEEE Applied Power Electronics Conference and Exposition - APEC 2014}, 
  title={A comprehensive cell capacitor energy control strategy of a modular multilevel converter (MMC) without a stiff DC bus voltage source}, 
  year={2014},
  volume={},
  number={},
  pages={602-609},
  doi={10.1109/APEC.2014.6803370}}

@ARTICLE{Rohner2010modulation,
  author={Rohner, Steffen and Bernet, Steffen and Hiller, Marc and Sommer, Rainer},
  journal={IEEE Transactions on Industrial Electronics}, 
  title={Modulation, Losses, and Semiconductor Requirements of Modular Multilevel Converters}, 
  year={2010},
  volume={57},
  number={8},
  pages={2633-2642},
  doi={10.1109/TIE.2009.2031187}}

@ARTICLE{Sekiguchi2014grid,
  author={Sekiguchi, Kei and Khamphakdi, Pracha and Hagiwara, Makoto and Akagi, Hirofumi},
  journal={IEEE Transactions on Industry Applications}, 
  title={A Grid-Level High-Power BTB (Back-To-Back) System Using Modular Multilevel Cascade Converters Without Common DC-Link Capacitor}, 
  year={2014},
  volume={50},
  number={4},
  pages={2648-2659},
  doi={10.1109/TIA.2013.2290867}}

@ARTICLE{Leon2017energy,
  author={Leon, Andres E. and Amodeo, Santiago J.},
  journal={IEEE Transactions on Power Electronics}, 
  title={Energy Balancing Improvement of Modular Multilevel Converters Under Unbalanced Grid Conditions}, 
  year={2017},
  volume={32},
  number={8},
  pages={6628-6637},
  doi={10.1109/TPEL.2016.2621000}}

@ARTICLE{Cui2018comprehensive,
  author={Cui, Shenghui and Lee, Hak-Jun and Jung, Jae-Jung and Lee, Younggi and Sul, Seung-Ki},
  journal={IEEE Journal of Emerging and Selected Topics in Power Electronics}, 
  title={A Comprehensive AC-Side Single-Line-to-Ground Fault Ride Through Strategy of an MMC-Based HVDC System}, 
  year={2018},
  volume={6},
  number={3},
  pages={1021-1031},
  doi={10.1109/JESTPE.2018.2797934}}

@ARTICLE{Yi2018nearest,
  author={Wang, Yi and Hu, Can and Ding, Ruoyu and Xu, Lie and Fu, Chao and Yang, Erlin},
  journal={IEEE Transactions on Power Electronics}, 
  title={A Nearest Level PWM Method for the MMC in DC Distribution Grids}, 
  year={2018},
  volume={33},
  number={11},
  pages={9209-9218},
  doi={10.1109/TPEL.2018.2792148}}

@ARTICLE{Suman2015operation,
  author={Debnath, Suman and Qin, Jiangchao and Bahrani, Behrooz and Saeedifard, Maryam and Barbosa, Peter},
  journal={IEEE Transactions on Power Electronics}, 
  title={Operation, Control, and Applications of the Modular Multilevel Converter: A Review}, 
  year={2015},
  volume={30},
  number={1},
  pages={37-53},
  doi={10.1109/TPEL.2014.2309937}}

@ARTICLE{Pengfei2015level,
  author={Hu, Pengfei and Jiang, Daozhuo},
  journal={IEEE Transactions on Power Electronics}, 
  title={A Level-Increased Nearest Level Modulation Method for Modular Multilevel Converters}, 
  year={2015},
  volume={30},
  number={4},
  pages={1836-1842},
  doi={10.1109/TPEL.2014.2325875}}

@ARTICLE{Lin2016improved,
  author={Lin, Lei and Lin, Yizhe and He, Zhen and Chen, Yu and Hu, Jiabing and Li, Wuhua},
  journal={IEEE Transactions on Power Electronics}, 
  title={Improved Nearest-Level Modulation for a Modular Multilevel Converter With a Lower Submodule Number}, 
  year={2016},
  volume={31},
  number={8},
  pages={5369-5377},
  doi={10.1109/TPEL.2016.2521059}}

@ARTICLE{Meshram2015simplified,
  author={Meshram, Prafullachandra M. and Borghate, Vijay B.},
  journal={IEEE Transactions on Power Electronics}, 
  title={A Simplified Nearest Level Control (NLC) Voltage Balancing Method for Modular Multilevel Converter (MMC)}, 
  year={2015},
  volume={30},
  number={1},
  pages={450-462},
  doi={10.1109/TPEL.2014.2317705}}

@ARTICLE{Yin2021variable,
  author={Yin, Jiapeng and Leon, Jose I. and Perez, Marcelo A. and Franquelo, Leopoldo G. and Marquez, Abraham and Li, Binbin and Vazquez, Sergio},
  journal={IEEE Transactions on Power Electronics}, 
  title={Variable Rounding Level Control Method for Modular Multilevel Converters}, 
  year={2021},
  volume={36},
  number={4},
  pages={4791-4801},
  doi={10.1109/TPEL.2020.3020941}}

@ARTICLE{Yin2024improved,
  author={Yin, Jiapeng and Dai, NingYi and Vazquez, Sergio and Marquez, Abraham and Leon, Jose I. and Perez, Marcelo A. and Franquelo, Leopoldo G.},
  journal={IEEE Transactions on Power Electronics}, 
  title={An Improved Indirect Pulsewidth Modulation Technique for Modular Multilevel Converters}, 
  year={2024},
  volume={39},
  number={1},
  pages={733-743},
  doi={10.1109/TPEL.2023.3321759}}

@ARTICLE{ansari2020mmc,
  author={Ansari, Jamshed Ahmed and Liu, Chongru and Khan, Shahid Aziz},
  journal={IEEE Access}, 
  title={MMC Based MTDC Grids: A Detailed Review on Issues and Challenges for Operation, Control and Protection Schemes}, 
  year={2020},
  volume={8},
  number={},
  pages={168154-168165},
  doi={10.1109/ACCESS.2020.3023544}}

@ARTICLE{sun2022beyond,
  author={Sun, Pingyang and Tian, Yumeng and Pou, Josep and Konstantinou, Georgios},
  journal={IEEE Open Journal of Power Electronics}, 
  title={Beyond the MMC: Extended Modular Multilevel Converter Topologies and Applications}, 
  year={2022},
  volume={3},
  number={},
  pages={317-333},
  doi={10.1109/OJPEL.2022.3175714}}

@ARTICLE{Nguyen2019phase,
  author={Nguyen, Minh Hoang and Kwak, Sangshin and Kim, Taehyung},
  journal={IEEE Access}, 
  title={Phase-Shifted Carrier Pulse-Width Modulation Algorithm With Improved Dynamic Performance for Modular Multilevel Converters}, 
  year={2019},
  volume={7},
  number={},
  pages={170949-170960},
  doi={10.1109/ACCESS.2019.2955714}}

@ARTICLE{Nguyen2020nearest,
  author={Nguyen, Minh Hoang and Kwak, Sangshin},
  journal={IEEE Access}, 
  title={Nearest-Level Control Method With Improved Output Quality for Modular Multilevel Converters}, 
  year={2020},
  volume={8},
  number={},
  pages={110237-110250},
  doi={10.1109/ACCESS.2020.3001587}}

@ARTICLE{Kim2025new,
  author={Kim, Jae-Myeong and Cui, Shenghui and Jung, Jae-Jung Jung},
  journal={IEEE Transactions on Power Electronics}, 
  title={A New E-STATCOM Topology Based on Single-Star Multilevel Converter with Centralized Energy Storage and Zigzag Transformer}, 
  year={2025},
  volume={},
  number={},
  pages={1-17},
  doi={10.1109/TPEL.2025.3618963}}

@INPROCEEDINGS{wei2014research,
  author={Zhang, Wei and Gao, Qiang and Su, Bonan and Jin, Miaoxin and Xu, Dianguo and Liu, Jianyu},
  booktitle={2014 International Power Electronics Conference (IPEC-Hiroshima 2014 - ECCE ASIA)}, 
  title={Research on the control strategy of STATCOM based on modular multilevel converter}, 
  year={2014},
  volume={},
  number={},
  pages={614-618},
  doi={10.1109/IPEC.2014.6869649}}

@ARTICLE{makato2010medium,
  author={Hagiwara, Makoto and Nishimura, Kazutoshi and Akagi, Hirofumi},
  journal={IEEE Transactions on Power Electronics}, 
  title={A Medium-Voltage Motor Drive With a Modular Multilevel PWM Inverter}, 
  year={2010},
  volume={25},
  number={7},
  pages={1786-1799},
  doi={10.1109/TPEL.2010.2042303}}

@ARTICLE{Jung2015control,
  author={Jung, Jae-Jung and Lee, Hak-Jun and Sul, Seung-Ki},
  journal={IEEE Journal of Emerging and Selected Topics in Power Electronics}, 
  title={Control Strategy for Improved Dynamic Performance of Variable-Speed Drives With Modular Multilevel Converter}, 
  year={2015},
  volume={3},
  number={2},
  pages={371-380},
  doi={10.1109/JESTPE.2014.2323955}}

@INPROCEEDINGS{antonopoulos2009dynamics,
  author={Antonopoulos, Antonios and Angquist, Lennart and Nee, Hans-Peter},
  booktitle={2009 13th European Conference on Power Electronics and Applications}, 
  title={On dynamics and voltage control of the Modular Multilevel Converter}, 
  year={2009},
  volume={},
  number={},
  pages={1-10},
  doi={}}

@INPROCEEDINGS{Tu2010parameter,
  author={Qingrui Tu and Zheng Xu and Huang, Hongyang and Jing Zhang},
  booktitle={2010 International Conference on Power System Technology}, 
  title={Parameter design principle of the arm inductor in modular multilevel converter based HVDC}, 
  year={2010},
  volume={},
  number={},
  pages={1-6},
  doi={10.1109/POWERCON.2010.5666416}}

@inproceedings{Jacobson2010VSCHVDC,
  title={VSC-HVDC Transmission with Cascaded Two-Level Converters},
  author={Bjorn Jacobson and P. Karlsson and Gunnar Asplund and Lennart Harnefors and Tomas U. Jonsson},
  year={2010},
  url={https://api.semanticscholar.org/CorpusID:114708164}
}

@ARTICLE{Tu2011reduced,
  author={Tu, Qingrui and Xu, Zheng and Xu, Lie},
  journal={IEEE Transactions on Power Delivery}, 
  title={Reduced Switching-Frequency Modulation and Circulating Current Suppression for Modular Multilevel Converters}, 
  year={2011},
  volume={26},
  number={3},
  pages={2009-2017},
  doi={10.1109/TPWRD.2011.2115258}}

@ARTICLE{Tu2012suppressing,
  author={Tu, Qingrui and Xu, Zheng and Chang, Yong and Guan, Li},
  journal={IEEE Transactions on Power Delivery}, 
  title={Suppressing DC Voltage Ripples of MMC-HVDC Under Unbalanced Grid Conditions}, 
  year={2012},
  volume={27},
  number={3},
  pages={1332-1338},
  doi={10.1109/TPWRD.2012.2196804}}

@INPROCEEDINGS{Heng22018ac, 
  author={Wu, Heng and Wang, Xiongfei and Kocewiak, Lukasz and Harnefors, Lennart},
  booktitle={2018 IEEE 19th Workshop on Control and Modeling for Power Electronics (COMPEL)}, 
  title={AC Impedance Modeling of Modular Multilevel Converters and Two-Level Voltage-Source Converters: Similarities and Differences}, 
  year={2018},
  volume={},
  number={},
  pages={1-8},
  doi={10.1109/COMPEL.2018.8459952}}

@ARTICLE{Luca2019method,
  author={Bessegato, Luca and Harnefors, Lennart and Ilves, Kalle and Norrga, Staffan},
  journal={IEEE Transactions on Power Electronics}, 
  title={A Method for the Calculation of the AC-Side Admittance of a Modular Multilevel Converter}, 
  year={2019},
  volume={34},
  number={5},
  pages={4161-4172},
  doi={10.1109/TPEL.2018.2862254}}

@article{Lyu2017subsynchronous,
author = {Lyu, Jing and Cai, Xu and Amin, Mohammad and Molinas, Marta},
year = {2017},
month = {10},
pages = {},
title = {Subsynchronous Oscillation Mechanism and Its Suppression in MMC-Based HVDC Connected Wind Farms},
volume = {12},
journal = {IET Generation, Transmission \& Distribution},
doi = {10.1049/iet-gtd.2017.1066}
}

@ARTICLE{Lyu2018optimal,
  author={Lyu, Jing and Cai, Xu and Molinas, Marta},
  journal={IEEE Journal of Emerging and Selected Topics in Power Electronics}, 
  title={Optimal Design of Controller Parameters for Improving the Stability of MMC-HVDC for Wind Farm Integration}, 
  year={2018},
  volume={6},
  number={1},
  pages={40-53},
  doi={10.1109/JESTPE.2017.2759096}}

@ARTICLE{Heng2020impedance,
  author={Wu, Heng and Wang, Xiongfei and Kocewiak, Lukasz Hubert},
  journal={IEEE Journal of Emerging and Selected Topics in Power Electronics}, 
  title={Impedance-Based Stability Analysis of Voltage-Controlled MMCs Feeding Linear AC Systems}, 
  year={2020},
  volume={8},
  number={4},
  pages={4060-4074},
  doi={10.1109/JESTPE.2019.2911654}}

@ARTICLE{Luca2019effects,
  author={Bessegato, Luca and Ilves, Kalle and Harnefors, Lennart and Norrga, Staffan},
  journal={IEEE Transactions on Power Electronics}, 
  title={Effects of Control on the AC-Side Admittance of a Modular Multilevel Converter}, 
  year={2019},
  volume={34},
  number={8},
  pages={7206-7220},
  doi={10.1109/TPEL.2018.2878600}}
%

%





\vfill


\end{document}